%% file: main.tex
\documentclass{aa}  

\usepackage{graphicx}
\usepackage{hyperref}
\usepackage[varg]{txfonts}
\usepackage{caption}
\usepackage{subcaption} % For subfigures
\usepackage{multicol}

\usepackage{mathtools} % For order of symbol
\DeclarePairedDelimiterXPP\BigOSI[2]%
  {\mathcal{O}}{(}{)}{}%
  {\SI{#1}{#2}}

\usepackage{float}
\usepackage{amsmath}
\usepackage{xcolor}

\usepackage{pifont} % Itemize bulletpoint options

\usepackage{lipsum} 

\hypersetup{
  colorlinks=true,   %% links colored instead of frames
  urlcolor=blue,     %% external hyperlinks
  linkcolor=blue,     %% internal latex links (eg Fig)
  citecolor=blue
}

\makeatletter
\newcommand{\bibnote}[2]{\global\@namedef{#1note}{#2}}
\newcommand{\biblink}[2]{\global\@namedef{#1link}{#2}}
\makeatother

\begin{document}

   \title{Statistics of magnification for extremely lensed high redshift stars}

  % \subtitle{Un paper de PM}

   \author{J. M. Palencia    \thanks{palencia@ifca.unican.es}
          \inst{1}                     
          \and
          J. M. Diego\inst{1}
          \and
          B. J. Kavanagh\inst{1}
           \and
          J. Martinez\inst{2}
          }

   \institute{Instituto de Física de Cantabria (CSIC-UC), Avda. Los Castros s/n, 39005 Santander, Spain\\
%              \email{palencia@ifca.unican.es}
      \and
      Facultad de Ciencias. Universidad de Cantabria. Avda. Los Castros s/n, 39005 Santander, Spain\\
             }
%   \date{Received September 15, 1996; accepted March 16, 1997}

\abstract{Microlensing of stars in strongly lensed galaxies can lead to temporary extreme magnification factors ($\mu\!>\!1000$), enabling their detection at high redshifts.
Following the discovery of Icarus, several stars at cosmological distances ($z\!>\!1$) have been observed using the Hubble Space Telescope (HST) and the James Webb Space Telescope (JWST).
This emerging field of gravitational lensing holds promise to study individual high redshift stars. Also offers the opportunity to study the substructure in the lens plane with implications for dark matter models, as more lensed stars are detected and analysed statistically.
Due to the computational demands of simulating microlensing at large magnification factors, it is important to develop fast and accurate analytical approximations for the probability of magnification in such extreme scenarios.
In this study, we consider different macro-model magnification and microlensing surface mass density scenarios and study how the probability of extreme magnification factors depends on these factors.
To achieve this, we create state of the art large simulations of the microlensing effect in these scenarios.
Through the analysis of these simulations, we derive analytical scaling relationships that can bypass the need for expensive numerical simulations. Our results are useful to interpret current observations of stars at cosmic distances which are extremely magnified and under the influence of microlenses. A public code based on our results for analytically computed magnification PDFs can be found here: (The link to the repository and the code will be made public once the paper is accepted).}
 
%  \abstract


   \maketitle
%
%-------------------------------------------------------------------
\section{Introduction}  \label{sec:intro}
%%%%%%%%%%%%%%%%%%%%%%%%%%%%
\input{sections/introduction}

\section{Simple notions of gravitational lensing}  \label{sec:formalism}
%%%%%%%%%%%%%%%%%%%%%%%%%%%%%%%%%%%%%%%%%
\input{sections/lensing_intro}

\section{Numerical simulations}  \label{sec:sims}
%%%%%%%%%%%%%%%%%%%%%%%%%%%%%%%%%%%%%%%%
\input{sections/simulations}
%\dummy{1}

\section{Single microlens near a CC}  \label{sec:1lens}
%%%%%%%%%%%%%%%%%%%%%%%%%%%%%%%%%%%%%%%%%%%%%%
\input{sections/1lens}
%\dummy{2}

\section{Generalisation to $N$ microlenses}  \label{sec:Nlenses}
%%%%%%%%%%%%%%%%%%%%%%%%%%%%%%%%%%%%%%%%%%%
\input{sections/Nlenses}
%\dummy{3}

\section{Analytical modelling}  \label{sec:modeling}
%%%%%%%%%%%%%%%%%%%%%%%%%%%%%%%%%%%%%%%%%%%
\input{sections/modeling}
%\dummy{4}

\section{Constraining compact DM with CCE}  \label{sec:constrains}
%%%%%%%%%%%%%%%%%%%%%%%%%%%%
\input{sections/constrains}
%\dummy{8}

\section{Discussion}  \label{sec:discussion}
%%%%%%%%%%%%%%%%%%%%%%%%%%%%
\input{sections/discussion}
%\dummy{5}

\section{Conclusions}  \label{sec:conclusions}
\input{sections/conclusion}
%\dummy{6}

\begin{acknowledgements}
      JMP acknowledges financial support from the Formaci\'on de Personal Investigador (FPI) programme, ref. PRE2020-096261, associated to the Spanish Agencia Estatal de Investigaci\'on project MDM-2017-0765-20-2.
      J.M.D. acknowledges the support of project PGC2018-101814-B-100 (MCIU/AEI/MINECO/FEDER, UE) Ministerio de Ciencia, Investigaci\'on y Universidades. 
      BJK acknowledges funding from the Ram\'on y Cajal Grant RYC2021-034757-I, financed by MCIN/AEI/10.13039/501100011033 and by the European Union ``NextGenerationEU"/PRTR.
      We acknowledge Santander Supercomputacion support group at the University of Cantabria who provided access to the supercomputer Altamira Supercomputer at the Institute of Physics of Cantabria (IFCA-CSIC), member of the Spanish Supercomputing Network, for performing simulations.
\end{acknowledgements}

\begin{appendix}
\section{Resolution effects}  \label{app:resolution}
%%%%%%%%%%%%%%%%%%%%%%%%%%%%
\input{appendixes/resolution_effects}
% \todo{Tamaño pixel}

% \todo{Mass of the microlenses}

% \todo{En general no suponen un problema}

% \dummy{7}

\section{Sub-structure in the central peak}  \label{app:peaks}
%%%%%%%%%%%%%%%%%%%%%%%%%%%%
\input{appendixes/peaks}
%\todo{Lows $\Sigma_{\rm eff}s$ muestran una patron raro y una subestructura entorno a la moda}

% \dummy{8}

\end{appendix}


%\begin{thebibliography}{}
%
%   \bibitem[Zheng(1997)]{zheng} Zheng, W., Davidsen, A. F., Tytler, D. \& Kriss, G. A. 1997, preprint
%\end{thebibliography}

\bibliographystyle{aasjournal} %use aasjournal.bst
\bibliography{Biblio} % References in MyBiblio.bib run bibtex after latex/pdflatex 

\end{document}

%% file: sections/introduction.tex
The discovery of Icarus \citep{Kelly2018}, a strongly magnified star at redshift z=1.49, recently opened the door to a new branch of gravitational lensing; the study of distant stars at extreme magnification factors ($\mu\!>\!1000$).
Icarus was being magnified by a large factor (several hundreds) by a galaxy cluster. In addition to the magnification from the cluster, a microlens in the cluster (most likely a star) boosted the amplification by another factor $\mathcal{O}$(10), bringing the net magnification to several thousands.
This large magnification was maintained only for several weeks, but it allowed to identify Icarus as a varying source, and monitor it further with additional Hubble Space Telescope (HST) observations that allowed to confirm its nature.  
After Icarus, a series of blue luminous stars have been discovered with similar magnification factors with the HST \citep{Chen2019,Kaurov2019,Welch2022a,Diego2022a,Meena2022a}, culminating with the discovery of Earendel, the farthest star ever observed at $z\!\approx\!6$ \citep{Welch2022a, Welch2022b}.
More recently, the James Webb Space Telescope (JWST) has expanded this list with additional blue supergiants \citep{Chen2022,Meena2022b}, and the first red supergiants, which are most luminous in the infrared (IR) bands beyond the reach of HST \citep{Diego2022b}.
Dedicated programs with HST to search for these extremely magnified stars near cluster critical curves, and undergoing microlensing events, such as FLASHLIGHTS \citep{Flashlights2019}, have recently proven to be very successful with the discovery of over a dozen new stars at cosmic distances \citep{Kelly2022}.
New programs with JWST, such as PEARLS \citep{Windhorst2023}, are taking over, and are finding new examples of extremely magnified stars, specially colder ones (T $\!<\!10000$ K).
With the number of individual stars beyond redshift $z=1$ rapidly increasing, we are approaching the point where the statistical significance of such detections may be sufficient to discriminate between different models of, i) the underlying population of luminous stars at cosmic distances and ii) the  amount of small scale substructure in the lens plane that is responsible for the microlensing fluctuations. 

Future observations, especially with JWST, will soon increase the number of strongly lensed stars to $\mathcal{O}(100)$, allowing studies that constrain not only the nature of the distant star, but also the intervening population of microlenses.
Of particular interest is the study of the most distant lensed stars, since at large redshifts the critical curves of gravitational lenses move outwards, where the role of stellar microlenses is smaller, but the role of possible compact dark  matter candidates (like primordial black holes, or PBHs) is, in relative terms, greatest.
In interacting clusters, critical curves can form in regions in between the different cluster components, where the stars responsible for the intracluster light have the smallest lensing effects.
In addition, if the cluster is relatively at large redshifts, the intracluster medium is still in its early stages of formation resulting in a relative pristine medium (in terms of stellar microlensing contamination), and hence offering a unique opportunity to constrain models of compact dark matter.
High redshift galaxies that are strongly lensed produce counter images in regions relatively far away from the centre of galaxy clusters, and where the abundance of stellar microlenses is smallest compared with the contribution from dark matter. Interacting clusters offer a unique opportunity to constrain models of compact dark matter due to the lower contamination from stellar microlensing in the region between the two clusters, thus offering a clearer view of the DM component.
Such arcs exist already in the literature, for instance the giant arc at $z>2$ nicknamed La Flaca, strongly magnified by the merging galaxy cluster El Gordo at $z=0.87$ \citep{Diego2022b}.
At least one lensed star candidate has been identified in this arc.
If confirmed, its interpretation will demand an accurate modelling of the probability of magnification, on a wide range of scenarios that involve different contributions to the microlensing signal from compact dark matter models.

To extract useful information (with cosmological significance) from these events, it is necessary to understand the role that microlenses play in the magnification for objects discovered near the critical curves of gravitational lenses.
This has been done so far by means of simulating the lens plane after populating it with microlenses from the gravitational lens.
These microlenses include stars and remnants form the lens itself but also sometimes more exotic microlenses such as primordial black holes \citep{Diego2018}.
Given the extreme values of the magnification factors, a target simulated area $A$ in the source plane needs to be simulated first in an area a factor $\mu$ times larger in the image plane.
Combined with the fact that the source to be resolved (a star) is very small in nature, this usually results in simulations that are very computationally demanding.
If one wants to perform a statistical study of the detected stars, a wide range of magnifications and microlens parameters needs to be explored making this task a daunting one if performed through simulations.
A more efficient approach is to approximate the probability of magnification by analytical forms that are flexible and accurate enough so they can reproduce a wide range of scenarios.
The present work aims at finding these analytical forms which will set the basis for future studies on the probability of finding these stars.
This probability involves several ingredients such as the number density of luminous stars, the probability of having magnification factors greater than 100 (by a galaxy or galaxy cluster)\footnote{In general, all lensed stars involve magnification factors, $\mu$, from galaxies or clusters of galaxies above $\mu=100$.}, and the probability of magnification when microlenses are present in a region magnified over a factor 100 by a galaxy or galaxy cluster. This paper hence focuses on this last ingredient, and we leave the much more complex full calculation of the probability of observing a particular type of star and at a given redshift for future work. 
%In this work we derive analytical scalings for the probability of magnification that will allow in the future to bypass the use of costly numerical simulations.

One direct application of our work is the study of certain models of dark matter that could act as microlenses. Among these, primordial black holes (PBHs) have gained significant attention as a potential candidate for dark matter (see \cite{Green2021} for a recent review).
These black holes are believed to have formed through the gravitational collapse of mass overdensities in the early Universe, predating the epoch of matter-radiation equality. 
Importantly, PBHs are non-baryonic and did not play a role in primordial nucleosynthesis.
PBHs can potentially interact with baryonic matter through two mechanisms.
Firstly, the lightest PBHs could undergo evaporation via Hawking radiation, leading to the release of energy and particles.
Secondly, the heaviest PBHs could interact with baryons through the transfer of energy from their accretion disks.
PBHs with initial masses $M_{\rm PBH}\!\gtrsim\!10^{11}$~kg are expected to have completely evaporated by the present time \citep{MacGibbon2008}.
On the other hand massive PBHs ($M_{\rm PBH}\!\gtrsim\!100~M_\sun$) would have left an inprint on the Cosmic Microwave Background \citep{Serpico2020} or interact with their surroundings heating the medium \citep{Lu2021}.
In the range of few tens of $M_{odot}$, PBHs could offer a natural explanation for the abundance of massive black holes fund by gravitational wave experiments \citep{LIGO2021}.
If PBHs with these masses exist in relatively large numbers, they will contribute to the microlensing signal.
Current constraints based on microlensing signatures set an upper limit on the abundance of PBHs at the 1\% level of the total amount of dark matter.
Future observations of distant lensed stars are expected to lower this upper limit even further.

This paper is organised as follows.
In section (\ref{sec:formalism}) we describe the basic properties of microlensing embedded in a region of high magnification.
In Section (\ref{sec:sims}) we introduce the methodology to follow and present the simulation procedure.
Section (\ref{sec:1lens}) presents results for isolated lenses simulations and a study on how a simple parameter scale in terms of the macro and micro-model parameters.
Section (\ref{sec:Nlenses}) generalise this to a set of $N$ microlenses and present our numerical simulations.
We present in section (\ref{sec:modeling}) the modelling of the magnification probability.
In section (\ref{sec:constrains}) we develop the methodology to constraint compact dark matter with caustic crossing events (CCEs) using our newly obtained analytical tool.
Our results and some prospects on the use of the approximations derived here are discussed in section (\ref{sec:discussion}), and we conclude in section (\ref{sec:conclusions}).

Unless noted otherwise, we adopt a flat cosmology with $\Omega_m=0.3$, $\Lambda=0.7$ and $h=0.7$ (100 km s$^{-1}$ Mpc$^{-1}$).
We use the term macrolens or macromodel when we refer to a lens with the mass of a galaxy or a galaxy cluster. The term microlens or micromodel is used to refer to much smaller lenses with masses comparable to stellar masses.
The caustics associated to microlenses are referred to as microcaustics.  

%% file: sections/lensing_intro.tex
In this section we provide a brief overview of the formalism of gravitational lensing, a fundamental tool in studying many astrophysical phenomena. 
We focus specifically on the gravitational lensing effect caused by the perturbation of a set of point-like lenses located in a region of high magnification near a critical curve (or CC).
This regions of extreme magnification map into the source plane into similar regions of extreme magnification known as caustics. 
A source with infinitesimal size has divergent magnification when is place at the exact position of the caustic.
Stars are very small but have a limited size.
In this case, the maximum magnification scales as $1\!/\!\sqrt{R_{\ast}}$, where $R_{\ast}$ is the star radius. 

As mentioned in the introduction, we refer to the components of the microlensing, the lensing effects caused by the point-like lenses, with the prefix ``micro-'' (e.g., micro-images, micro-caustics, etc.). 
In contrast, we will use the prefix ``macro-'' for those effects associated with the galaxy cluster.

The position of a lensed image $\vec{\theta}$, and its actual position in the sky $\vec{\beta}$ are related through the lens equation (\cite{Schneider1992}, \cite{Narayan1996})
\begin{equation}    \label{eq:lens_eq}
    \vec{\beta} = \vec{\theta} - \vec{\alpha}(\Sigma,\vec{\theta}),
\end{equation}
where $\vec{\alpha}$ is the deflection angle at the position $\vec{\theta}$ produced by a lens of surface mass density $\Sigma(\vec{\theta})$.
As $\vec{\alpha}$ is a function of $\vec{\theta}$, Eq. (\ref{eq:lens_eq}) is generally non-linear leading to multiple solutions $\vec{\theta}$ for the same $\vec{\beta}$, and usually forbidding an analytical solution. 
The deflection angle can be obtained from the derivatives of the effective lensing potential
\begin{equation}    \label{eq:eff_potential}
    \psi(\vec{\theta}) = \frac{D_{\rm ds}}{D_{\rm d}D_{\rm s}}\frac{2}{c^2}\int\phi(D_{\rm d}\vec{\theta}, z)\,\rm{d} \textit{z},
\end{equation}
where $D_{\rm d}$, $D_{\rm s}$ and $D_{\rm ds}$ are the angular diameter distances to the lens, the source, and between the lens and the source, respectively, and $\phi$ is the Newtonian potential of the lens. The deflection angle is the gradient of $\psi$ with respect to $\vec{\theta}$
\begin{equation}
    \vec{\alpha}=\vec{\nabla}_{\vec{\theta}} \psi, %=D_{\rm d}\vec\nabla_{\xi} \psi = \frac{D_{\rm ds}}{D_{\rm s}}\frac{2}{c^2}\int\vec\nabla_{\perp} \phi\rm d z 
\end{equation}
%where $\vec\xi$ represents a 2D vector within the lens plane.
while its Laplacian is related to $\Sigma(\vec{\theta})$ as
\begin{equation}
    \vec{\nabla}_{\theta}^2 \psi = %\frac{D_{\rm d}D_{\rm ds}}{D_{\rm s}}\frac{2}{c^2}\int\vec\nabla_{\xi}^2 \phi\rm d z = 
    2\frac{D_{\rm d}D_{\rm ds}}{D_{\rm s}}\frac{4\pi G}{c^2}\Sigma(\vec{\theta}) = 2\frac{\Sigma(\vec{\theta})}{\Sigma_{\rm crit}}\equiv 2\kappa(\vec{\theta}),
\end{equation}
where $\kappa$ is know as the convergence and is the ratio between the surface mass density and the critical surface mass density, $\Sigma_{\rm crit}$.

The lensing distortion can be described by the Jacobian matrix
\begin{equation}    \label{eq:jacobian1}
    \tens A \equiv \frac{\partial\vec{\beta}}{\partial\vec{\theta}} = \left( \delta_{ij}-\frac{\partial\alpha_i(\vec{\theta})}{\partial\theta_j}\right) = \left(\delta_{ij}-\frac{\partial^2\psi(\vec{\theta})}{\partial\theta_i\partial\theta_j}\right) = \left(\delta_{ij}-\psi_{ij}\right) = \tens M^{-1},
\end{equation}
which is the inverse of the magnification tensor $\tens M$.
Since the convergence is half of the Laplacian of $\psi$, we have that
\begin{equation}
    \kappa(\vec{\theta}) = \frac{1}{2}(\psi_{11}+\psi_{22})=\frac{1}{2}\rm tr\, \psi_{ij}.
\end{equation}
Another important quantity is the shear tensor $\gamma(\vec{\theta})$, whose components can be also expressed as linear combinations of $\psi_{ij}$,
\begin{align}
    &\gamma_1(\vec{\theta}) = \frac{1}{2}(\psi_{11}-\psi_{22}),\\\nonumber
    &\gamma_2(\vec{\theta}) = \psi_{12} = \psi_{21},
\end{align}
the modulus of the shear tensor is $\gamma^2 = (\gamma_1^2 + \gamma_2^2)^{1/2}$, where $\gamma$ is known as the shear. We can rewrite the Jacobian matrix in terms of the convergence and the shear
\begin{equation}    \label{eq:jacobian2}
    \tens A = \begin{pmatrix}
                1-\kappa-\gamma_1 & -\gamma_2 \\
                -\gamma_2 & 1-\kappa+\gamma_1
              \end{pmatrix}.
\end{equation}

Gravitational lensing preserves the surface brightness of an object while altering its apparent size.
The magnification, which represents the ratio between the total observed flux of a lensed image and that of the unaltered image, is determined by the ratio between the solid angles of the image and the original, unlensed source.
The CC of a lens system refers to a collection of positions in the lens plane where the magnification diverges.
By utilising Eq. (\ref{eq:lens_eq}), we can trace the origins of these critical curves back to the source plane, thereby obtaining their corresponding complementary caustic curves.

Combining Eqs. (\ref{eq:jacobian1}, \ref{eq:jacobian2}) the magnification $\mu$ is
\begin{equation}
    \mu = \mu_{\rm t}\mu_{\rm r} = \rm det\, \tens M = \frac{1}{\rm det\, \tens A} = \frac{1}{(1-\kappa)^2-\gamma^2},
\end{equation}
where $\mu_{\rm t}$ and $\mu_{\rm r}$ are the tangential and radial components of the magnification, $\mu_{\rm t}^{-1} = 1-\kappa-\gamma$, and $\mu_{\rm r}^{-1} = 1-\kappa+\gamma$. 
Note that like $\tens A$ and $\tens M$, $\mu$ is a function of the position $\vec\theta$ in the lens plane.

Despite the non-linearity of Eq. (\ref{eq:lens_eq}), the deflection angle remains linear. Thus, we can express $\vec\alpha$ as the sum of the individual components of the lens system.
In our case of interest
\begin{equation}    \label{eq:alpha_tot}
    \vec\alpha = \sum_{i=0}^{\rm N} \vec\alpha_{\ast,\,i} + \vec\alpha_{\rm m} = \vec\alpha_{\ast} + \vec\alpha_{\rm m},
\end{equation}
where $\vec\alpha_{\ast, \, i}$ is the deflection angle caused by each of the $N$ point-like lenses, $\vec\alpha_{\ast}$ is their combined deflection angle, and $\vec\alpha_{\rm m}$ is the contribution from the galaxy cluster. 
For a point-like lens substituting the corresponding Newtonian potential into Eq. (\ref{eq:eff_potential}), the deflection angle for a lens of mass $M_i$ at the position $\vec\theta_i$ is
\begin{equation}    \label{eq:alpha_pl}
    \vec\alpha_{\ast, \, i} = \frac{D_{\rm ds}}{D_{\rm d}D_{\rm s}}\frac{4GM_i}{c^2} \frac{\vec\theta-\vec\theta_i}{|\vec\theta-\vec\theta_i|^2}.
\end{equation}
To obtain $\vec\alpha_{\rm m}$ we consider a region near the tangential CC of the cluster.
By definition a tangential CC forms where $1-\kappa-\gamma\!=\!0$ and $|1-\kappa+\gamma|\!>\!0$.
For simplicity we assume $\gamma_2\!=\!0$ and so $\gamma\!=\!\gamma_1$. This is equivalent to considering a portion of the CC where the deflection field is aligned in a direction perpendicular to the CC, which is a common situation in many real lenses.
The high magnification arises from the tangential component of the macro-magnification $\mu_{\rm t}\!\gg\!\mu_{\rm r}$.
Depending on which side of the CC we consider we might have a positive or negative parity.
The convergence and shear values of the macro-lens are given by
\begin{align}\label{eq:macro_conv_shear}
    &\gamma_{\rm m}=\frac{1}{2}\left(\frac{1}{\mu_{\rm r}} \mp \frac{1}{\mu_{\rm t}}\right),\\\nonumber
    &\kappa_{\rm m} = 1-\gamma_{\rm m} \mp \frac{1}{\mu_{\rm t}},
\end{align}
where the minus and plus sign refer to the positive and negative parities respectively.
If $(1-\kappa)^2-\gamma^2\!<\!0$ the parity is positive, otherwise if this factor is negative we are in a negative parity region.
This subtle change in the form of the convergence and the shear allows for completely different phenomena occurring at each parity.
The deflection angle caused by a smooth potential such as that of the macrolens increases linearly as we approach the macro-CC, thus we have that the deflection angles from the macro model in the tangential and radial directions is
\begin{align}\label{eq:alpha_macro}
    &\alpha_{\rm m,\,t} \propto (\tilde\kappa+\gamma_{\rm m}),\\\nonumber
    &\alpha_{\rm m,\,r} \propto (\tilde\kappa-\gamma_{\rm m}).
\end{align}
It is important to note that we have not used the macro-convergence but its modified value $\tilde\kappa\!=\!\kappa_{\rm m}-\kappa_{\ast}$, taking into account the modification of the point-like lenses on the macro-potential \citep{Diego2018}.
As the point-like lenses are isotropically (and randomly) distributed they do not contribute a shear, so there is no need for a correction in the macro-shear.
Including Eqs. (\ref{eq:alpha_pl}, \ref{eq:alpha_macro}) in Eq. (\ref{eq:alpha_tot}) we obtain the total deflection angle, thus allowing us to solve numerically any lens problem related to our given configuration.

%% file: sections/simulations.tex
Extreme magnification near a CC gives rise to very crowded source planes. The microlens surface mass density, $\Sigma_\ast$, is effectively re-scaled by the large macromodel magnification $\mu$, resulting in a large effective surface mass density $\Sigma_{\rm eff}=\mu_{\rm t}\Sigma_\ast$.
The value of the effective surface mass density often surpasses the critical density, leading to optically thick lenses where continuous microlensing events occur.
Earlier work has relied on expensive simulations to study the statistical properties of the observed flux \citep{Diego2018,Venumadhav2017}. 

To approximate the changes in flux properties over time for an extremely lensed source, caused by a specific microlens population with surface mass density $\Sigma_{\ast}$ at a given distance from a tangential macro-CC ($\mu_m\!=\!\mu_{\rm t}\mu_{\rm r}$), a series of simulations must be conducted by varying these parameters.
Subsequently, the $\rm d(\rm{area})/\rm d\mu$ or probability distribution function of the total magnification needs to be modelled, which will be discussed in more detail in Section (\ref{sec:modeling}).
The model function's parameters will be studied in relation to the physical parameters, establishing their underlying relation.
Once this step is completed, it becomes possible to bypass the numerical simulations that currently consume the majority of time in these analysis, thereby eliminating this  bottleneck.
Previous works have obtained analytical approximations for the magnifications (\cite{Deguchi1987}; \cite{Seitz1994a}; \cite{Seitz1994b}; \cite{Neindorf2003}; \cite{Tuntsov2004}).
However, these methods fail at large macro-magnifications and high surface mass densities, where this work is focused.
Thus the need to derive analytical methods in this regime.

The computation of magnification in the source plane is accomplished using the inverse ray shooting technique \citep{Kayser1986}.
This method uses the lens equation as shown in Eq. (\ref{eq:lens_eq}) to trace back an image observed in the lens plane to its original position in the source plane.
By discretising the lens plane over a sufficiently large area and performing this tracing, a reasonably accurate approximation of the magnification in the source plane can be obtained.
The accuracy of this magnification probability with respect to pixel size and other resolution effects is extensively discussed in Appendix (\ref{app:resolution}).

In order to utilise the inverse ray shooting technique effectively, it is necessary to determine the deflection angle at each pixel of the simulated lens plane, considering both the contributions from the micro and macro components.
Initially, we compute the deflection angle generated by the microlensing population, denoted as $\vec\alpha_{\ast}$.
To accomplish this, we randomly distribute a set of point-like lenses within a circular region with a radius of $N_{\rm x}$, where $N_{\rm x}$ corresponds to the size in pixels of our simulated lens plane in the tangential direction of the macro-magnification.
The masses of these lenses are randomly picked from a Spera initial–final mass function \citep{Spera2015}, with the constraint that they do not exceed 1.5$M_\sun$.
We only compute the deflection angle from the lenses within a circle of radius $N_{\rm x}$
By utilising Eq. (\ref{eq:alpha_pl}) and considering the position and mass of each microlens, we can calculate the corresponding deflection angle $\vec\alpha_{\ast, , i}(\vec\theta)$ for each individual microlens.
Next, we proceed to calculate the corresponding deflection angle associated with the galaxy cluster $\vec\alpha_{\rm m}(\vec\theta)$.
To begin, we calculate the convergence and shear of the macro-model based on the macro-magnification, as described in Eq. (\ref{eq:macro_conv_shear}).
The specific choice of parity will depend on the simulation being conducted.
Subsequently, we proceed to correct the convergence, $\kappa$, by considering the influence of the microlenses,that is the effective convergence is $\kappa_{\rm eff}=\kappa-\kappa_{*}$.
Although this correction may be negligible in cases where the surface mass density of microlenses is small, it becomes crucial in crowded regions characterised by large stellar populations.
In such scenarios, the effect of microlenses becomes significant and must be accounted for in the analysis.
Using Eqs. (\ref{eq:alpha_macro}) and (\ref{eq:alpha_tot}), we can proceed to calculate the total deflection angle at each position required for the inverse ray shooting method.
These equations allow us to combine the contributions from the macromodel and the microlenses to obtain the complete deflection angle necessary for the analysis.
Since the total deflection angle is the sum of independent contributions from each component, we can exploit this property by utilising the same deflection angle obtained from a set of microlenses with different macro-model configurations.
This approach allows us to save both time and computational resources by avoiding redundant calculations of the microlens deflection angles for each specific macro-model configuration.
In this study, we conducted simulations using a computational grid consisting of approximately $N_{\rm x}\!\approx\!3 \times10^6$ pixels in the tangential direction and $N_{\rm y}\!\approx\!0.15 \times10^6$ pixels in the radial direction. This area gets remapped into a region $(3 \times10^6/\mu_t)\times(0.15 \times10^6/\mu_r)$ in the source plane. Each pixel in the simulation corresponds to a size of $2~$nas.
Consequently, the total size of the simulation area amounted to $6\times0.3$~mas$^2$, with a lens plane redshift of $z_{\rm d}\!=\!0.7$ and a source plane redshift of $z_{\rm s}\!=\!1.3$.
For this particular system, the critical surface mass density was estimated to be approximately $\Sigma_{\rm crit}\!\approx\!3.1\times 10^3 M_\sun/\rm pc^2$.
We conducted 39 distinct simulations for each parity, where we varied the values of $\Sigma_{\ast}$, $\mu_{\rm t}$, and $\mu_{\rm r}$.
As a result, we obtained a range of effective surface mass densities spanning from 0.01 to 66.5 times the critical density.
Furthermore, in order to gain a deeper understanding of the statistics related to the negative parity regime and investigate additional effects associated with low effective surface mass densities, we conducted a few additional realisations alongside the main simulations.
On average, it takes 32 CPU-hours to simulate the deflection angle for a single microlens of the aforementioned size and resolution that occupies 1.4TB of disk space.
Complex simulations with thousands of microlenses can reach hundreds of thousands of CPU-hours.
Detailed explanations of these supplementary simulations can be found in Section (\ref{sec:modeling}) and Appendix (\ref{app:peaks}).
As far as we are aware, these state-of-the-art simulations represent the largest scale simulations conducted at this resolution level for the purpose of studying microlensing by compact lenses around a cluster-CC.
The substantial size of the simulations provides us with a significant amount of statistical data, enabling a comprehensive analysis of the probability of magnification for background sources across a wide range of values.

To simplify the modelling of magnification probabilities and ease a direct comparison between different simulations, the final magnification is normalised by a factor of $1000/\mu_{\rm m}$. Unless specifically stated otherwise, this normalisation has been consistently applied throughout our analysis. As a result, a pixel in our simulations with a magnification value of $\mu$ will now be represented as $\tilde\mu\!=\!1000\times\mu/\mu_{\rm m}$. The PDF of the magnification after being re-scaled this way peaks around values close to $\mu=1000$, but not exactly at this value as we describe below. For ease of notation, we will continue to refer to the re-scaled magnification as $\mu$ in our discussions.

The inverse ray shooting technique is a computationally intensive numerical method that relies on the total number of pixels (a combination of the simulated area and pixel size) and the number of microlenses.
Simulating scenarios with high surface mass densities can be exceptionally expensive, necessitating the development of tools that offer an alternative to the costly numerical simulations.
The main objective of this work is to develop and present a tool that serves as a shortcut from the lens parameters to the magnification probability. 
Providing an alternative approach that eliminates the requirement for time-consuming and computationally expensive simulations.
By offering this tool, researchers can efficiently estimate the magnification probability based on the lens parameters, streamlining their analyses and reducing the reliance on extensive numerical simulations.

It is crucial to emphasise that despite the development of shortcut tools, the spatial distribution of the micro-caustic network remains an important factor in many analyses.
Therefore, numerical simulations continue to play a vital role in understanding and studying the intricate details of the micro-caustic structure.

To facilitate the development of this work, a parallelised version of the inverse ray shooting technique was employed, thus allowing a fully exploit of the resources of the computer cluster ALTAMIRA.

%% file: sections/1lens.tex
\begin{figure}[tpb]
  \centering
  \includegraphics[width=\columnwidth]{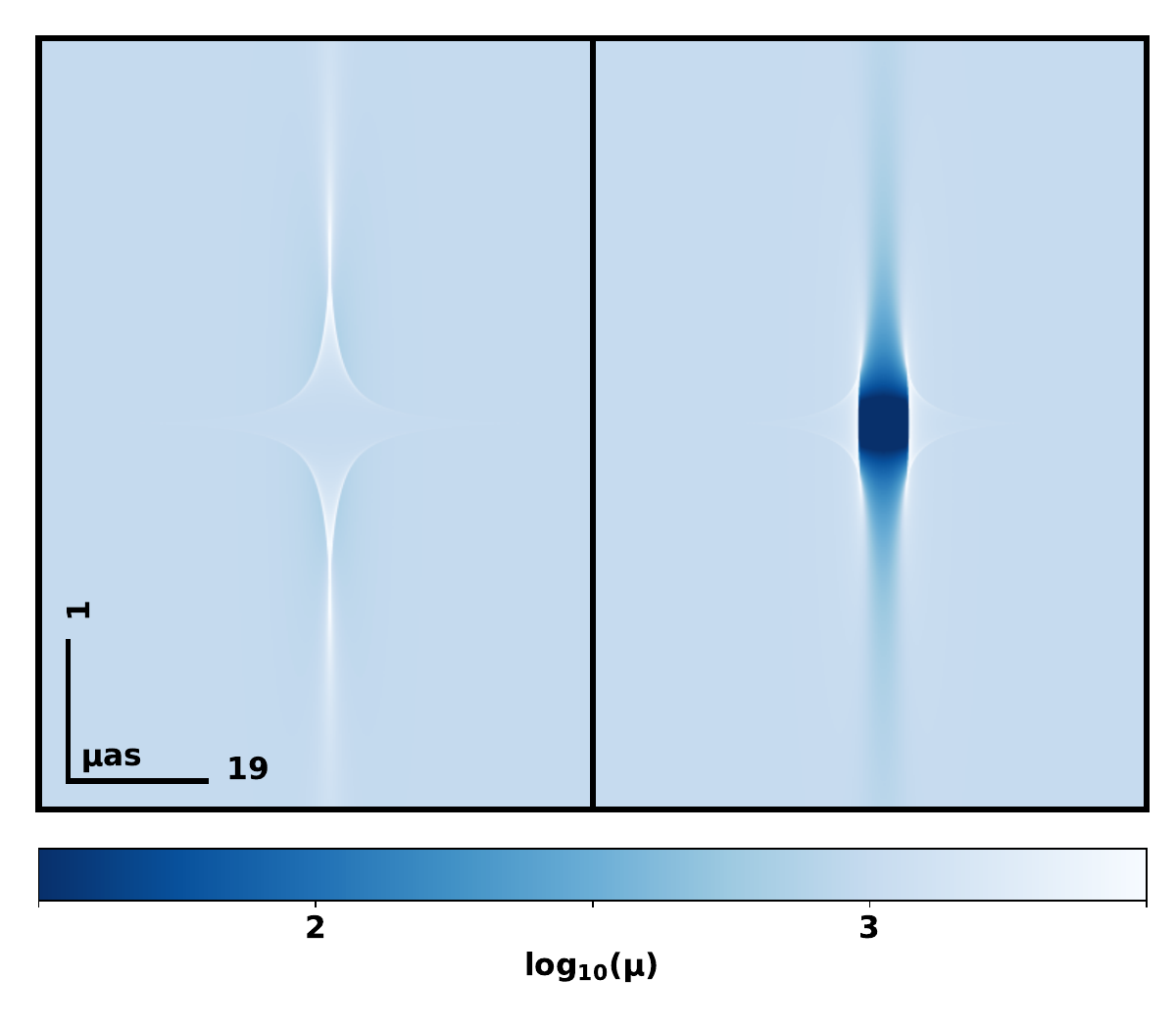}
  \caption[]{\label{fig:1lens} %
  Magnification in the source plane of a single microlens near a macro-CC where $\mu_{\rm m}\!=\!1000$.
  The darkest blue indicates a lower magnification, while the whites show where the magnification diverges (i.e. the micro-caustics).
  {\em Left:\/} Single microlens on the positive parity side of the CC.
  The caustics have the familiar diamond shape.  
  {\em Right:\/} The same, but in the negative parity region, crossing the macro CC.
  Now the shape of the micro-caustics is broken into two distorted triangles with a demagnification area in between, but increasing the maximum achievable magnification at the tips of these triangles.
  Outside the caustics, the average magnification is that of the macro model $\mu_{\rm m}$.
  The vertical direction has been scaled down a factor $\sim$ 10 for better visualisation.
  }
\end{figure}

The primary objective of this work is to model the statistical properties of the flux emitted by a background source as it traverses the micro-caustic network created by microlenses, such as stars from the intracluster light (ICL) and compact dark matter candidates like PBHs, in a region of high magnification near a CC associated with a galaxy cluster.
This problem is inherently complex due to the non-linear nature of the lens equation.
To gain insights into the flux statistics arising from the perturbation of the galaxy cluster's smooth potential by numerous microlenses, it is beneficial to analyse the case of an isolated microlens near a CC.
This analysis provides valuable information for scenarios characterised by low optical depths, where a microlens can be treated as an isolated entity.
This was the case for Icarus and its counter image Iapyx \citep{Kelly2018}, and generally applies for images not too close to the macro-CC and within regions of low $\Sigma_{\ast}$ ($\Sigma_{\rm eff}\!\ll\!\Sigma_{\rm crit}$).
In addition, the single microlens case can be used to test limitations of the numerical simulation, such as the limited spatial resolution, or pixel size. 

\begin{figure}[tpb]
  \centering
  \includegraphics[width=\columnwidth]{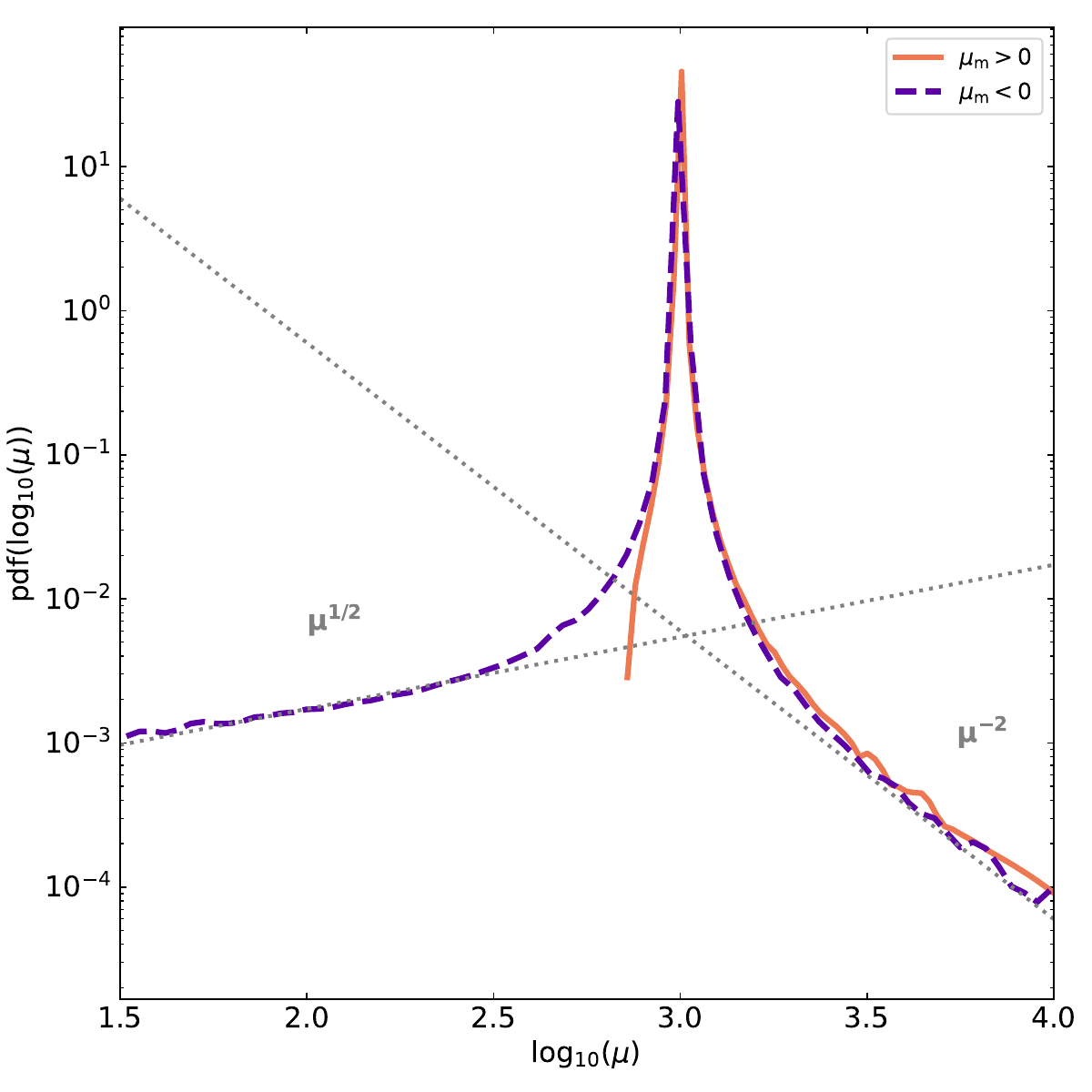}
  \caption[]{\label{fig:1lens_pdf} %
  Probability of magnification (in log$_{10}$ bins) for the two single lenses.
  The purple solid line represents the microlens on the negative parity side of the macro-CC, while the orange solid line represents the positive parity case on the right and left panels of Fig. (\ref{fig:1lens}), respectively.
  The dotted lines represents the power-laws that always appear for isolated microlenses near macro-CCs, and for N microlenses with low $\Sigma_{\rm eff}$.
  The variation in the value of the $\mu^{-2}$ power-law amplitude is shown in Fig. (\ref{fig:1_2_lenses_scalings}).
  }
\end{figure}

The outcome of the inverse ray shooting technique applied to a single microlens situated near a macro-CC is presented in Fig. (\ref{fig:1lens}).
This illustration showcases the results for both positive and negative parities. 
To exemplify the magnification probability for a single microlens within positive and negative macro-model regions, we refer to Fig. (\ref{fig:1lens_pdf}). 

This visualisation offers insights into several key observations.
Firstly, the size of the critical curves is scaled down by factors $\mu_{\rm t}$ and $\mu_{\rm r}$ in the tangential and radial directions, respectively (refer to \citep{Oguri2018} for further details). Furthermore, both parities exhibit a prominent peak in the probability of magnification around the value of the macro-model magnification $\mu_{\rm m}$. As magnification increases beyond this peak, the probability decreases following a power-law scaling of $\mu^{-3}$ (or $\mu^{-2}$ if logarithmic bins are considered, as in our analysis).
This behaviour is well-established in the literature (see \citep{Rauch1991}) and serves as a means to assess the resolution of our simulations (See Appendix \ref{app:resolution}).
When the probability of magnification no longer adheres to this power-law decline, it indicates the limit of our simulation's resolution.
Additionally, the negative parity introduces a demagnification relative to $\mu_{\rm m}$, with the probability of demagnification exhibiting a decreasing slope of -1/2 for lower magnifications.
These findings shed light on the behaviour and characteristics of magnification probabilities associated with a single microlens near a macro-CC at both positive and negative parities.

\begin{figure*}[tpb]
\centering
   \includegraphics[width=17cm]{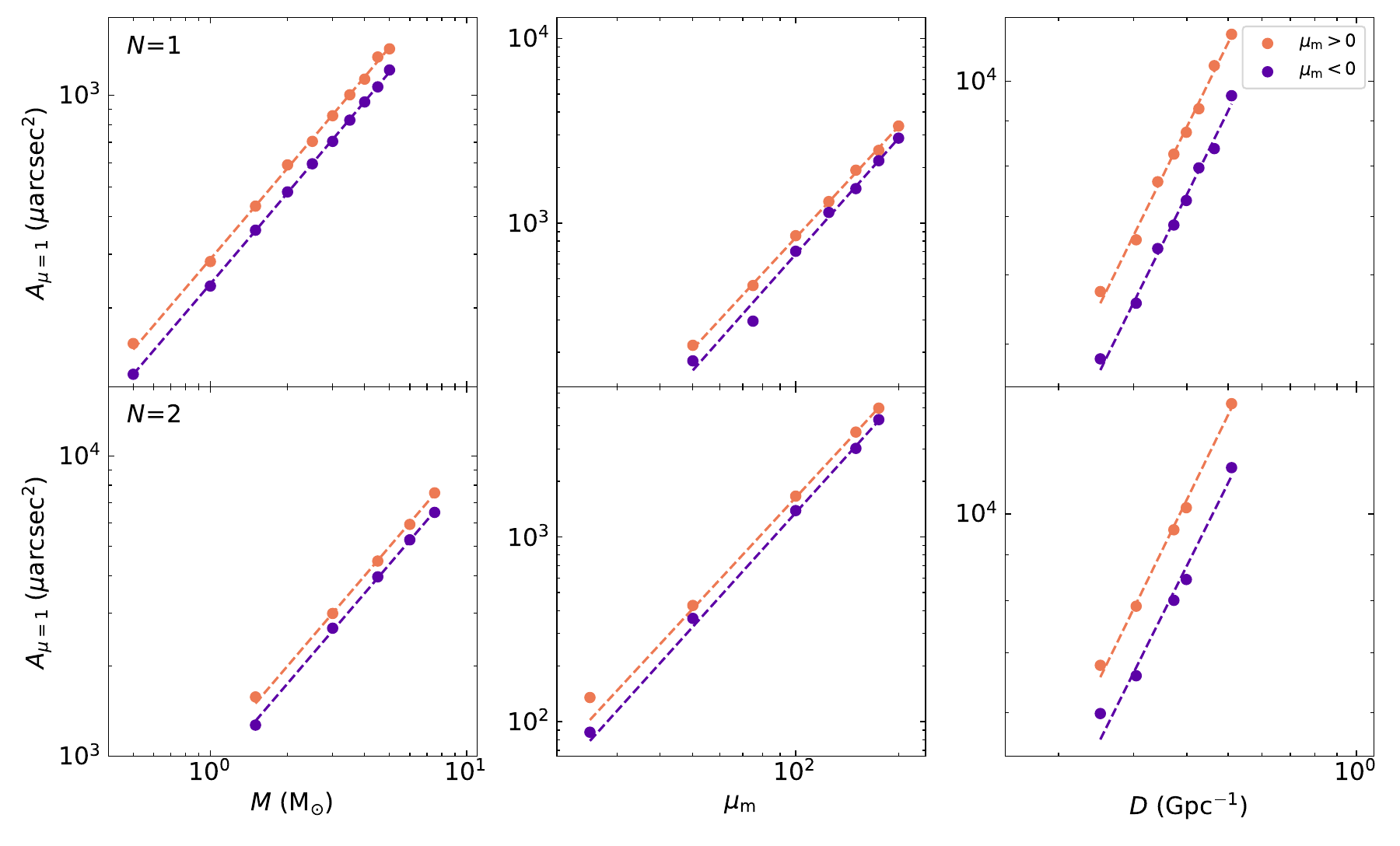}
     \caption{Extrapolated value of the power-law $\propto \mu^{-2}$ at $\mu=1$ from the probability distribution function of a system of $N$ microlenses in the vicinity of a macro CC. The case of one microlens is shown in the top row. The case for two microlenses is shown in the bottom row.
     We have changed the total mass of the microlenses, the magnification from the macro model and the reduced distance in the first, second and third columns respectively, while keeping the other parameters fixed.
     Note that for both systems the amplitude scales at the same rate for each parameter.
     All these lens systems have $\Sigma_{\rm eff}\!\ll\!\Sigma_{\rm crit}$.}
     \label{fig:1_2_lenses_scalings} %
\end{figure*}

This straightforward scenario provides valuable insights into the impact of various parameters on the magnification probability.
Parameters such as the lens mass, the macro-model magnification, and the reduced distance $D=D_{\rm ds}/(D_{\rm s}D_{\rm d})$ play crucial roles in shaping the magnification probability.
To examine these effects, we perform a fitting procedure to the power-law function, $\mu^{-2}$, on the right tail of the PDF of the magnification, and investigate the relationship between the amplitude of the power-law and these variables.
The results of this analysis are illustrated in Fig. (\ref{fig:1_2_lenses_scalings}).
The amplitude of the power-law (dotted lines in Fig. (\ref{fig:1_2_lenses_scalings}) increases cuadratically with the mass and the macromagnification, and it increases linearly with the reduced distance.

% \begin{figure}[tpb]
%   \centering
%   \includegraphics[width=.8\columnwidth]{figures/mu_scaling_1lens.pdf}
%   \caption[]{\label{fig:basic_scaling_1lens_mu} %
%   Extrapolated value of the power-law $\propto \mu^{-2}$ at $\mu=1$ from the probability distribution function of a single microlens near a macro-CC varying the magnification from the macro-model.
%   }
% \end{figure}

% \begin{figure}[tpb]
%   \centering
%   \includegraphics[width=.8\columnwidth]{figures/mass_scaling_1lens.pdf}
%   \caption[]{\label{fig:basic_scaling_1lens_mass} %
%   Extrapolated value of the power-law $\propto \mu^{-2}$ at $\mu=1$ from the probability distribution function of a single microlens near a macro-CC varying the mass of such lens.
%   }
% \end{figure}

% \begin{figure}[tpb]
%   \centering
%   \includegraphics[width=.8\columnwidth]{figures/mu_distance_1lens.pdf}
%   \caption[]{\label{fig:basic_scaling_1lens_D} %
%   Extrapolated value of the power-law $\propto \mu^{-2}$ at $\mu=1$ from the probability distribution function of a single micro-ens near a macro-CC varying the reduced distance of the lens system.
%   }
% \end{figure}

\subsection{Double microlens case}
%%%%%%%%%%%%%%%%%%%%%%%%%%%%%%%%%%%
Before delving into the realistic scenario of a distribution of $N$ microlenses, it is insightful to examine the case of two microlenses.
This analysis will elucidate the fundamental scalings that will be further explored and elucidated throughout the remainder of this paper.
We consider a second microlens with half the mass of the first microlens, centred at a distance equal to half the size of the original microcaustic in each direction, exactly on top of the microcaustic, maximising the overlapping (If placed closer it would behave as a single microlens whose mass is the combination of each sub-mass.

The scenario of two nearby microlenses in a high magnification region exhibits similarities to the single lens scenario.
The linearity shown in Figure~\ref{fig:1_2_lenses_scalings} persists even for larger number of microlenses as long as the effective surface mass density remains below  0.1$~\Sigma_{\rm crit}$.
However, once the micro-caustics start to overlap, resulting in $\Sigma_{\rm eff}$ values beyond this threshold, the magnification probability undergoes a transformation.
At this point, the power-law decrease in the probability for higher magnifications vanishes, and a log-normal behaviour starts to emerge. In the most extreme scenarios of $\Sigma_{\rm eff}>>1$, the log-normal shape of the PDF of the magnification gets narrower the larger the value of $\Sigma_{\rm eff}$.  
%This log-normal behaviour further reduces the probability of higher magnifications.
This phenomenon, known as the ``more is less effect'', has been previously observed and studied (see \citep{Welch2022a}, \citep{Diego2018}) and is studied in more detail in Section (ref{sec:modeling}).
The scalings of the probability of magnification are exactly the same as the ones observed for the case of a single microlens.
These scalings are shown in the bottom row of Fig. (\ref{fig:1_2_lenses_scalings}).

% \begin{figure}[tpb]
%   \centering
%   \includegraphics[width=.8\columnwidth]{figures/mu_scaling_2lens.pdf}
%   \caption[]{\label{fig:basic_scaling_2lens_mu} %
%   Extrapolated value of the power-law $\propto \mu^{-2}$ at $\mu=1$ from the probability distribution function of two nearby microlenses near a macro-CC varying the magnification from the macro-model.
%   }
% \end{figure}

% \begin{figure}[tpb]
%   \centering
%   \includegraphics[width=.8\columnwidth]{figures/mass_scaling_2lens.pdf}
%   \caption[]{\label{fig:basic_scaling_2lens_mass} %
%   Extrapolated value of the power-law $\propto \mu^{-2}$ at $\mu=1$ from the probability distribution function of a single microlens varying the mass of such lens.
%   }
% \end{figure}

% \begin{figure}[tpb]
%   \centering
%   \includegraphics[width=.8\columnwidth]{figures/mu_distance_2lens.pdf}
%   \caption[]{\label{fig:basic_scaling_2lens_D} %
%   Extrapolated value of the power-law $\propto \mu^{-2}$ at $\mu=1$ from the probability distribution function of a single microlens varying the reduced distance of the lens system.
%   }
% \end{figure}

This simple test demonstrates that increasing the mass of the lens/lenses, increasing the magnification of the macro-model (i.e., reducing the distance to the macro-CC), or increasing the reduced distance all lead to an increase in the probability of magnification at higher values.

Building upon these insights, a similar procedure is applied to the general case of $N$ microlenses in Section (\ref{sec:modeling}), which is the main objective of this work.

%% file: sections/Nlenses.tex
Here we consider a more realistic scenario that is more likely to be found in real observations.
In this scenario, we consider a large population of microlenses with different surface mass densities and macro-model magnifications.
The simulations consist of nearly 90 cases with different effective surface mass densities and parities, and the computational resources required for these simulations amount to approximately $0.5 \times 10^6$ CPU-hours.

\begin{figure}[tpb]
  \centering
  \includegraphics[width=\columnwidth]{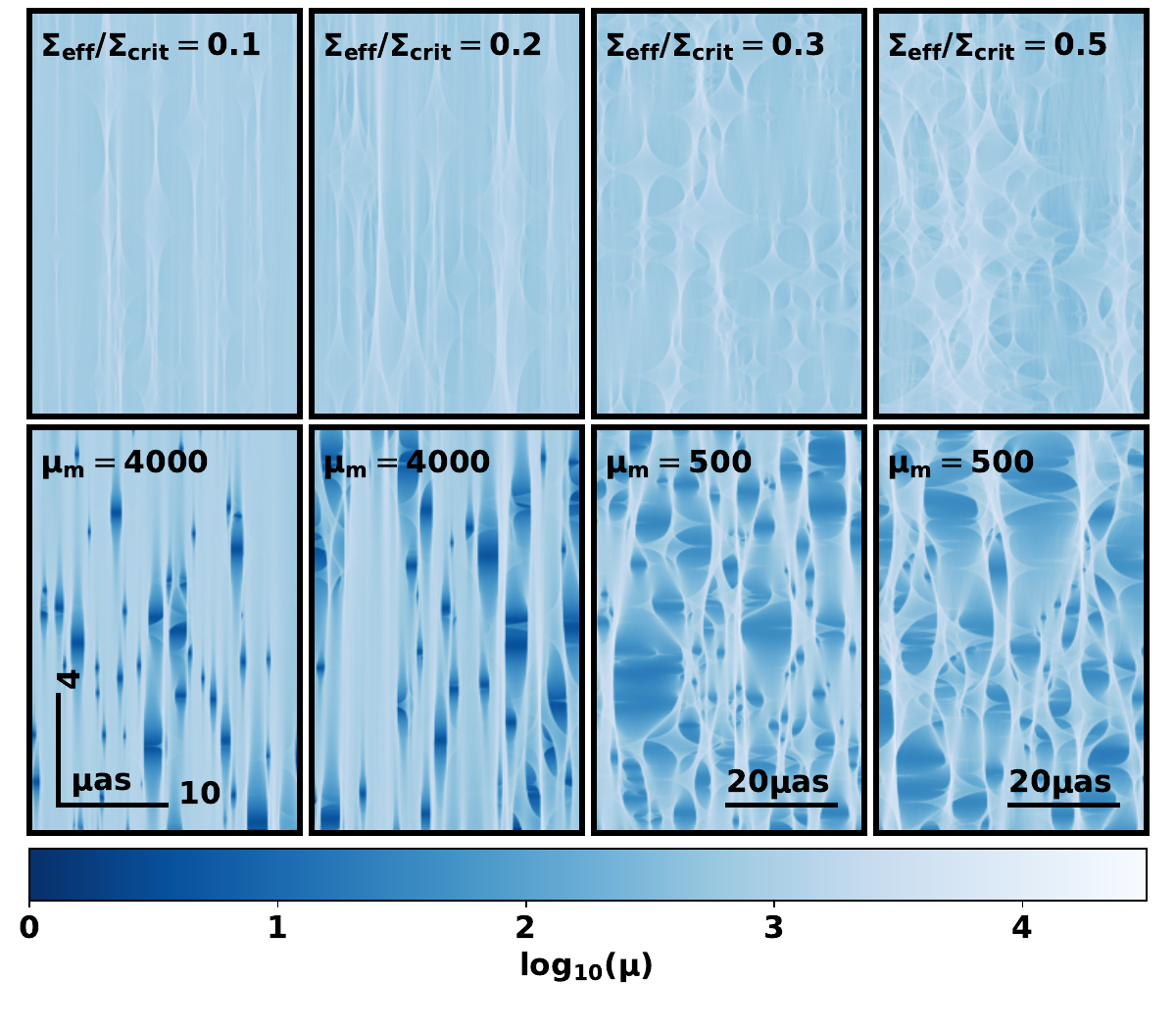}
  \caption[]{\label{fig:caustics_low_sigma} %
  High resolution view of the magnification at the source plane.
  {\em Top panels: \/} Positive parity side of the macro CC.
  {\em Bottom panels: \/} Negative parity side of the macro CC.
  The effective surface mass density increases from left to right.
  The value of the macro magnification and the ratio of the effective surface density to the critical value are shown for each column.
  Note that the y-direction has been compressed a factor 10 for better visualisation.
  For the same reason, the two rightmost columns have been compressed a factor 2 in the x-direction.
  }
\end{figure}

A significant finding is that the magnification statistics exhibit distinct behaviours depending on whether the ratio $\Sigma_{\rm eff}/\Sigma_{\rm crit}$ is smaller or greater than one.
We refer to these cases as the low surface mass density and medium-high surface mass density scenarios, respectively.
These two cases display different characteristics in terms of the magnification statistics.
Figs. (\ref{fig:caustics_low_sigma}) and (\ref{fig:caustics_high_sigma}) show some of the simulated magnifications at the source plane for each surface mass density regime, low and high respectively, increasing the surface mass density towards the right at each parity (column).

To increase the effective surface mass density, there are two approaches.
Firstly, one can increase the surface mass density of the microlenses by either including more microlenses or by having more massive ones.
This results in larger micro-caustics, which in turn contribute to an increase in $\Sigma_{\rm eff}$.
Secondly, one can approach the macro-CC more closely, effectively increasing the macro-magnification.
This mapping of micro-caustics into a smaller region enhances the likelihood of overlap between micro-caustics, thereby increasing the probability of a microlensing event due to caustic crossings.

The overlapping micro-caustics distorts the probability of magnification in the high-end. A simple way to understand this is by realising that locally, each microlens can perturb the magnification acting over nearby microlenses, perturbing the magnification from the macromodel. Alternatively, the maximum magnification is inversely proportional to the smoothness of the deflection field. Very corrugated deflection fields naturally result in smaller maximum magnifications. 
In the negative parity case, there is still a tail of demagnification, but as the surface mass density increases, this tail disappears, and the magnification probability becomes similar to that of the positive case. This effect is evident in Fig. (\ref{fig:caustics_high_sigma}), where for an effective surface mass density 48 times larger than the critical value, the micro-caustics for the negative parity are indistinguishable from those of the positive case. In these high-density scenarios, the areas of demagnification have been completely eliminated. 
This more is less effect leads to a decrease in the probability of higher magnifications.
The effects of micro-caustic overlap and the more is less effect can be observed in Figs. (\ref{fig:caustics_low_sigma}) and (\ref{fig:caustics_high_sigma}).

\subsection{Low surface mass density}
%%%%%%%%%%%%%%%%%%%%%%%%%%%%%%%%%%%
% \begin{figure}[tpb]
%   \centering
%   \includegraphics[width=\columnwidth]{figures/pdfs_low_pos.pdf}
%   \caption[]{\label{fig:pdfs_low_sigma_pos} %
%   Probability of magnification for a set of simulations with positive parity and different $\Sigma_{\rm eff}$ in the low surface mass density regime.
%   }
% \end{figure}
% \begin{figure}[tpb]
%   \centering
%   \includegraphics[width=\columnwidth]{figures/pdfs_low_neg.pdf}
%   \caption[]{\label{fig:pdfs_low_sigma_neg} %
%   Probability of magnification for a set of simulations with negative parity and different $\Sigma_{\rm eff}$ in the low surface mass density regime.
%   }
% \end{figure}
\begin{figure*}
\centering
   \includegraphics[width=17cm]{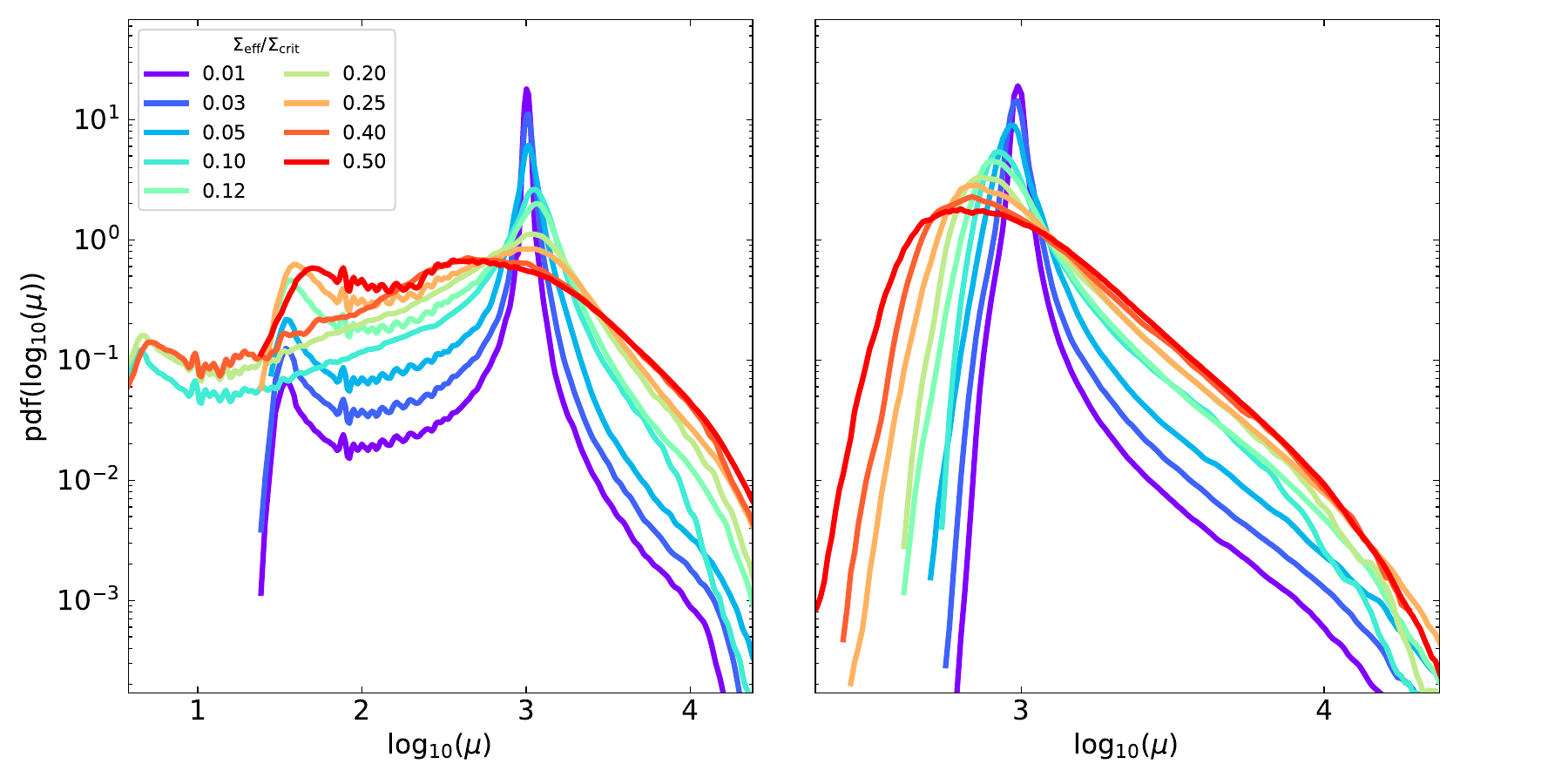}
     \caption{Probability of magnification for a set of simulations with different $\Sigma_{\rm eff}$ in the low surface mass density regime.
     {\em Left panel: \/} Negative parity side of the macro-CC.
     {\em Right panel: \/} Positive parity side of the macro-CC.}
     \label{fig:pdfs_low_sigma}
\end{figure*}

In the low surface mass density case, where the combined effect of the macro-model and microlenses does not exceed the critical surface mass density, the micro-caustics do not heavily overlap.
As a result, they maintain a similar shape to the isolated microlens case, and the magnification probability resembles that shown in Fig. (\ref{fig:1lens_pdf}).
As the effective surface mass density increases, both positive and negative parity cases exhibit a broadening of the main peak and a shift towards smaller magnification values.
Additionally, the power-law tails of the probability distributions increase in amplitude and begin to transition to a log-normal distribution at smaller magnifications due to the increased overlap of micro-caustics.
In the case of the negative parity, there is an excess probability of lower magnifications compared to the macro-magnification value.
This demagnification effect follows a power-law distribution, with an increasing amplitude and a broadening peak structure as the surface mass density increases.
This behaviours can be observed in Fig. (\ref{fig:pdfs_low_sigma}) for the positive (left panel) and negative (right panel) macro-magnification parity cases.

\subsection{Medium-high surface mass density}
%%%%%%%%%%%%%%%%%%%%%%%%%%%%%%%%%%%
\begin{figure}[tpb]
  \centering
  \includegraphics[width=\columnwidth]{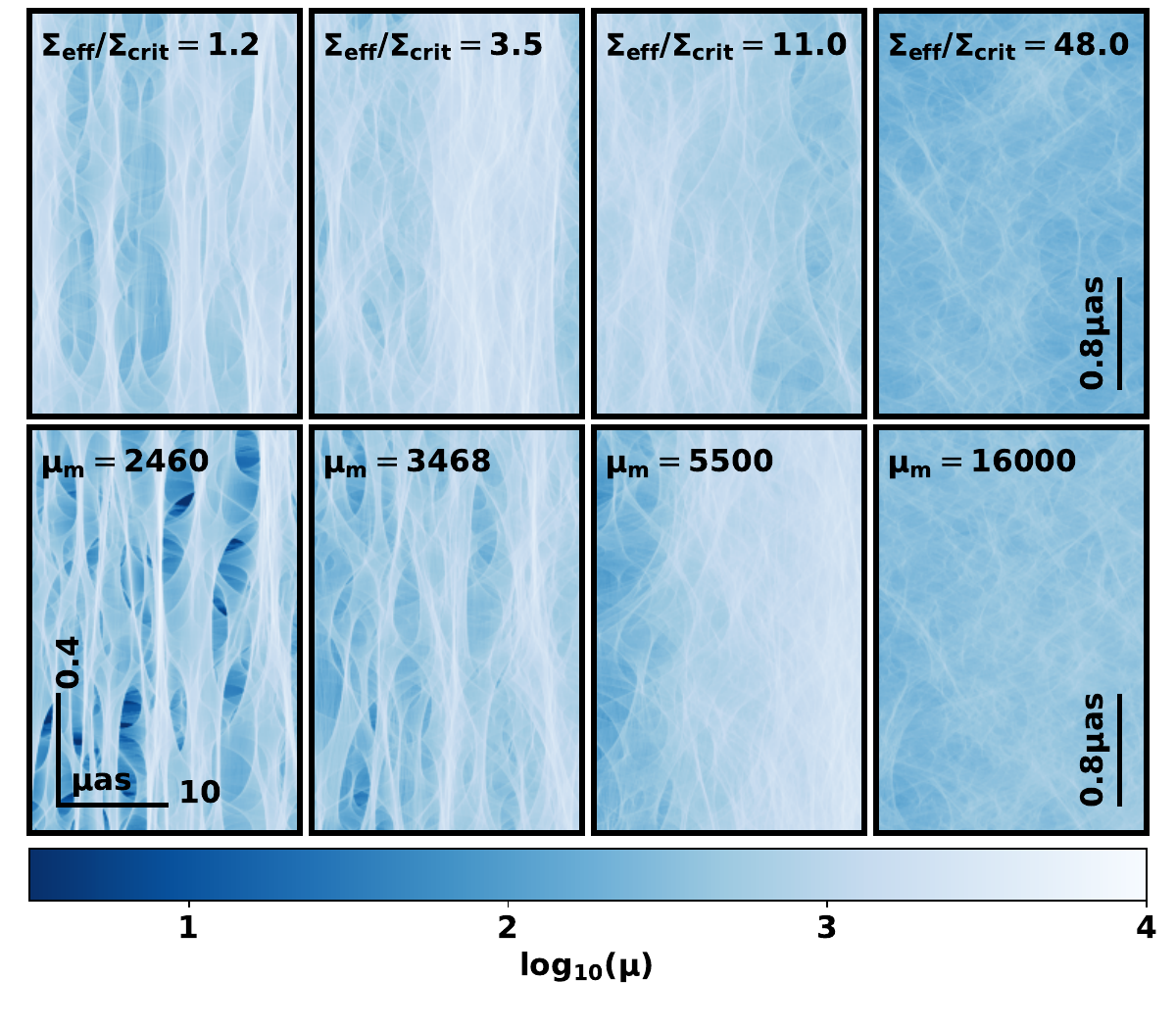}
  \caption[]{\label{fig:caustics_high_sigma} %
  High resolution view of the magnification at the source plane.
  {\em Top panels: \/} Positive parity side of the macro CC.
  {\em Bottom panels: \/} Negative parity side of the macro CC.
  The effective surface mass density increases from left to right.
  The value of the macro magnification and the ratio of the effective surface density to the critical value are shown for each column.
  Note that the y-direction has been compressed a factor 25 (50 for the last column) for better display.
  }
\end{figure}
% \begin{figure}[tpb]
%   \centering
%   \includegraphics[width=\columnwidth]{figures/pdfs_huge_pos.pdf}
%   \caption[]{\label{fig:pdfs_huge_sigma_pos} %
%   Probability of magnification for a set of simulations with positive parity and different $\Sigma_{\rm eff}$ in the medium-high surface mass density regime.
%   }
% \end{figure}
% \begin{figure}[tpb]
%   \centering
%   \includegraphics[width=\columnwidth]{figures/pdfs_huge_neg.pdf}
%   \caption[]{\label{fig:pdfs_huge_sigma_neg} %
%   Probability of magnification for a set of simulations with negative parity and different $\Sigma_{\rm eff}$ in the medium-high surface mass density regime.
%   }
% \end{figure}
\begin{figure*}
\centering
   \includegraphics[width=17cm]{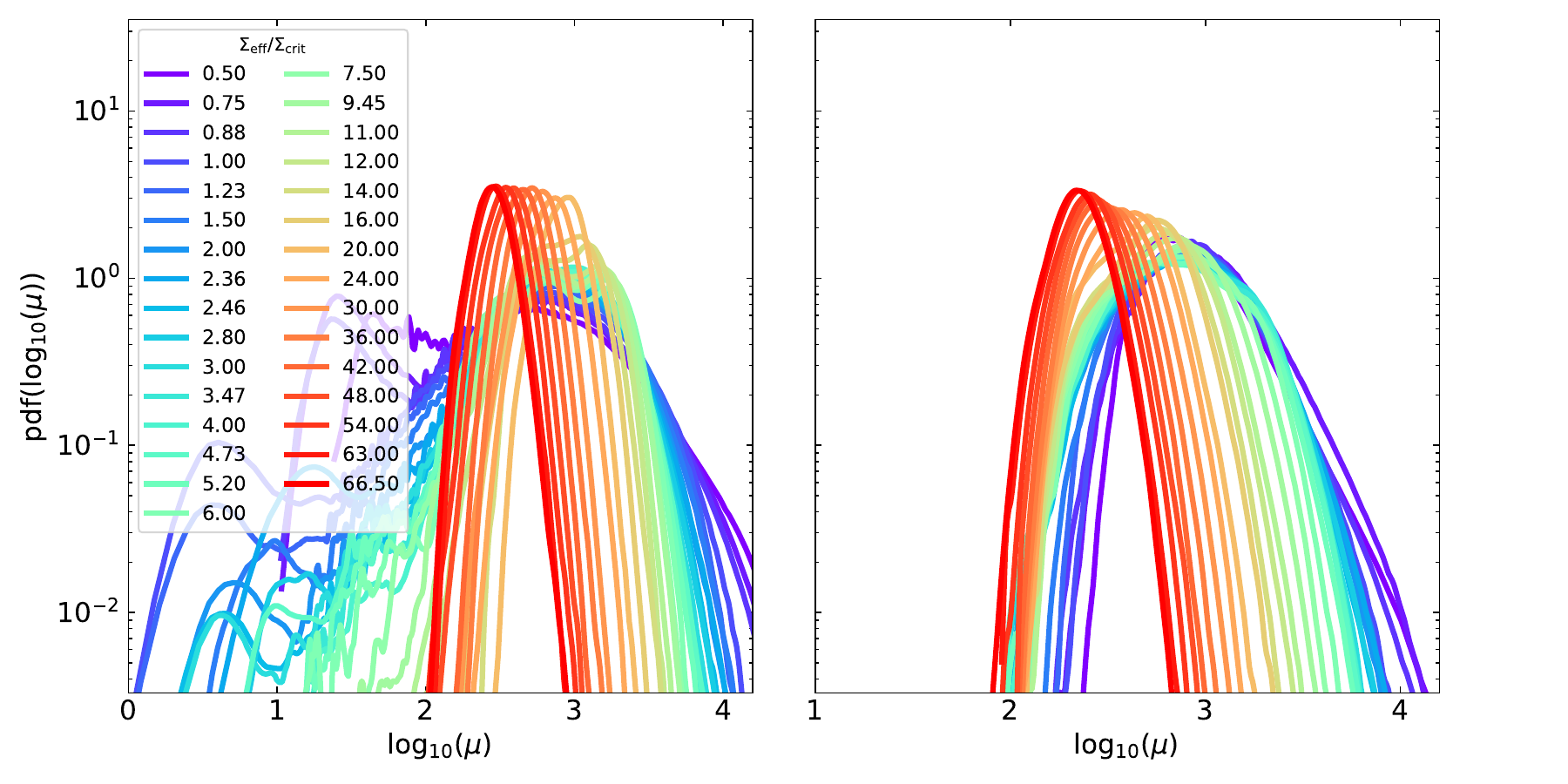}
     \caption{Probability of magnification for a set of simulations with different $\Sigma_{\rm eff}$ in the medium-high surface mass density regime.
     {\em Left panel: \/} Negative parity side of the macro-CC.
     {\em Right panel: \/} Positive parity side of the macro-CC.}
     \label{fig:pdfs_huge_sigma}
\end{figure*}

In the medium-high surface mass density case, the effective surface mass density exceeds the critical value, leading to the inevitable overlap of micro-caustics.
The microlenses affect the macro-model magnification considerably and form a complex web of deformed micro-caustics that covers the entire plane, as depicted in Fig. (\ref{fig:caustics_high_sigma}).
The magnification probabilities for the medium-high surface mass density simulations are shown in Fig. (\ref{fig:pdfs_huge_sigma}) for the positive (right panel) and negative (left panel) macro-magnification parity cases.

As $\Sigma_{\rm eff}$ increases, the probability distribution of the re-scaled magnification becomes more concentrated within a smaller range and around smaller magnification values\footnote{We recall that all magnification values in this plot have been re-scaled by the factor 1000$/\mu_m$}. Both parities exhibit a peak-shaped distribution that saturates to a constant width.
In the negative parity case, there is still a tail of demagnification, but as the surface mass density increases, this tail becomes less relevant and eventually it disappears. For the most extreme values of $\Sigma_{\rm eff}$, the magnification probability becomes similar to that of the positive case.
This effect is evident in Fig. (\ref{fig:caustics_high_sigma}), where for an effective surface mass density 48 times larger than the critical value, the micro-caustics for the negative parity are indistinguishable from those of the positive case.
In these high-density scenarios, the areas of demagnification have been completely eliminated.

%% file: sections/modeling.tex
\begin{figure}[tpb]
  \centering
  \includegraphics[width=\columnwidth]{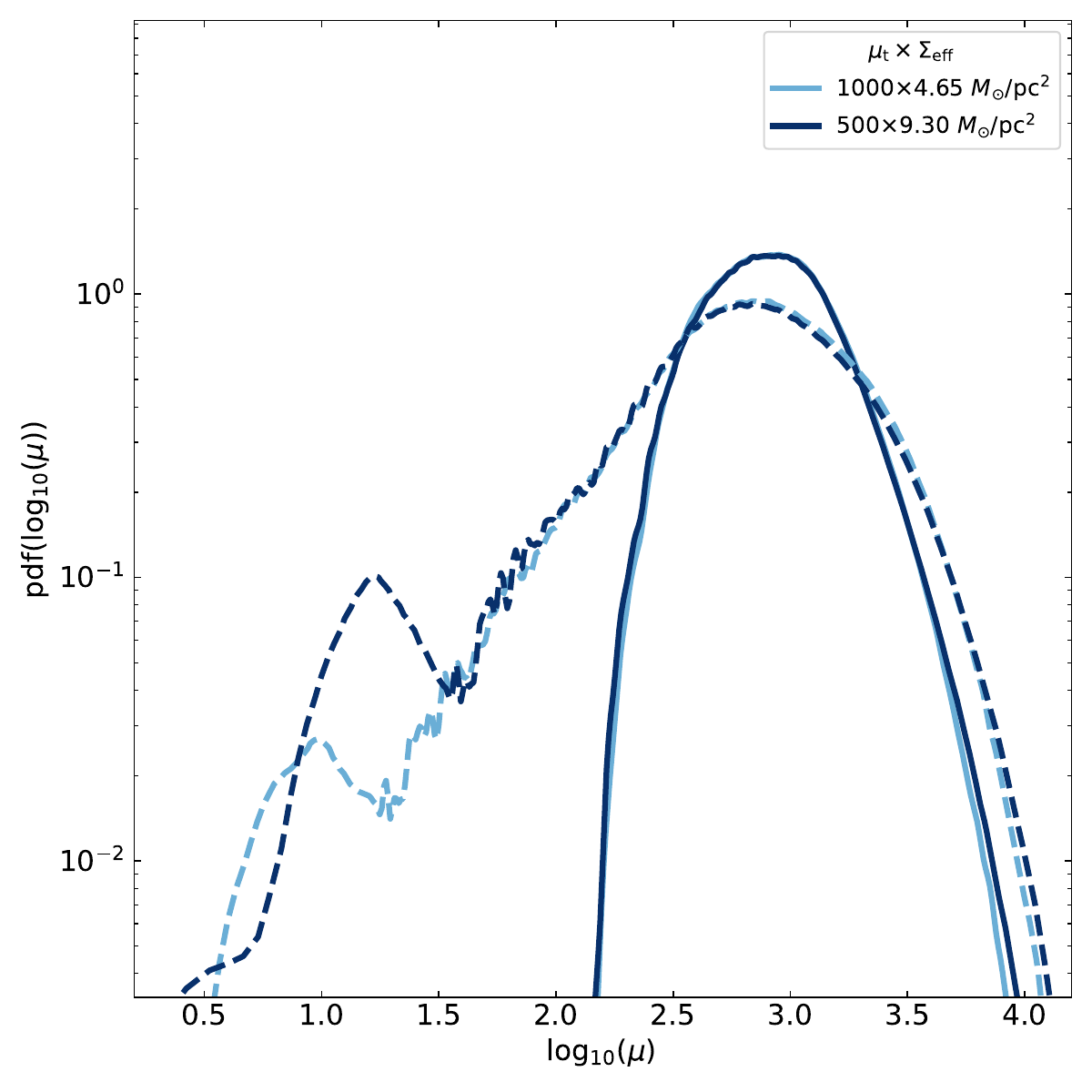}
  \caption[]{\label{fig:pdfs_same_sigma_eff} %
  Probability of magnification for two simulations with different $\Sigma_{\ast}$ and $\mu_{\rm t}$ but the same $\Sigma_{\rm eff}$. 
  Positive parities are shown as solid lines, while dashed lines represent negative parities.
  A difference in the position of the probability excess in the low magnification regime is due to a difference in $\mu_{\rm r}$ between the simulations.
  }
\end{figure}

In this section, we describe the analytical functions employed to fit the magnification probabilities in each simulation.
We differentiate between the low and medium-high surface mass density regimes as well as the positive and negative parities.
Additionally, we demonstrate how each model parameter scales in relation to the physical properties of the lens system.
To account for resolution effects, as described in Appendix (\ref{app:resolution}), we have carefully considered them in each simulation.

The process we follow is straightforward.
For each case study, we fitted the simplest analytical functions that not only offer a good approximation to the magnification probability distribution but also exhibit function parameters that demonstrate a discernible trend with respect to the lens model parameters.
These trends can be easily extrapolated.
As a result, we can work in the inverse direction: by utilising the lens characteristics, we can determine their corresponding model parameters.
These parameters can then be incorporated to obtain a analytical approximation of the magnification probability distribution.
This approach allows us to bypass the need for extensive numerical simulations, saving us time and computational resources.

Before delving into the analytical modelling that links the properties of the microlenses and the macro-model to the magnification probability, it is important to acknowledge a degeneracy present in the magnification probability with respect to $\Sigma_{\rm eff}$.
Specifically, two different simulations with the same $\Sigma_{\rm eff}$ but different values of $\Sigma_{\ast}$ and $\mu_{\rm t}$ can yield the same magnification probability.

Under certain circumstances, as described in Appendix (\ref{app:peaks}), this degeneracy can be broken.
Additionally, if we undo the previous normalisation, different values of $\mu_{\rm r}$ can also break this degeneracy.
Furthermore, we observe that, in general, $\mu_{\rm r}$ does not significantly influence the magnification probability.
In the most extreme cases of surface mass density, and for high magnifications regardless of the surface mass density, this parameter does not have a significant impact on the probability of attaining such magnifications.

\subsection{Low effective surface mass density}
%----------------------------------
This case is similar to the isolated microlens studied previously.
It describes the probability of magnification, which is important in scenarios with a low density of microlenses, important for high redshift stars whose CC will be in the outermost parts of the cluster, where stellar microlenses are reduced.
It also describes sources at lower macro-magnifications, or larger distances to the macro-CC.
Furthermore, it serves as a useful bridge, linking the single microlens case to the high-density regime scenarios that are computationally more demanding.
In this regime, we observe slight differences in magnifications between the positive and negative parities.
As mentioned earlier, we will now present the modelling of each case to provide a comprehensive understanding of the magnification behaviour in this regime.

The magnification probability functions in this study are characterised by a combination of power-law and log-normal functions.
To facilitate the understanding of the parameter scalings in each case, we adopt the following notation:
\begin{itemize}
    \item[$\bullet$] Capital letters are used to represent the parameters of the log-normal functions.
    \item[$\bullet$] Lowercase letters are used to denote the parameters of the power-law functions.
    \item[$\bullet$] The alphabetical ordering of the parameters corresponds to their appearance in the probability distribution function from lower to higher magnification values, progressing from left to right in the PDF figures.
    \item[$\bullet$] A superscript ``+'' or ``-'' is added to indicate the positive or negative parity case, respectively.
\end{itemize}
By employing this notation, we can succinctly represent and communicate the parametrisation of the magnification probability functions in a clear and organised manner.

\subsubsection{Positive parity}
A positive parity is obtained when $(1-\kappa)^2-\gamma^2\!>\!0$, indicating that the image retains the same parity as the source.
This regime has been extensively investigated in previous studies.
However, the understanding of probabilities at the highest magnification factors has been limited by the constraints of past simulations.
The large field of view and high resolution of our simulations enable us to provide a precise description of the highest magnification factors, where the more is less effect comes into play.

The modelling of the magnification probability in this regime is straightforward and resembles that of a single microlens embedded in a highly magnified region.
The probability distribution exhibits a peak structure centred around values similar to the magnification of the macro-model at that specific distance from the macro-CC.
With our re-scaling of the magnification, these peaks are centred at $\mu\!=\!1000$.
This central peak smoothly transitions into the well-known power-law decay that characterises the probability distribution over the remaining range of magnification values.
However, the presence of overlapping micro-caustics leads to a decrease in the probability at the highest magnification values.
This decrease occurs more rapidly than the power-law decay, resulting in a distinctive behaviour in the tail of the magnification probability distribution.

As the effective surface mass density increases, several notable changes occur in the magnification probability distribution.
Firstly, the peaks in the distribution decrease in amplitude, become broader, and shift towards values smaller than the macromodel magnification.
Additionally, these initially symmetrical peaks begin to exhibit a skewness towards larger magnification values.

Conversely, the power-law components of the distribution experience an increase in amplitude as the effective surface mass density increases.
However, the range of magnifications described by these power-laws gets reduced, and the suppression of extreme magnification factors starts at smaller magnification values compared to lower surface mass density regimes.

To effectively model these magnification probabilities, we have divided the probability density functions into three distinct magnification regimes: i) The low magnification regime, which is characterised by the presence of peaks, including the main one at the mode of the distribution; 
ii) An intermediate region, which is described by the classic $\mu^{-2}$ power-law; 
iii) The saturation extreme regime, where the power-law gets suppressed by microlenses.
By breaking down the PDFs into these three regimes, we can provide a comprehensive and accurate representation of the magnification probabilities across the entire range of magnification values.

    \begin{itemize}
        \item Low magnification regime:
        
        For the main peak of the PDF, we utilise a skewed log-normal distribution to capture the shape and characteristics of the magnification probabilities (part of the equation~\ref{Eq_14} with normalisation factor $\mathrm{B^{+}}$).
        In addition to the skewed log-normal distribution for the peak, we incorporate another log-normal distribution to account for the slight excess of probability at magnification values smaller than the peak($\mathrm{A^{+}}$ term in the same equation).
        \begin{align}
            \mathrm{PDF(log}_{10}(\mu))=&\,\mathrm{A^{+}}\exp{\left(-\frac{\left(\log_{10}(\mu) - \log_{10}(\mu_\mathrm{A}^{+})\right)^2}{2\sigma_\mathrm{A}^{+\,2}}\right)}\,+\nonumber\\
            &\, \mathrm{B^{+}}\exp{\left(-\frac{\left(\log_{10}(\mu) - \log_{10}(\mu_\mathrm{B}^{+})\right)^2}{2\sigma_\mathrm{B}^{+\,2}}\right)}\,\times \nonumber\\
            &\, \left[1 + \mathrm{erf}\left(\alpha_\mathrm{B}^{+}\frac{\log_{10}(\mu) - \log_{10}(\mu_\mathrm{B}^{+})}{\sqrt{2}\sigma_\mathrm{B}^{+}}\right)\right].
            \label{Eq_14}
        \end{align}
        By combining these analytical functions, we can effectively model the peak structure and the slight excess of probability at lower magnification values observed in the PDFs for the lowest surface mass densities.
        From now on, in the context of this paper, a capital letter ``X'' refers specifically to the amplitude of the X-log-normal distribution, where X stands for A, B, C, etc. $\mu_{\rm X}$ represents the centre of distribution without skewness, $\sigma_{\rm X}$ represents its width, and $\alpha_{\rm X}$ represents its skewness amplitude.
        In Fig. (\ref{fig:scale_low_pos_1}), we present the scaling of these model parameters with respect to the effective surface mass density.
        \begin{figure}[tpb]
          \centering
          \includegraphics[width=\columnwidth]{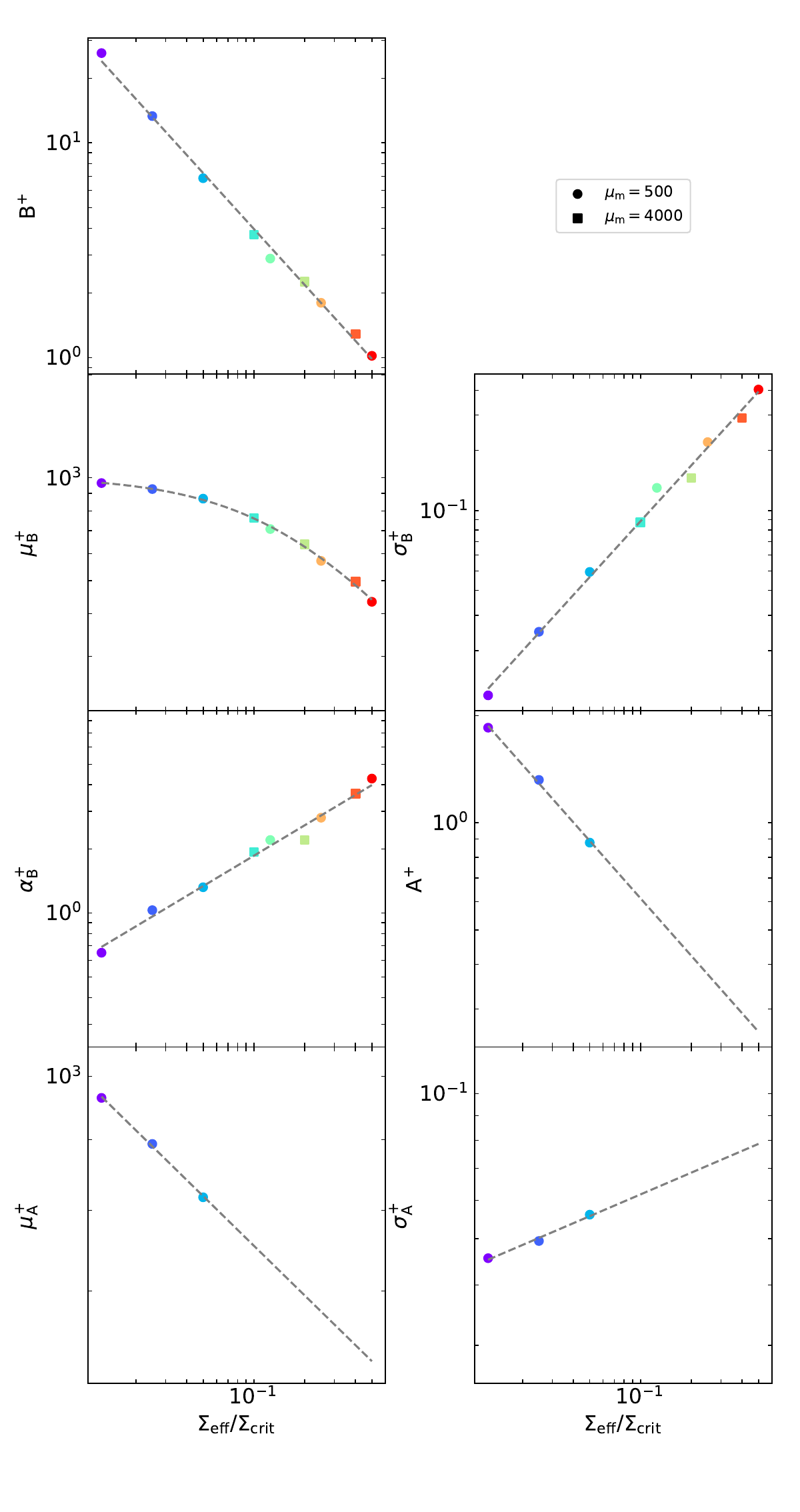}
          \caption[]{\label{fig:scale_low_pos_1} %
          Scaling of the model parameters for the low magnification regime at the low surface mass density and positive parity scenario.
          }
        \end{figure}
        \item Intermediate magnification regime:

        In the intermediate magnification regime, we model the magnification probability using a combination of two power-law functions with a fixed relation between their amplitudes.
        These power-laws capture the smooth transition from the peak region to the power-law decay with a fixed index of -2.
        The amplitude of the free index power-law is obtained by fitting the ratio of the amplitudes from a first fit of two independent power-laws.
        We use this model function to reduce the number of parameters.
        This procedure is repeated for the power-laws in the negative parity case.
        \begin{equation}
           \mathrm{PDF(log}_{10}(\mu)) = \mathrm{a^{+}}\left(\frac{\sqrt{\Sigma_{\rm eff}}}{591}\left(\frac{\mu}{10^{3.5}}\right)^{\beta_{\rm a}^+}+\left(\frac{\mu}{10^{3.5}}\right)^{-2}\right).
        \end{equation}
        \begin{figure}[tpb]
          \centering
          \includegraphics[width=\columnwidth]{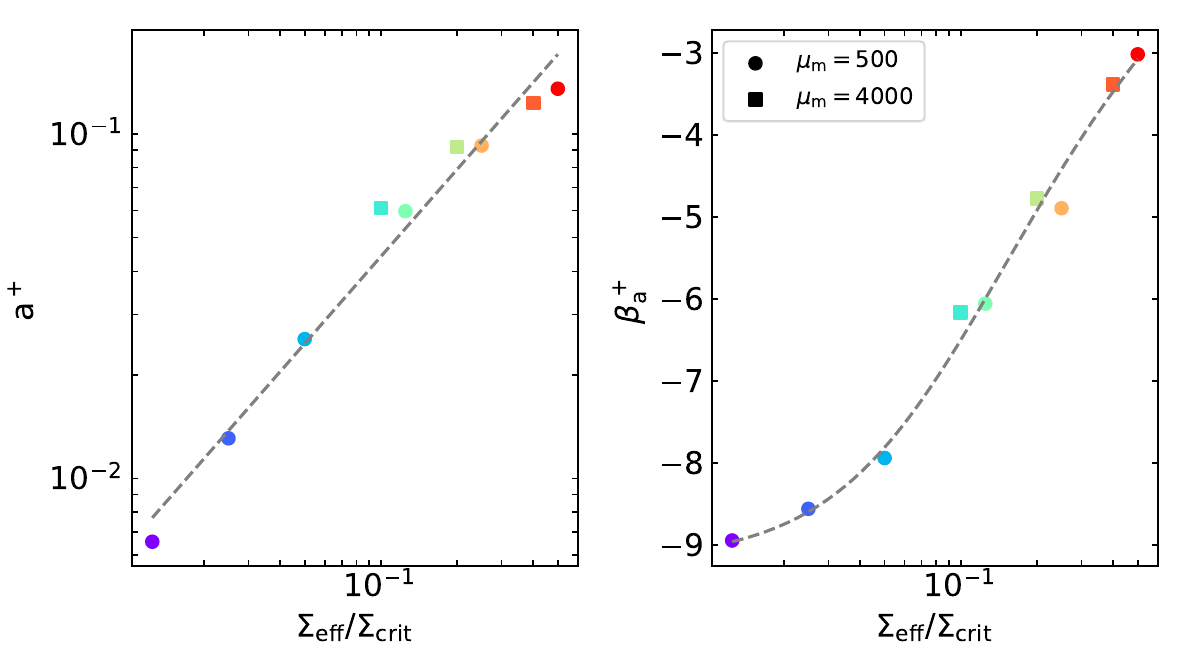}
          \caption[]{\label{fig:scale_low_pos_2} %
          Scaling of the model parameters for the intermediate magnification regime at the low surface mass density and positive parity scenario.
          }
        \end{figure}
        This analytical model provides a convenient way to approximate the magnification probability in the intermediate magnification regime.
        Similarly to our previous notation for log-normal functions, we adopt a notation for power-law distributions.
        The lowercase letter "x" represents the amplitude of the x-power-law distribution, while $\beta_{\rm x}$ denotes the index or exponent of the power-law.
        Fig. (\ref{fig:scale_low_pos_2}) shows the scaling of these model parameters with $\Sigma_{\rm eff}$.
        
       \item Extreme magnification regime:
        The suppression of the magnification probability at extreme magnification factors due to the microlenses can be described by a log-normal,
        \begin{equation}
            \mathrm{PDF(log}_{10}(\mu))= \mathrm{C}^{+}\exp{\left(\frac{-\left(\log_{10}(\mu) - \log_{10}(\mu_\mathrm{C}^{+})\right)^2}{2\sigma_\mathrm{C}^{+\,2}}\right)}.
        \end{equation}
        \begin{figure}[tpb]
          \centering
          \includegraphics[width=\columnwidth]{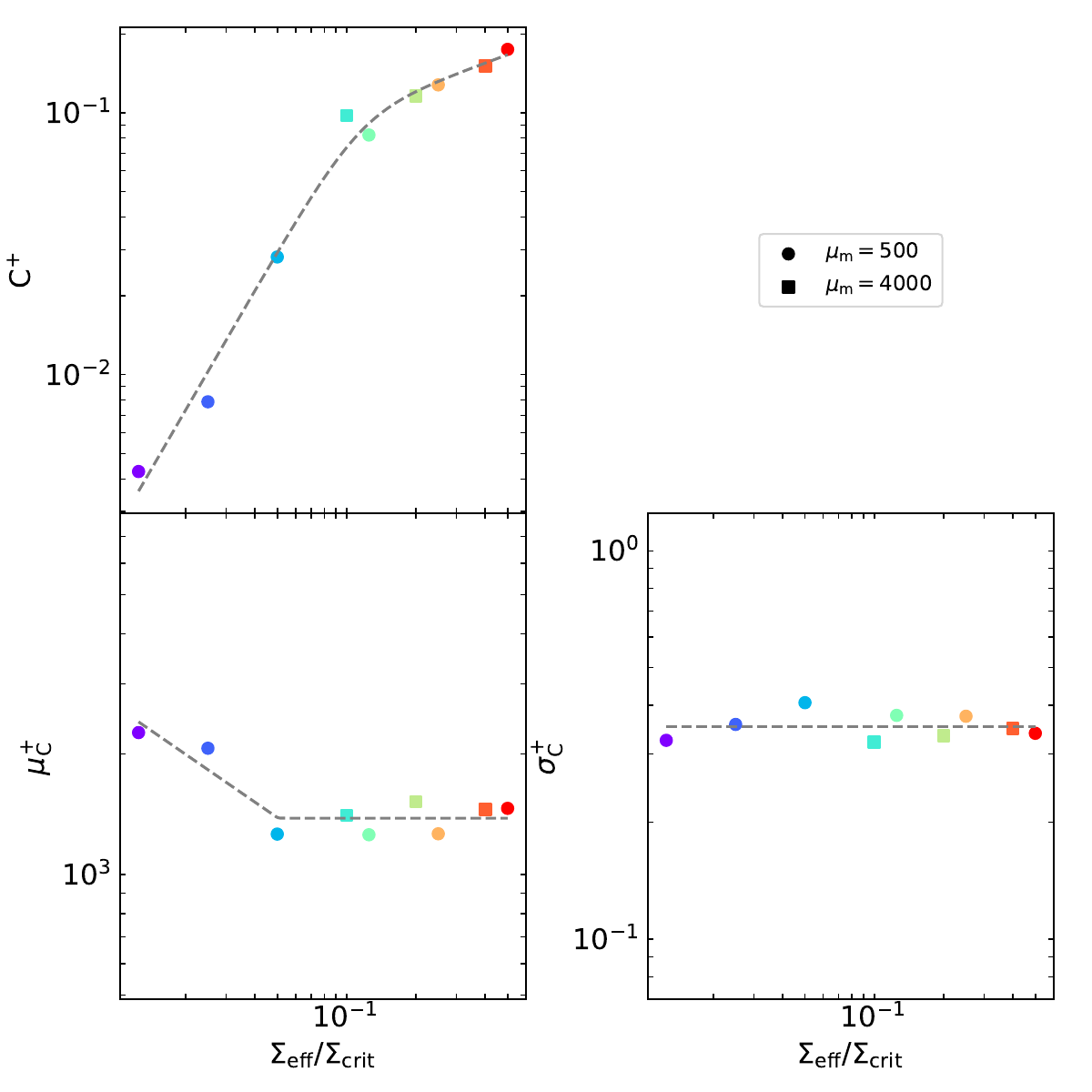}
          \caption[]{\label{fig:scale_low_pos_3} %
          Scaling of the model parameters for the extreme magnification regime at the low surface mass density and positive parity scenario.
          }
        \end{figure}
        These parameter scaling are shown at Fig. (\ref{fig:scale_low_pos_3}).
    \end{itemize}

\begin{figure}[tpb]
  \centering
  \includegraphics[width=\columnwidth]{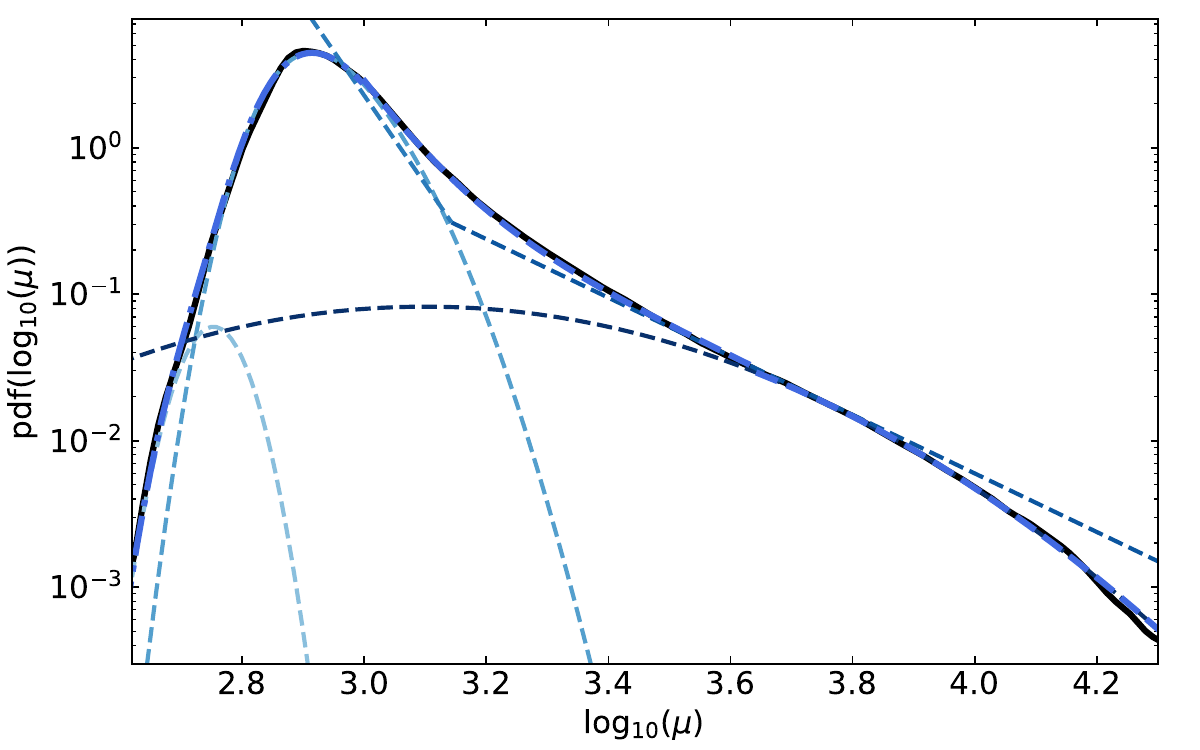}
  \caption[]{\label{fig:fit_compound_low_pos} %
  Magnification probability function for a simulation within the low surface mass density regime and positive parity. The dashed lines represents each of the fitting model components and their total contribution.
  }
\end{figure}
An example of a magnification probability curve fitted using this model is shown in Fig. (\ref{fig:fit_compound_low_pos}).
The figure showcases the agreement between the fitted model (dot-dashed line) and the simulated data (solid line).
This exemplifies how our analytical approach can accurately capture the magnification probability distribution over a wide range of magnifications.

\subsubsection{Negative parity}
The negative magnification regime, where $(1-\kappa)^2-\gamma^2\!<\!0$, corresponds to cases where the magnification $\mu$ is negative, resulting in an image with reversed parity compared to the source.
While this regime has often been overlooked in the literature, it plays a significant role in understanding certain features observed in caustic crossing events.
The unique characteristics of the negative parity regime can offer a better explanation for some observed events (such as the non-observation of Icarus' counterimage), making a thorough modelling of this regime crucial for a comprehensive understanding of extreme magnification events.  

In the negative parity regime, the magnification probability exhibits similar characteristics to the positive parity case at the highest magnifications.
However, there are unique features observed at lower magnifications.
We still observe a peak close to the macromodel value ($\mu=1000$ after re-scaling), followed by a power-law decrease that transitions to a log-normal behaviour at the highest magnification values.
At the lowest magnifications, we observe a mild increase in the probability from a power-law component, which is much less steep compared to the power-law at higher magnifications.
Additionally, there is a ``bump'' or excess in the probability of the lowest magnifications.
Interestingly, the position of this bump does not seem to follow a specific dependence on $\Sigma_{\rm eff}$.
These characteristics of the magnification probability in the negative parity regime provide important insights into the unique features and behaviours of caustic crossing events in gravitational lensing.

In the negative parity regime, similar to the positive regime, the width of the log-normals increases with increasing effective surface mass density.
The range of the power-law component decreases, and the amplitude of the central peak decreases while the amplitudes of the other log-normals increase, similar to the amplitudes of the power-law components.
However, there is a slight difference in the behaviour of the central peaks.
Initially, they shift towards higher magnification values, but as the surface mass density further increases, they start moving towards lower magnification values, mimicking the trend observed in the positive parity case.
This behaviour, alongside the demagnification components, reflects the unique characteristics of the negative magnification regime and provides important insights into the magnification probability in caustic crossing events with negative parities.

Here we also break the modelling into three regimes based on magnification values.
Thus, providing a comprehensive description of the magnification probability distribution in the negative magnification regime: i) The low magnification or demagnification regime featuring the main differences with respect to the positive parity;
ii) The intermediate magnification, covering the central peak;
iii) The high magnification regime, spawning the power-law decrease and log-normal decrease at the highest magnification factors.

    \begin{itemize}
        \item Low magnification regime:
        In the negative magnification regime, we model the probability distribution using a combination of power-law functions and a skewed log-normal distribution.
        An interesting finding is the presence of a constant index of 0.5 in the low magnification regime power-law, similar to the index of -2 observed in the high magnification regime.
        This modelling approach allows us to capture the unique characteristics of the negative magnification regime, including a smooth transition to the peaks.
        The skewed log-normal appears as a bump or excess of probability on top of the power-law at the lowest magnifications.
        To the left of this excess, the decrease is only that of the log-normal. Analytically this model is described as:

        \begin{align}
           \mathrm{PDF(log}_{10}(\mu))=\,&\mathrm{a^{-}}\left(\left(\frac{\Sigma_{\rm eff}}{1928}\right)^{1.7678}\left(\frac{\mu}{10^{2.4}}\right)^{\beta_{\rm a}^-}+\left(\frac{\mu}{10^{2.4}}\right)^{0.5}\right)\,+\nonumber\\
            &\, \mathrm{A^{-}}\exp{\left(-\frac{\left(\log_{10}(\mu) - \log_{10}(\mu_\mathrm{A}^{-})\right)^2}{2\sigma_\mathrm{A}^{-\,2}}\right)}\,\times \nonumber\\
            &\, \left[1 + \mathrm{erf}\left(\alpha_\mathrm{A}^{-}\frac{\log_{10}(\mu) - \log_{10}(\mu_\mathrm{A}^{-})}{\sqrt{2}\sigma_\mathrm{A}^{-}}\right)\right]
        \end{align}
        \begin{figure}[tpb]
          \centering
          \includegraphics[width=\columnwidth]{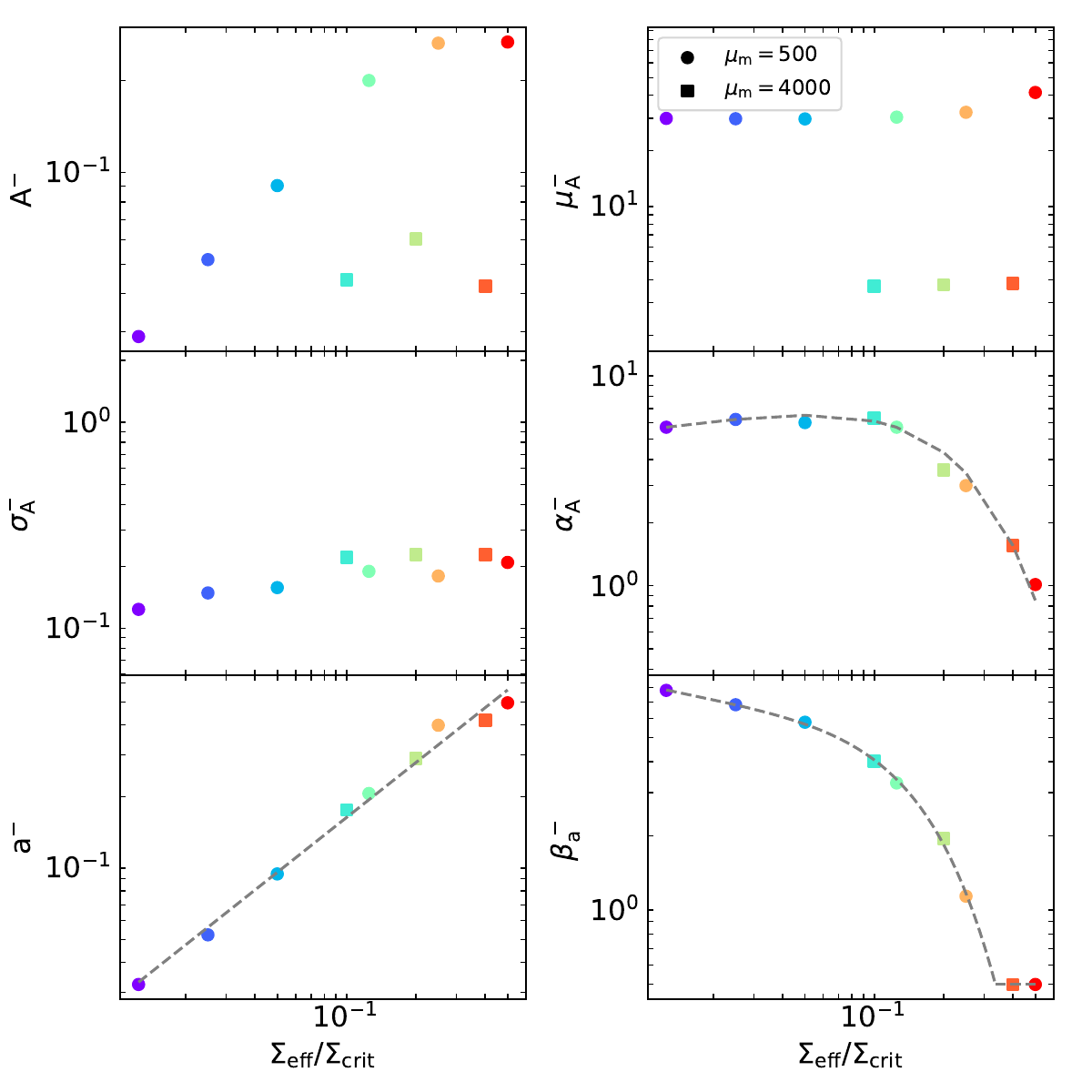}
          \caption[]{\label{fig:scale_low_neg_1} %
          Scaling of the model parameters for the low magnification regime at the low surface mass density and negative parity scenario.
          }
        \end{figure}
        These parameter scaling are shown at Fig. (\ref{fig:scale_low_neg_1}).
        We have observed that unlike other parameters, the characteristics of this excess in the probability distribution, that appear at low magnification factors for the negative parity only, do not scale directly with $\Sigma_{\rm eff}$.
        This distinction arises because the position of the excess is solely determined by the $\mu_{\rm r}$ parameter, independent of $\Sigma_{\rm eff}$.
        Consequently, the normalisation of the distribution places this peak at different positions relative to $\mu_{\rm m}$.
        This observation becomes apparent when examining Fig. (\ref{fig:pdfs_same_sigma_eff}), where the negative parity probability distributions are identical for the entire range of magnifications except for the lowest factors where these bumps occur.
        To gain further insight into these peaks at low magnification factors, we conducted an additional set of 12 simulations with varying $\mu_{\rm t}$, $\mu_{\rm r}$, and $\Sigma_{\ast}$, in addition to the previous simulations.
        This comprehensive data-set helps us better understand the characteristics of these peaks.
        We proceed to fit the position of the peak before normalisation as a function of $\mu_{\rm r}$:
        \begin{equation}
            \tilde\mu_{\rm A}^- = \mu_{\rm A}^- \frac{\mu_{\rm m}}{1000}.
        \end{equation}
        Similarly, we perform a fitting procedure to determine the width of the peak at each $\Sigma_{\rm eff}$ and $\mu_{\rm r}$. 
        We use the same scaling of the width with respect to $\Sigma_{\rm eff}$, but with different scaling factors determined by $\mu_{\rm r}$.
        We define this scaling factor in terms of $\mu_{\rm r}$ as follows:
        \begin{equation}
            R_{\sigma_{4}}(\,\mu_{\rm r}) = \frac{\sigma_{\rm A}^-\left(\,\mu_{\rm r}\right)}{\sigma_{\rm A}^-\left(4\right)}.
        \end{equation}
        The dependence of this quantity with respect to $\mu_{\rm r}$ is shown in Fig. (\ref{fig:scale_low_neg_extra_2}).
        Furthermore, we observe that the amplitude does not exhibit a simple scaling with $\Sigma_{\rm eff}$, regardless of the peak position corrections.
        To characterise the amplitude, we introduce a new quantity:
        \begin{equation}
            \tilde{\rm A}^- = \rm A^-\mu_{\rm m},
        \end{equation}
        that shows a clear trend with $\Sigma_{\rm eff}$. 
        Fig. (\ref{fig:scale_low_neg_extra_1}) illustrates the scaling of the peak parameters, with the corrections applied, as a function of $\Sigma_{\rm eff}$.
        \begin{figure}[tpb]
              \centering
              \includegraphics[width=\columnwidth]{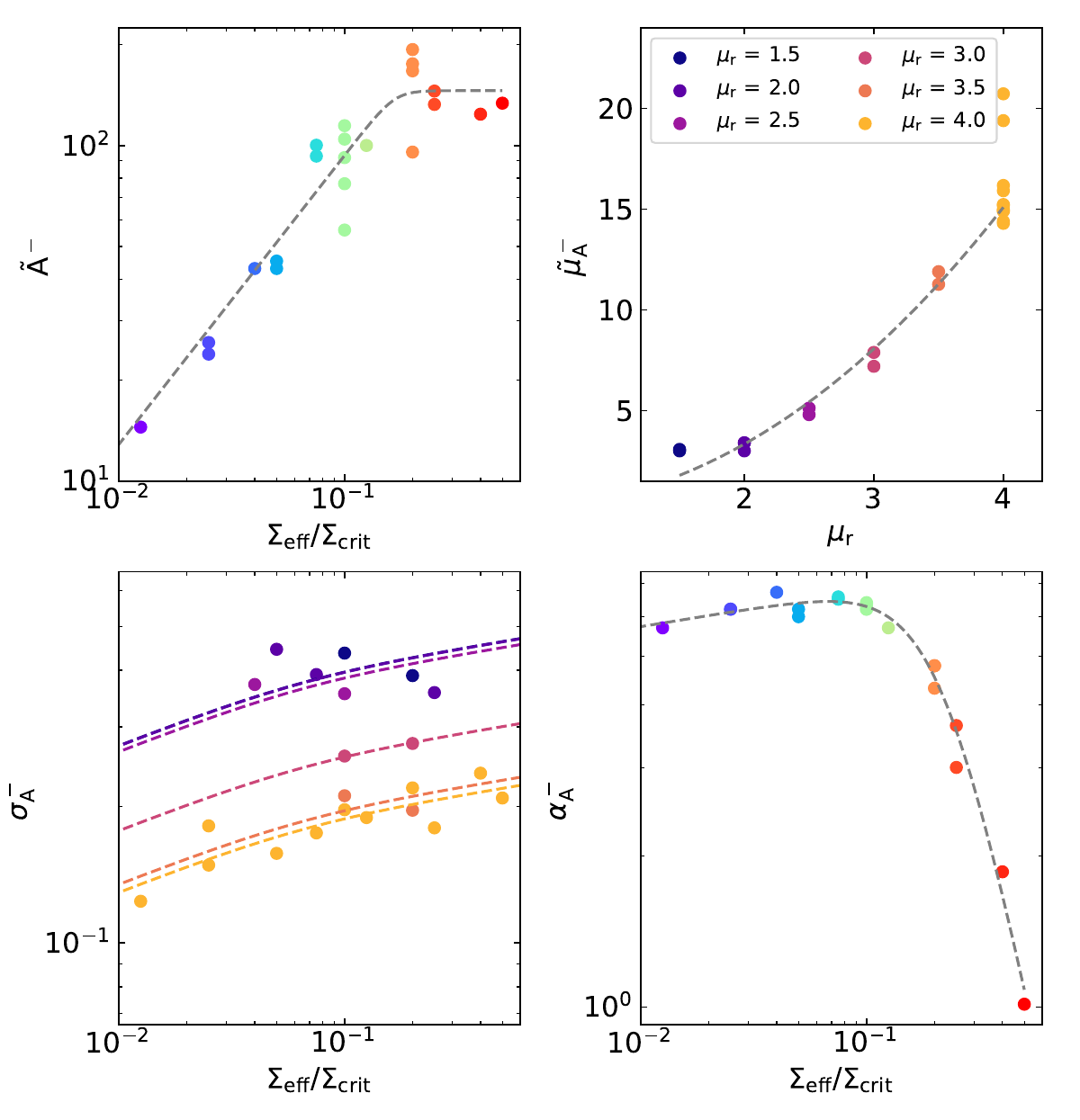}
              \caption[]{\label{fig:scale_low_neg_extra_1} %
              Scaling of the model parameters for the low magnification excess at the low magnification regime for the low surface mass density and negative parity scenario.
              }
        \end{figure}
        \begin{figure}[tpb]
              \centering
              \includegraphics[width=0.8\columnwidth]{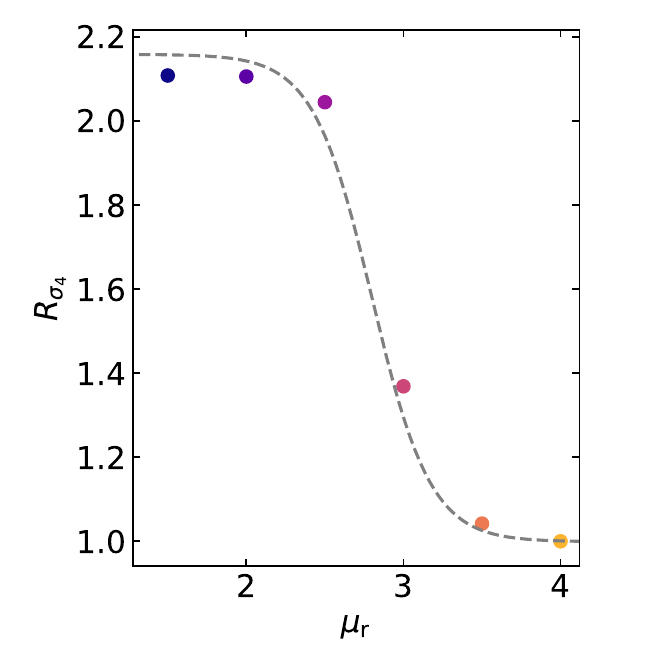}
              \caption[]{\label{fig:scale_low_neg_extra_2} %
              Ratio between the width of the peak (at low magnifications for low surface mass density and negative parity) at a given $\mu_{\rm r}$ and its corresponding width at $\mu_{\rm r}\!=\!4$.
              }
        \end{figure}

    \item Intermediate magnification regime

        This regime encompasses the central peak, which can be effectively modelled as a log-normal distribution:
        \begin{equation}
            \mathrm{PDF(log}_{10}(\mu))=\mathrm{B^{-}}\exp{\left(-\frac{\left(\log_{10}(\mu) - \log_{10}(\mu_\mathrm{B}^{-})\right)^2}{2\sigma_\mathrm{B}^{-\,2}}\right)}.
        \end{equation}
        The scaling for these parameters is illustrated in Fig. (\ref{fig:scale_low_neg_2}).
        \begin{figure}[tpb]
          \centering
          \includegraphics[width=\columnwidth]{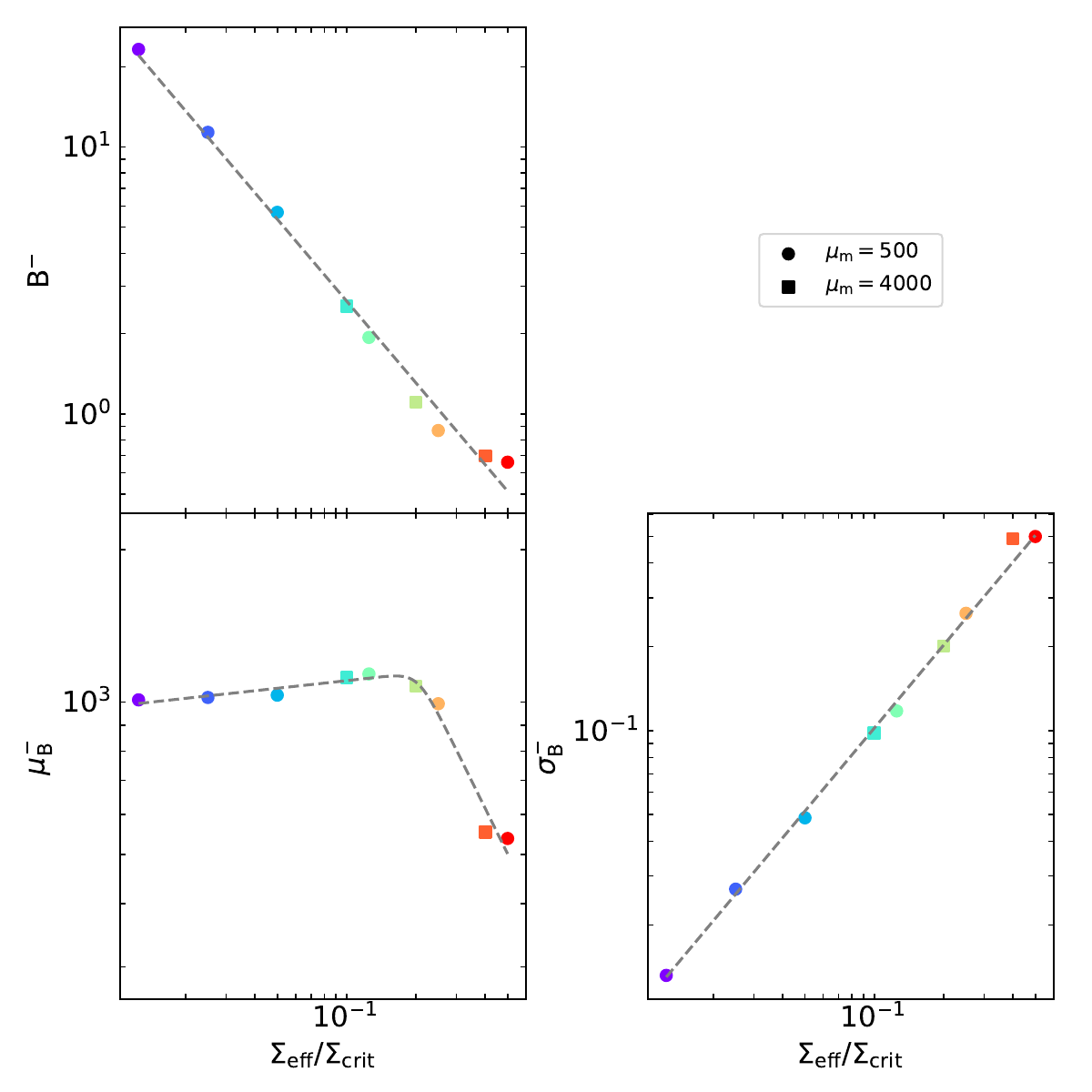}
          \caption[]{\label{fig:scale_low_neg_2} %
          Scaling of the model parameters for the intermediate magnification regime at the low surface mass density and negative parity scenario.
          }
        \end{figure}

    \item High magnification regime

    This regime can be modelled similarly to the positive parity case, with two power-laws and a log-normal.
    However, in this case, the log-normal is added to the power-laws instead of being a piecewise function.
    At the right end of the log-normal, the contribution of the power-laws becomes negligible.
    This analytical model is:
    \begin{align}
       \mathrm{PDF(log}_{10}(\mu))=\,&\mathrm{b^{-}}\left(\left(\frac{\Sigma_{\rm eff}}{995}\right)^2\left(\frac{\mu}{10^{3.5}}\right)^{\beta_{\rm b}^+}+\left(\frac{\mu}{10^{3.5}}\right)^{-2}\right)\,+\nonumber\\
       &\,\mathrm{C^{-}}\exp{\left(-\frac{\left(\log_{10}(\mu) - \log_{10}(\mu_\mathrm{C}^{-})\right)^2}{2\sigma_\mathrm{C}^{-\,2}}\right)}.
    \end{align}    
    \begin{figure}[tpb]
      \centering
      \includegraphics[width=\columnwidth]{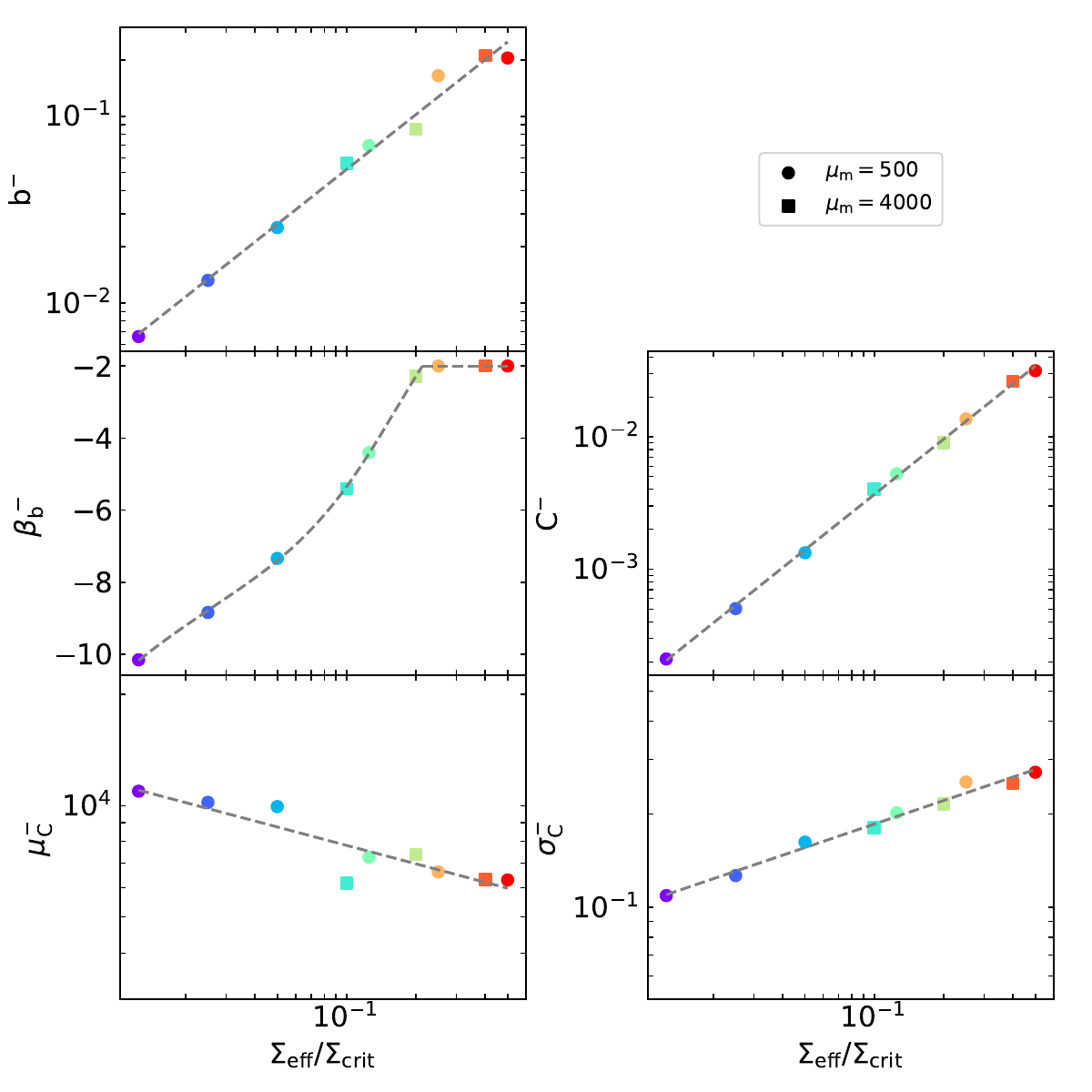}
      \caption[]{\label{fig:scale_low_neg_3} %
      Scaling of the model parameters for the extreme magnification regime at the low surface mass density and negative parity scenario.
      }
    \end{figure}
    Fig. (\ref{fig:scale_low_neg_3}) shows the scaling of this parameters with respect to $\Sigma_{\rm eff}$.
    \end{itemize}
\begin{figure}[tpb]
  \centering
  \includegraphics[width=\columnwidth]{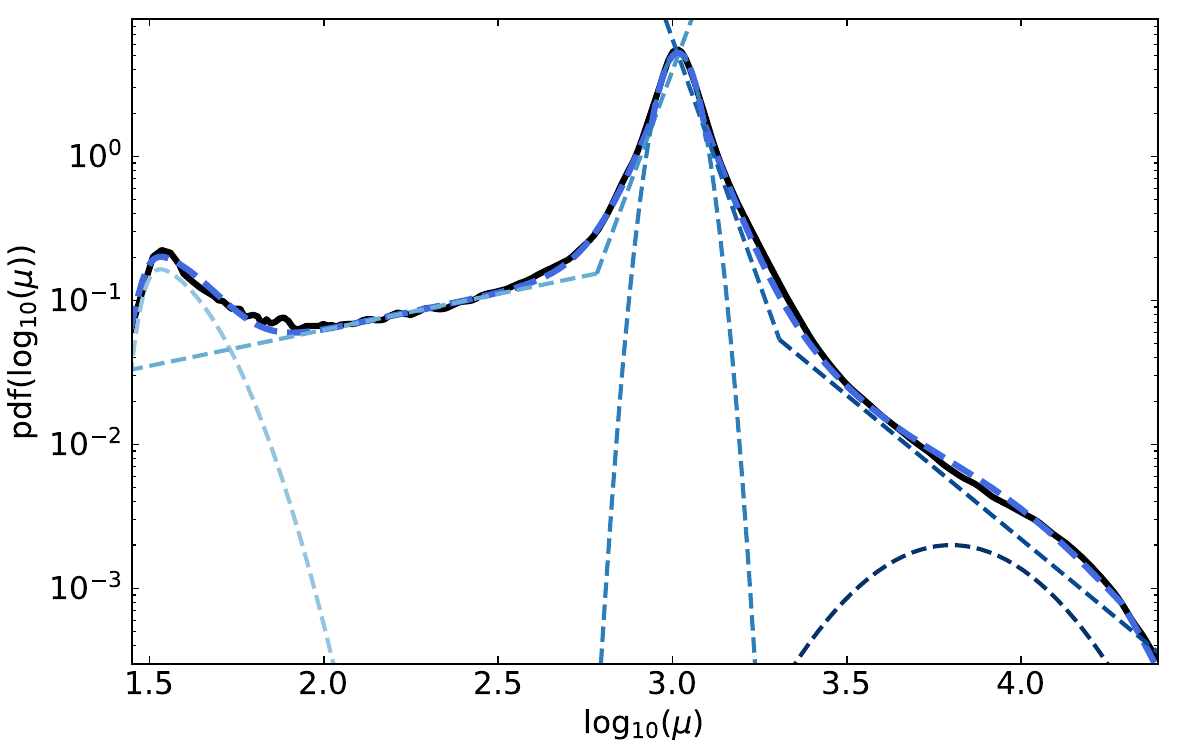}
  \caption[]{\label{fig:fit_compound_low_neg} %
  Magnification probability function for a simulation within the low surface mass density regime and negative parity. The dashed lines represents each of the fitting model components and their total contribution.
  }
\end{figure}
Fig. (\ref{fig:fit_compound_low_neg}) showcases the modelling of the magnification probability in the negative parity regime, as previously developed.

\subsection{High effective surface mass density}
%----------------------------------
This regime is expected to be more prevalent in realistic scenarios, where the combination of factors such as the distance to the macro-CCs, the surface mass densities of stars derived from the ICL, and the presence of a hypothetical population of compact dark matter can lead to higher values of $\Sigma_{\rm eff}$.
These elevated values often exceed the critical surface mass density, making the occurrence of this regime more likely.

In this regime, the magnifications for positive and negative parity differ, similar to what is observed in the low surface scenario.
However, as the effective surface mass density increases, the magnification probabilities for both parity regimes tend to become more similar.
In fact, for larger values of surface mass density, the modelling for both parity scenarios becomes identical.
Following the previous approach, we will now explain the modelling for both parity scenarios.

It is worth mentioning that even though this scenario is relatively straightforward in terms of modelling, the scaling of parameters exhibits more complexity and displays distinct regime behaviours.
Nonetheless, we have successfully captured the entire range of effective surface mass densities included in our simulations.
Moreover, we observe a saturation regime in the magnification probability at the highest densities, allowing for straightforward extrapolation to larger values.
This enables us to fully describe the magnification statistics and behaviour across the entire range of effective surface mass densities in our study.

\subsubsection{Positive parity}
\begin{figure}[tpb]
  \centering
  \includegraphics[width=\columnwidth]{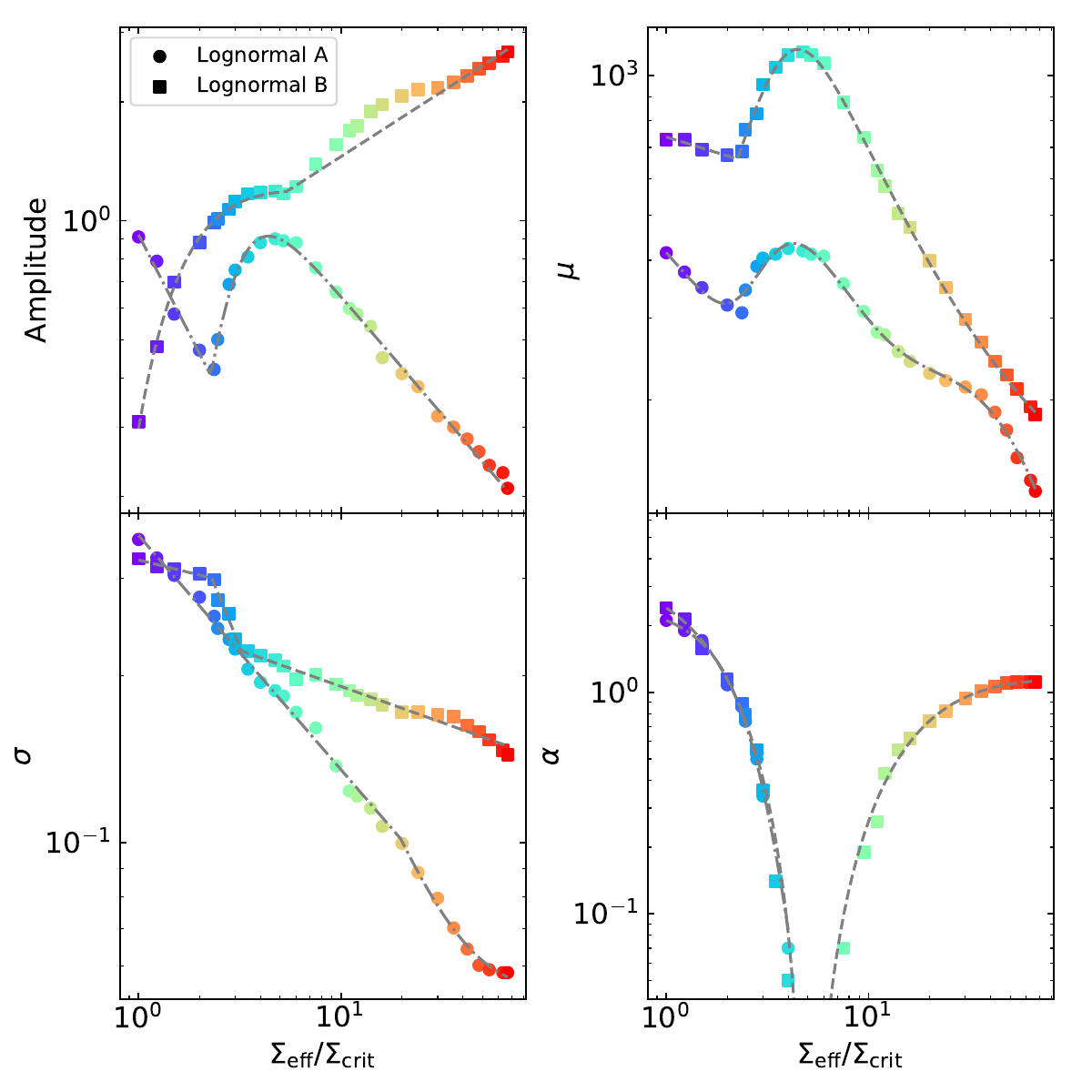}
  \caption[]{\label{fig:scale_high_pos} %
  Scaling of the model parameters at the high surface mass density and positive parity scenario.
  }
\end{figure}

\begin{figure}[tpb]
  \centering
  \includegraphics[width=\columnwidth]{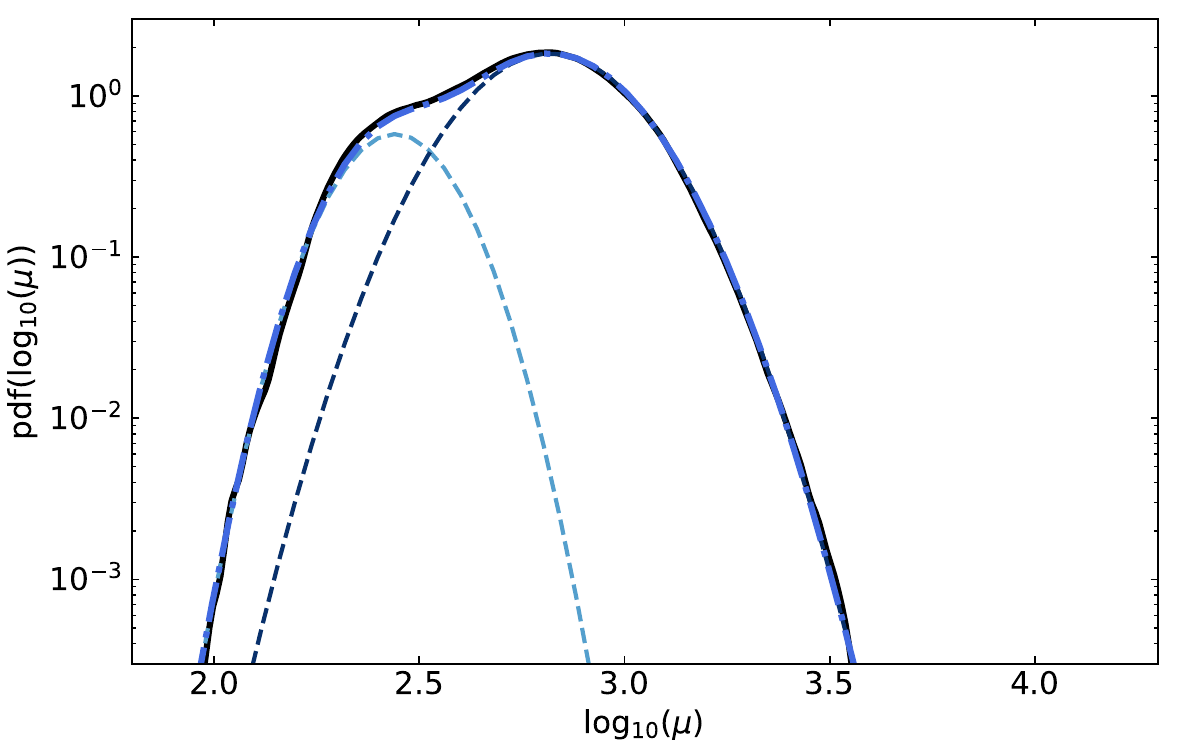}
  \caption[]{\label{fig:fit_compound_high_pos} %
  Magnification probability function for a simulation within the high surface mass density regime and positive parity. The dashed lines represents each of the fitting model components and their total contribution.
  }
\end{figure}

In this regime, we observe a transition from the low surface mass density regime, where the power-law distribution is dominant, to a regime characterised by a high-magnification log-normal distribution.
The magnification probability in this regime can be effectively modelled as the sum of two log-normal distributions, with both distributions exhibiting skewness.
Analytically, this can be represented as the sum of two skewed log-normal distributions:

\begin{align}
   \mathrm{PDF(log}_{10}(\mu))=\,&\mathrm{A^{+}}\exp{\left(-\frac{\left(\log_{10}(\mu) - \log_{10}(\mu_\mathrm{A}^{+})\right)^2}{2\sigma_\mathrm{A}^{+\,2}}\right)}\,\times \nonumber\\
    &\, \left[1 + \mathrm{erf}\left(\alpha_\mathrm{A}^{+}\frac{\log_{10}(\mu) - \log_{10}(\mu_\mathrm{A}^{+})}{\sqrt{2}\sigma_\mathrm{A}^{+}}\right)\right]\,+ \nonumber\\
    &\, \mathrm{B^{+}}\exp{\left(-\frac{\left(\log_{10}(\mu) - \log_{10}(\mu_\mathrm{B}^{+})\right)^2}{2\sigma_\mathrm{B}^{+\,2}}\right)}\,\times \nonumber\\
    &\, \left[1 + \mathrm{erf}\left(\alpha_\mathrm{B}^{+}\frac{\log_{10}(\mu) - \log_{10}(\mu_\mathrm{B}^{+})}{\sqrt{2}\sigma_\mathrm{B}^{+}}\right)\right]
\end{align}
The scaling of the parameters in this model with respect to $\Sigma_{\rm eff}$, as well as an example of a probability density function obtained from our simulations and fitted using this model, are depicted in Figs. (\ref{fig:scale_high_pos}) and (\ref{fig:fit_compound_high_pos}), respectively. These figures provide a visual representation of how the model parameters evolve with increasing $\Sigma_{\rm eff}$ and demonstrate the individual components of the fitted compound distribution.

\subsubsection{Negative parity}
In this regime, the probability distribution at the highest magnification factors is conceptually similar to that of the positive parity regime.
The power-law distribution is no longer an accurate description of the probability, and instead, the sum of two log-normal distributions provides a more appropriate magnification probability approximation.
However, at the lowest magnification factors, the power-law distribution still persists.
The crucial difference is that the power-law index is no longer fixed at 0.5.
As the effective surface mass density increases, the power-law index grows, while the amplitude of the distribution remains constant.
This variation in the power-law index reflects the changing behaviour of the magnification probability distribution compared to the low surface mass density regime.
The range of magnification values for which the power-law distribution remains a valid description covers all values up to the magnification at which the log-normal and power-law distributions yield the exact same probability.
However, as $\Sigma_{\rm eff}$ approaches a value of approximately $5~\Sigma_{\rm crit}$, the power-law distribution returns extremely low probabilities, rendering its contribution negligible.
In contrast to the positive parity regime, the most suitable model that we have found to fit our simulations in this regime consists of three log-normal distributions, which are fully symmetrical and do not exhibit any skewness. 

Additionally, we have observed a peak at the lowest magnification factors in this regime, which is likely the same excess that we observed in the low density regime.
However, due to limitations in our simulation statistics, we do not have sufficient data to fully describe this peak in detail. 
Given that the excess peak at the lowest magnification factors is unlikely to be observed due to the known luminosity functions of stars, high redshift considerations, and limitations of current telescopes, we can omit the modelling of this peak and approximate the magnification as a power-law distribution in this regime.
Qualitatively, we have observed that these peaks exhibit growth in both width and amplitude as well as a shift towards larger magnification values.
Additionally, there is a correlation between the position of the peak and $\mu_{\rm r}$, similar to what was observed in the low density regime.
Further investigations and simulations would be required to better understand and characterise this feature at the lowest magnification values, we left this task for a future work.
The analytical model we have employed for the probability distribution is expressed as follows:
\begin{align}
   \mathrm{PDF(log}_{10}(\mu))=\,&\mathrm{A^{-}}\exp{\left(-\frac{\left(\log_{10}(\mu) - \log_{10}(\mu_\mathrm{A}^{-})\right)^2}{2\sigma_\mathrm{A}^{-\,2}}\right)}\,+ \nonumber\\
    &\, \mathrm{B^{-}}\exp{\left(-\frac{\left(\log_{10}(\mu) - \log_{10}(\mu_\mathrm{B}^{-})\right)^2}{2\sigma_\mathrm{B}^{-\,2}}\right)}\,+ \nonumber\\
    &\, \mathrm{C^{-}}\exp{\left(-\frac{\left(\log_{10}(\mu) - \log_{10}(\mu_\mathrm{C}^{-})\right)^2}{2\sigma_\mathrm{C}^{-\,2}}\right)}.
\end{align}
The power-law distribution, which is valid only for lower magnification factors until it becomes overtaken by the log-normal distributions, is described by the equation:
\begin{equation}
    \mathrm{PDF(log}_{10}(\mu))=\mathrm{a^{-}}\left(\frac{\mu}{10^{2.4}}\right)^{\beta_{\rm a^-}}
\end{equation}

\begin{figure}[tpb]
  \centering
  \includegraphics[width=\columnwidth]{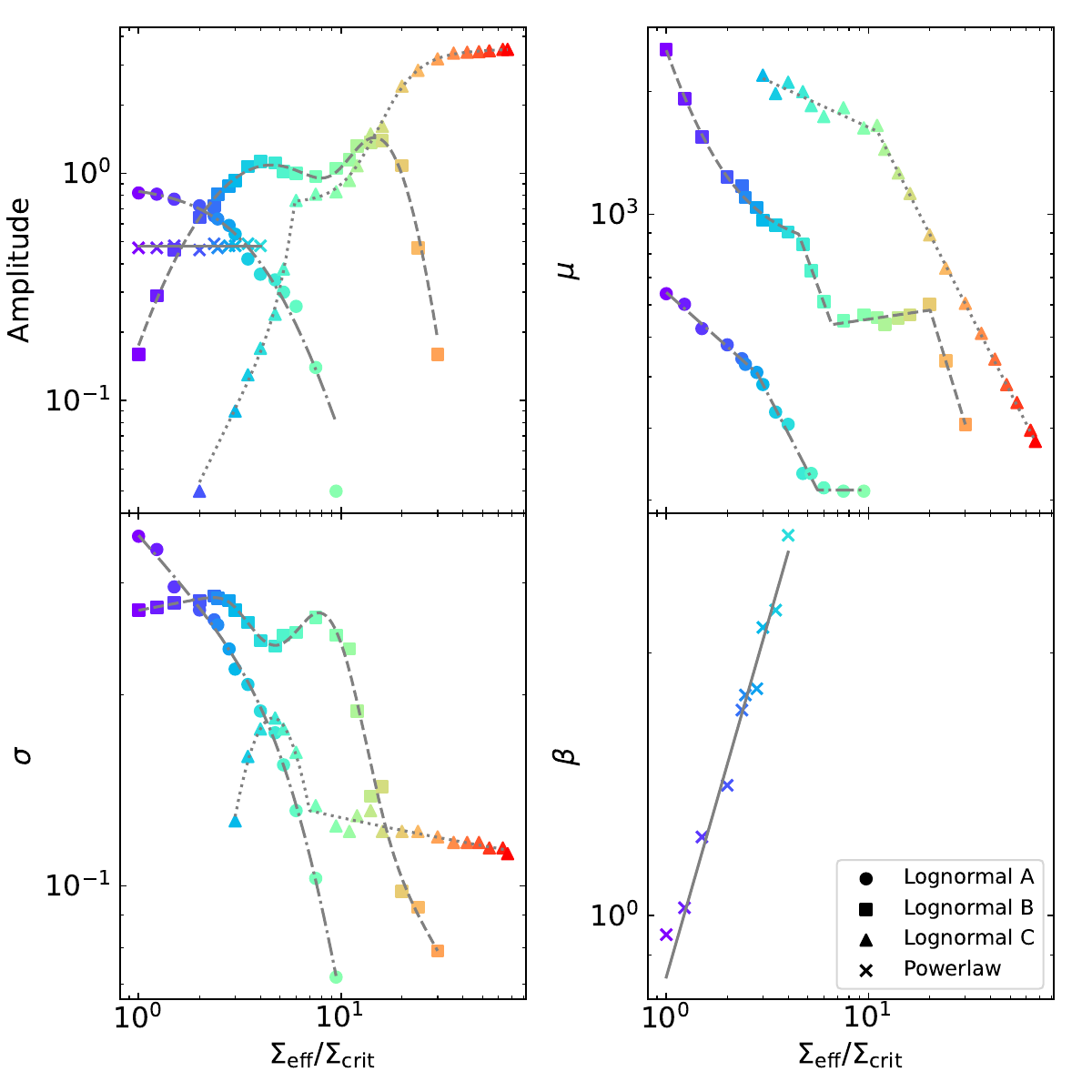}
  \caption[]{\label{fig:scale_high_neg} %
  Scaling of the model parameters at the high surface mass density and negative parity scenario.
  }
\end{figure}
\begin{figure}[tpb]
  \centering
  \includegraphics[width=\columnwidth]{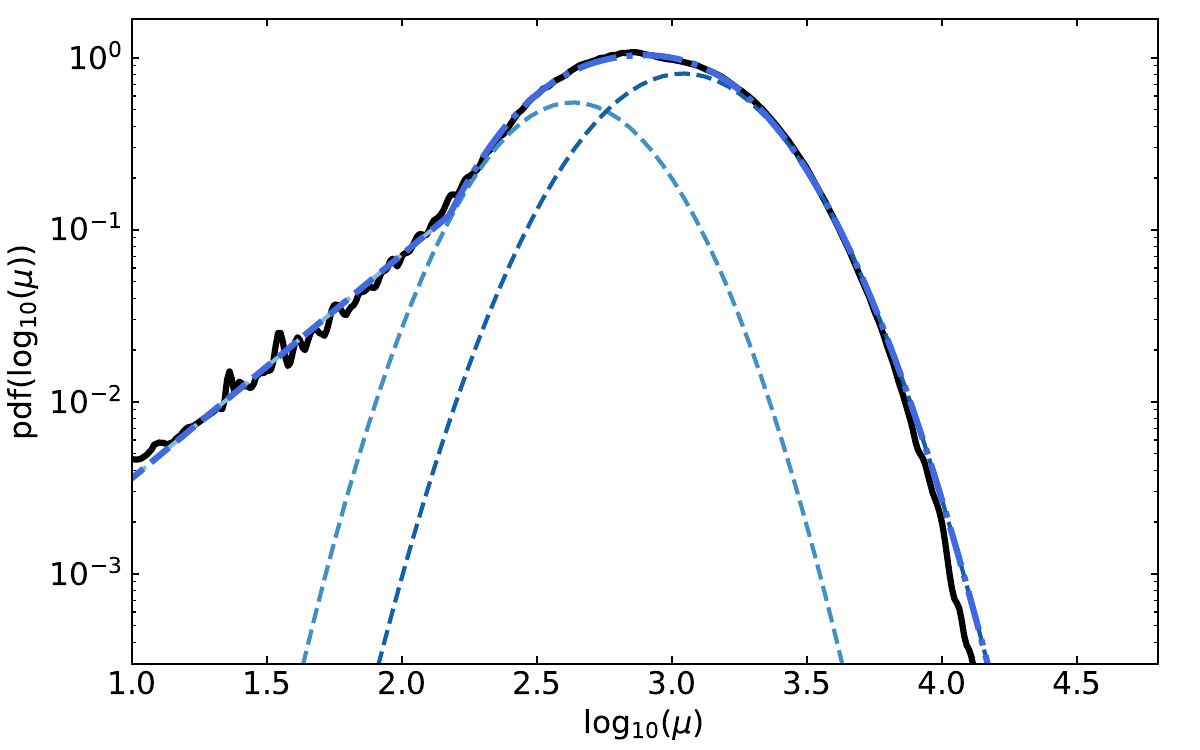}
  \caption[]{\label{fig:fit_compound_high_neg} %
  Magnification probability function for a simulation within the high surface mass density regime and negative parity. The dashed lines represents each of the fitting model components and their total contribution.
  }
\end{figure}
Regarding the negative parity regime, we have also depicted the dependence of the model parameters on $\Sigma_{\rm eff}$ in Fig. (\ref{fig:scale_high_neg}), and we have also provided an example of a simulation magnification probability fitted according to this models in Fig. (\ref{fig:fit_compound_high_neg}). 

%% file: sections/constrains.tex
During caustic crossing events, the flux of the background stars can increase by factors of several thousands, to tens of thousands. This translates in magnitude changes of $\mathcal{O}$(10) compared to fluxes without magnification. 
This change in their apparent magnitude allows the detection of these sources, as well as a characterisation of the underlying microlensing population causing such events.
Here we outline how the tools we have developed can be used to constrain compact DM with caustic crossing events, and provide an illustration of the expected magnification as a function of time, which illustrates how we can estimate the magnification probability.

Microlensing of small sources within a strongly lensed galaxy is responsible for the extremely lensed events such as Icarus, Earendel, or Godzilla (\cite{Kelly2018}, \cite{Welch2022a}, and  \cite{Diego2022a}).
These lensing events result in a significant enhancement of the observed fluxes from these objects, enabling their detection and facilitating detailed studies of their properties.
This scientific scenario in itself presents an intriguing subject of investigation and has led to the development of dedicated scientific programs like FLASHLIGHTS \citep{Flashlights2019}.
However, these extremely lensed events hold particular importance in the field of dark matter studies.
They offer unique opportunities to probe the nature and distribution of DM, as the lensing effects depend on the underlying DM distribution within the lensing system \citep{Oguri2018}.

As seen in the previous section the magnification probability in microlensing events is primarily determined by two factors: $\Sigma_{\ast}$ and $\mu_{\rm t}$.
The latter can be obtained from the strong-lensing models of the cluster, taking into account the distance of the image to the macro-CC. But even when the magnification can not be accurately described, the scaling of the magnification with distance, $d$,  to the CC, $\mu \propto d^{-1}$ allows to study the relative detection rate as a function of distance.  
On the other hand, $\Sigma_{\ast}$ can be further divided into two components: $\Sigma_{\rm b}$, which accounts for baryonic matter such as stars, and $\Sigma_{\rm DM}$, which represents the contribution from dark matter.

The estimation of $\Sigma_{\rm b}$ can be derived from the ICL, which provides valuable information about the distribution of stars within the cluster.
The remaining unknown parameter is $\Sigma_{\rm DM}$, which represents the contribution of dark matter.
This parameter can be inferred from the observed rate of  microlensing events in a statistical sense, after a sufficiently large number of lensed stars has been observed. Of particular interest are lensed stars observed in regions where the contribution from  $\Sigma_{\rm b}$ is lowest. This includes the CC forming between two merging clusters, specially during the first core passage, since the intermediate region in between the two clusters is expected to have very small contributions from stars in the intracluster light.

Monitoring the change of flux over time in the magnified images as the source moves in the source plane provides valuable information about the magnification probability.
By comparing these observed flux variations with the analytical models derived from the previous section, we can perform statistical inference to estimate the parameter $\Sigma_{\rm DM}$ that best fits the data.
This statistical analysis allows us to obtain an approximation of the density of compact dark matter in the microlensing population.
Specifically, we can update the fraction of dark matter made up of our model of interest, denoted as ``X'', by determining the ratio of its density to the total density of dark matter, represented by $f_{\rm X}\!=\!\Omega_{\rm X}/\Omega_{\rm DM}$.
This provides insights into the contribution of the specific model of interest to the overall dark matter content in the microlensing population.

\begin{figure}[tpb]
  \centering
  \includegraphics[width=\columnwidth]{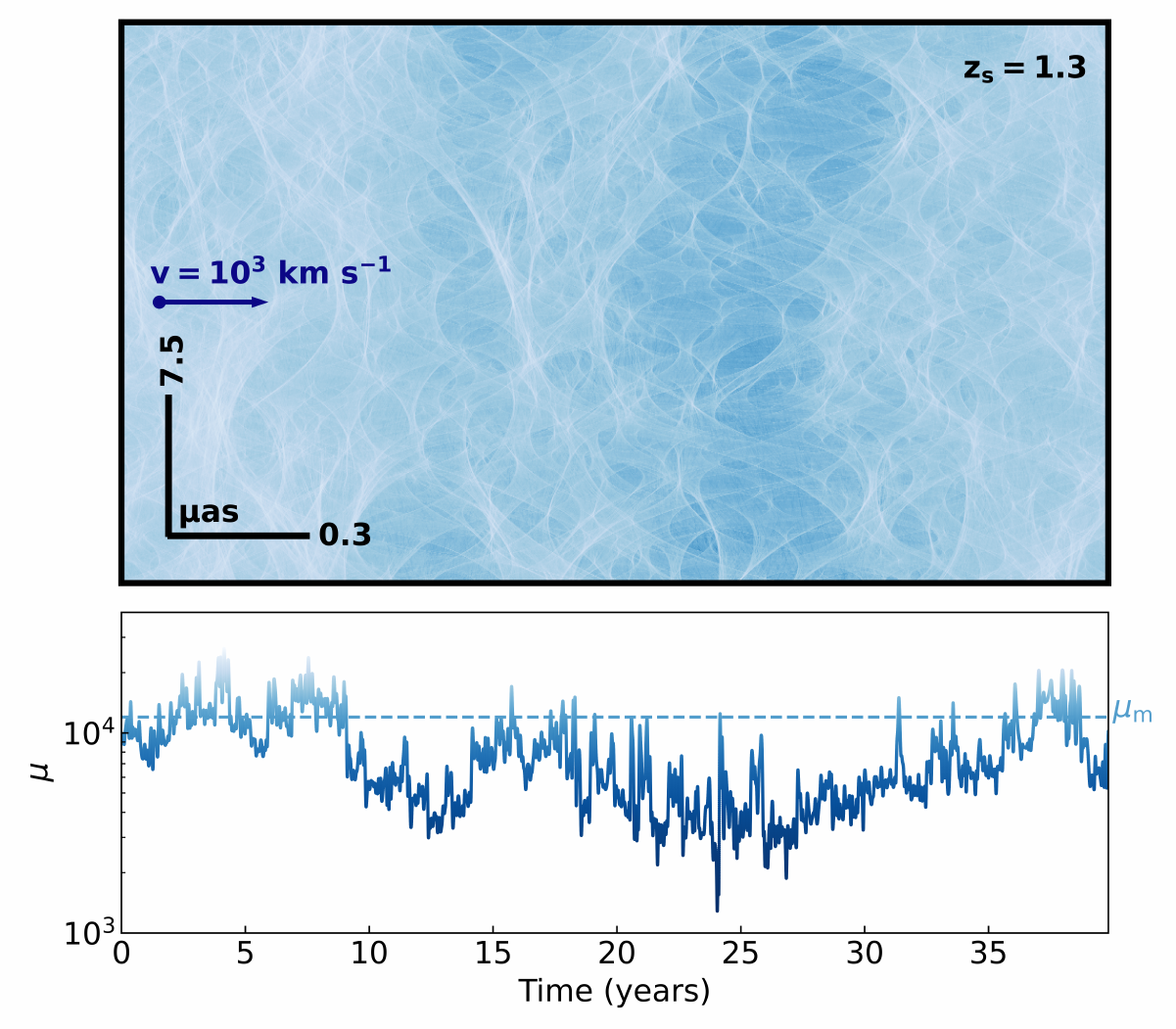}
  \caption[]{\label{fig:mag_over_time} %
  Magnification of a source at redshift 1.3 and angular size 2 nas sweeping the micro-caustic web at a constant speed of 1000 km s$^{-1}$.
  {\em Top panel: \/} Magnification pattern in the source plane.
  {\em Bottom panel: \/} Magnification over time.
  The dashed line represents the magnification from the macro model.
  }
\end{figure}

The characteristic timescales of microlensing events are influenced by several factors, including the distance to the source and its transverse motion relative to the micro-caustics.
To illustrate this, we can consider a simulated caustic pattern in which the source is located at a redshift of $z_{\rm s}\!=\!1.3$.
The transverse velocity of a star with respect to the caustics at this distance can be estimated using the formula (see \cite{Miralda-Escude1991})
\begin{equation}
    v_\perp = \left|\frac{v_{\rm s}-v_{\rm o}}{(1+z_{\rm s})} - \frac{D_{\rm s}(v_{\rm d}-v_{\rm o})}{D_{\rm d}(1+z_{\rm d})}\right|.
\end{equation}
The transverse velocity depends on the distances to the lens plane, and the source plane through their angular diameter distances, and their redshifts, and depends also on the transverse velocities of the cluster, the source, and the observer, $v_{\rm d}$, $v_{\rm s}$, and $v_{\rm o}$, respectively.
To fix the transverse velocity $v_\perp$, a typical value of $\sim\!1,000$~km/s can be used as suggested by \citep{Kelly2018}. 
Finally, the angular transverse velocity at the source plane (see \cite{Oguri2018}) is
\begin{equation}
    u_{\rm s} = \frac{v_{\rm s}}{1+z_{\rm s}}\frac{1}{D_{\rm s}}.
\end{equation}
Plugging the transverse velocity value of $\sim\!1,000$~km/s into the equation, assuming a distance corresponding to a redshift of $\sim 0.3$, we can calculate the transverse velocity in angular units. 
This yields a value of approximately $52$~nas per year.
In the context of our simulation, this corresponds to a crossing time of approximately 40 years. 
Fig. (\ref{fig:mag_over_time}) illustrates an example of such a microlensing event, showing the evolution of the magnification over time due to the motion of the source.

Monitoring an arc close to a macro-CC allows us to observe flux changes in the sources within the arc, which can be attributed to microlensing events.
By characterising the source properties, such as its intrinsic flux, we can translate the observed flux changes into magnification values.
With longer exposure times, we can monitor the flux changes of more sources over extended periods, which improves the accuracy of the magnification probability estimation.
By combining the flux changes from multiple sources, we can obtain a better estimate of the magnification probability distribution.
Using the analytical approximations that we have establish in the previous section and the estimated magnification probabilities, we can quickly obtain an estimate of the magnification probability in terms of parameters such as $\Sigma_{\ast}$, $\mu_{\rm t}$, and $\mu_{\rm r}$.
This approach allows us to bypass the need for time-consuming simulations, which would require significant computational resources and subsequent analysis to obtain the magnification probability.
Furthermore, by employing Markov Chain Monte Carlo methods, we can perform parameter estimation and identify the most favoured combination of parameters that best explain the observed data.
This enables us to derive constraints on the physical properties of the lensing system, such as the surface mass density of the microlenses and the macro-magnification factors, by comparing the observed flux variations with the estimated magnification probabilities.
Independent estimations of the macro-magnification factors $\mu_{\rm t}$ and $\mu_{\rm r}$, as well as the baryon surface mass density $\Sigma_{\rm b}$constrain the parameter space reducing the number of free parameters to one, $\Sigma_{\rm DM}$, the contribution from compact DM.

Studying the outskirts of clusters is a promising approach for probing the role of dark matter in microlensing effects.
In these regions, the contribution of stellar populations to the microlensing population is diminished, allowing the potential dominance of DM objects, if present.
This offers an opportunity to investigate the properties of DM in the form of compact objects.
Arcs farther from the cluster centre are associated with source planes at larger redshifts.
To probe larger redshifts, it becomes necessary to observe intrinsically brighter objects.
By comparing the observed probabilities of magnification to what would be expected solely from the stellar populations, we can set lower limits on the contribution of compact DM.
Setting lower limits on the fraction of compact DM, $f_{\rm X}$, requires careful analysis and comparison between observed magnification probabilities and our analytical predictions.
By constraining the minimum amount of compact DM necessary to explain the lack of higher magnification events, we gain valuable insights into the presence and properties of DM in the form of compact objects within galaxy clusters.

\begin{figure}[tpb]
  \centering
  \includegraphics[width=\columnwidth]{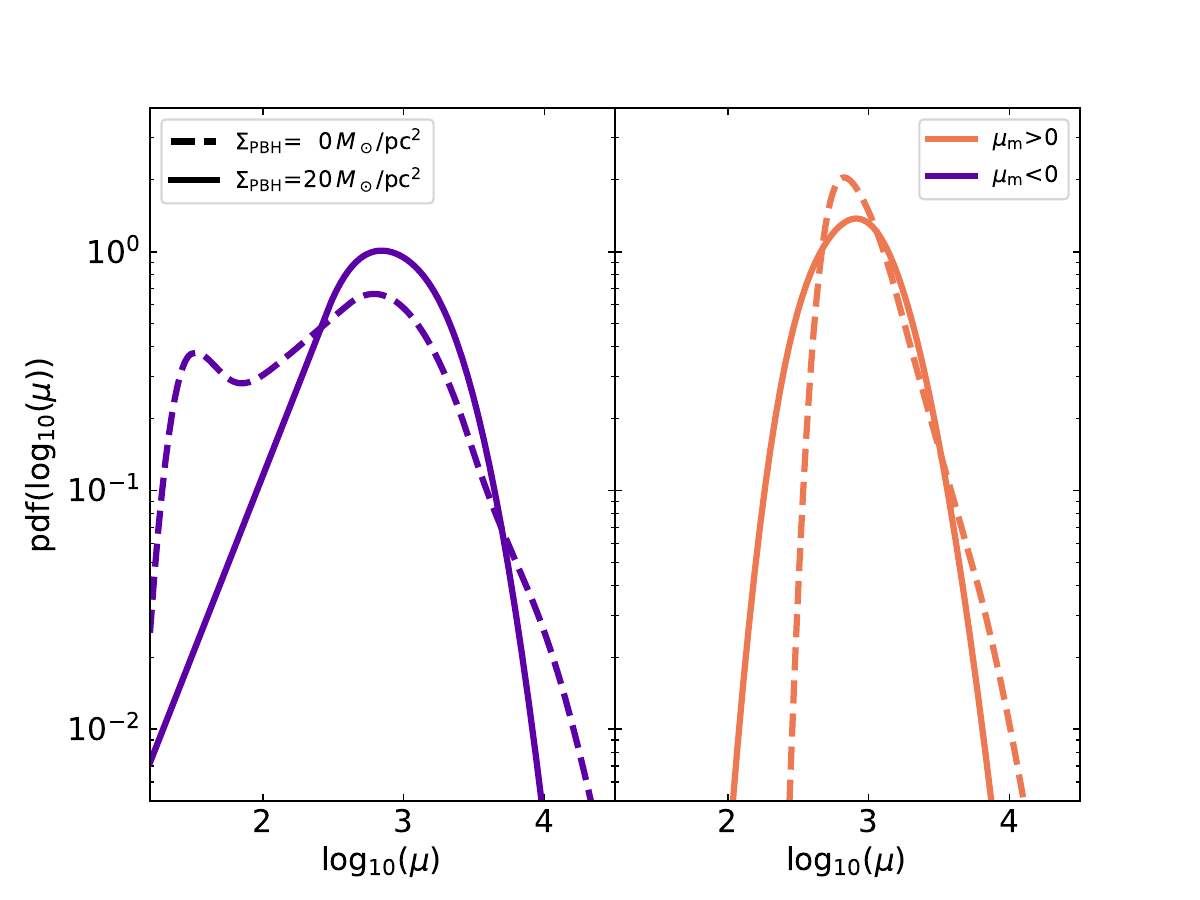}
  \caption[]{\label{fig:analytic_pdf_2sigma_crit} %
  Magnification probability for a case with $\Sigma_{\rm eff}/\Sigma_{\rm crit}\!=\!2$, in both parity regimes.
  This mimics the magnification probability of a hypothetical microlensing population consisting of 5~$M_{\sun}/{\rm pc}^2$ stars and 20~$M_{\sun}/{\rm pc}^2$ PBHs.
  The critical density is $\Sigma_{\rm crit}\!=\!2500~M_{\odot}/{\rm pc}^2$.
  $\mu_{\rm t}\!=\!200$ and $\mu_{\rm r}\!=\!4$.
  {\em Left panel: \/} Negative macro-magnification.
  {\em Right panel: \/} Positive macro-magnification.
  The dashed curve represents the magnification probability for the same system with no PBHs.
  }
\end{figure}

Fig. (\ref{fig:analytic_pdf_2sigma_crit}) shows the magnification probability from a hypothetical population of microlenses made of 5~$M_{\sun}/{\rm pc}^2$ in the form of stars, and 20~$M_{\sun}/{\rm pc}^2$ made of PBHs which corresponds to approximately a 1\% of the total DM in realistic scenarios.
For completeness the magnification probability for the same scenario with no PBHs is displayed as a dashed line.
The macro-model tangential magnification, $\mu_{\rm t}$, is 200 which makes $\Sigma_{\rm eff}\!=\!2\Sigma_{\rm crit}\!=\!5000~M_{\sun}$.
Being the macro-images at a distance of approximately 0.2 arcseconds from a macro-CC, a powerful telescope like the JWST would have the capability to resolve both the positive parity image and the negative parity image.
By comparing the flux differences between these two images, we can estimate the probability distribution functions of magnification at both parities.
If we can observe both parities we have even more constraining power, usually the negative parity image is more constraining.
In this particular scenario, the observed flux differences can be used to infer the presence of an underlying surface mass density of approximately 20 $M_{\sun}/{\rm pc}^2$ composed of primordial black holes (PBHs) or other compact objects.
This analysis can be carried out with only one image, but adding a second image of the same source in the reverse parity regime greatly increases the constraining power on the surface mass density of the microlensing population.

%% file: sections/discussion.tex
In this paper, we have used the inverse ray shooting technique to generate very large simulations of extragalactic microlensing phenomena caused by compact objects within galaxy clusters, and near the caustic region.
These microlensing effects occur alongside the strong lensing effects generated by the gravitational field of the cluster, resulting in a combined lensing scenario.
This emerging field of gravitational lensing provides a unique opportunity to investigate and study background sources.
Furthermore these microlensing events allow us to constrain the surface mass density of the microlenses and provide valuable information about their luminous and non-luminous constituents.
The non-luminous components are particularly interesting as they are directly linked to the presence and properties of compact dark matter.

In our study, we have focused on the simulation of microlensing near tangential CCs, where $\mu_{\rm t}\!\gg\!\mu_{\rm r}$.
This choice is motivated by the fact that radial CCs are located closer to the centre of the cluster, where the influence of dark matter compact objects is comparatively smaller.
In contrast, tangential CCs offer a more favourable scenario for studying the effects of dark matter, as the population of microlenses in this region is expected to be predominantly composed of dark matter.
Furthermore, higher redshift sources provide an additional advantage in probing the dark matter population associated with microlensing.
As the redshift of the source increases, the tangential CCs move further outwards from the centre of the cluster.
Consequently, the contribution from the cluster's light becomes less significant, enhancing the relative impact of the dark matter population.
This makes higher redshift sources more effective in isolating and studying the effects of dark matter compact objects within the microlensing framework.

In our simulations, we replicate the magnification effects experienced by stars or compact objects passing through an arc that intersects the perturbed gravitational potential of the cluster.
This occurs due to the presence of a population of microlenses near a macro-caustic, considering both parity configurations.
We have obtained the magnification probabilities for each combination of $\mu_{\rm t}$, $\mu_{\rm r}$, and $\Sigma_{\ast}$ simulated for both parity configurations, considering scenarios where $\Sigma_{\rm eff}$ is below the critical value and scenarios where it is equal to or greater than this value.
In the presence of increasing surface mass density and a stronger influence of perturbations from microlenses, the most frequently observed magnification deviates from the macro-model's magnification at the simulated distance to the macro-CC.
Specifically, it tends to shift towards smaller values as the disruption of the cluster's smooth potential becomes more dominated by the perturbation effects of microlenses.
We have found previously unknown effects such as the power-law decay in magnification at negative parities, along with the bump or excess at the lowest magnifications.
Using our newly developed tool, we have the capability to generate probability density functions for magnification within the range of effective surface mass densities $\Sigma_{\rm eff}$ spanning from $0.01$ to $66.5~M_\sun/\rm{pc}^2$.
Furthermore, we can extrapolate these magnification probabilities for even larger effective surface mass densities as needed.

\subsection{Saturation regime and log-normal behaviour}
 %--------------------------------------------------------
The extreme saturation regime, characterised by $\Sigma_{\rm eff}/\Sigma_{\rm crit}\gg 1$, poses significant computational demands and thus it is the more interesting to describe with analytical models. 
Remarkably, we observe that in this regime, the well-established power-law behaviour that typically characterises the magnification probability for sparsely overlapping caustics gets completely eradicated.
Instead, we find that the magnification probability follows a log-normal distribution.
While the log-normal behaviour has been previously noted in the literature (\cite{Diego2018}, \cite{Welch2022a}), its complete understanding or comprehensive explanation has remained elusive due to limitations in microlensing simulations. 
Our simulations provide a sufficiently large area in the source plane with high resolution, enabling us to place confidence in the statistical properties of magnification in this regime.

\begin{figure}[tpb]
  \centering
  \includegraphics[width=\columnwidth]{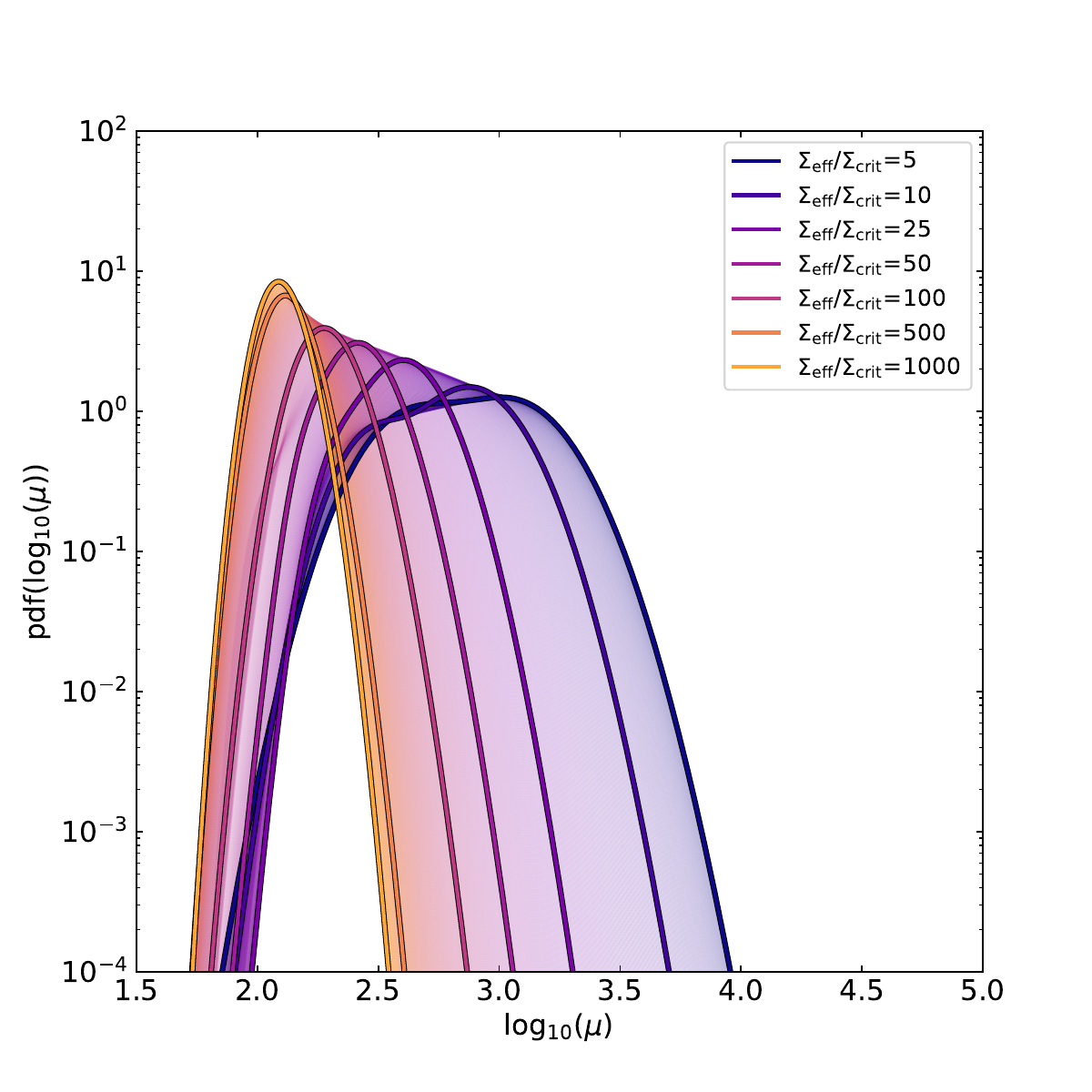}
  \caption[]{\label{fig:more_is_less} %
  Analytical approximation for the probability of magnification for a fixed $\mu_{\rm m}\!=\!1000$, and varying $\Sigma_\ast$.
  By increasing $\Sigma_{\rm eff}$ the more is less and saturation effects appear, obtaining a thicker log-normal centred at lower magnifications.
  }
\end{figure}

Fig. (\ref{fig:more_is_less}) shows how increasing $\Sigma_{\rm eff}$ results in the more is less effect, that reduces the spread of variability in flux when more and more microlenses are added. Such small variability in flux has been already observed in highly magnified stars such as Earendel \citep{Welch2022a,Welch2022b}. 
The lack of variability due to microlensing in Earendel can be explained if the probability of magnification is already well into the log-normal regime. \\

%The log-normal behaviour arises from the multiple overlapping of micro-caustics when $\Sigma_{\rm eff}$ increases.
Log-normal distributions are common in nature, in particular in multiplicative processes that transform into small additive increments in log-scale. The superposition of many microcaustics mimics a multiplicative process in which each microcaustic multiplies the flux of the background source by some amount.  \\
An increase in $\Sigma_{\rm eff}$ can be attributed to two factors: either an increase in $\mu_{\rm t}$ or in $\Sigma_{\ast}$. 
At higher macro-magnifications, the lens plane is mapped onto smaller regions in the source plane, so the probability of caustics overlapping increases.
To better understand the log-normal behaviour and the saturation regime, it is useful to understand how the micro-caustics are shaped and how increasing $\Sigma_{\ast}$ and $\mu_{\rm m}$ affects in the probability of overlap.
In their paper, \cite{Oguri2018} derive an analytical approximation for the shape of a micro-caustic by solving the equation where the inverse of the magnification equals zero.
They determined the sizes of a micro-caustic located near the tangential CC, in the positive regime, in both the tangential and radial directions to be
\begin{align}
    &\beta_{\rm {caus, t}} \approx \frac{\sqrt{\mu_{\rm t}}}{\mu_{\rm r}}\theta_{\rm E},\nonumber \\
    &\beta_{\rm {caus, r}} \approx \frac{\theta_{\rm E}}{\sqrt{\mu_{\rm r}}},
\end{align}
respectively. For the negative parity the caustic size is similar but reduced by a factor of $2\sqrt{2}$ in the tangential direction, and it exhibits a demagnification region in the centre of the caustics.
$\theta_{\rm E}$ is the Einstein radius of the lens defined as
\begin{equation}
    \theta_{\rm E} = \left(\frac{4GM}{c^2}\frac{D_{\rm ds}}{D_{\rm s}D_{\rm d}}\right).
\end{equation}
A similar solution for the sizes of the micro-CCs was also obtained by \citep{Diego2018}.

Increases in $\mu_{\rm m}$ result in larger caustics concentrated in smaller regions, while larger $\Sigma_{\ast}$ results from more microlenses, more massive microlenses resulting in larger micro-caustics, or a combination of both effects.
These aforementioned effects promote the overlapping of micro-caustics, particularly at the tips of the caustics where higher magnification is concentrated.
Consequently, these regions are the first to experience distortion as a result of overlapping.
This phenomenon explains why the more is less effect becomes prominent at larger values of $\Sigma_{\rm eff}$.
The log-normal behaviour of the magnification probability can be understood as the manifestation of the central limit theorem in the logarithmic domain, arising from the overlapping of caustics.

\begin{figure}[tpb]
  \centering
  \includegraphics[width=\columnwidth]{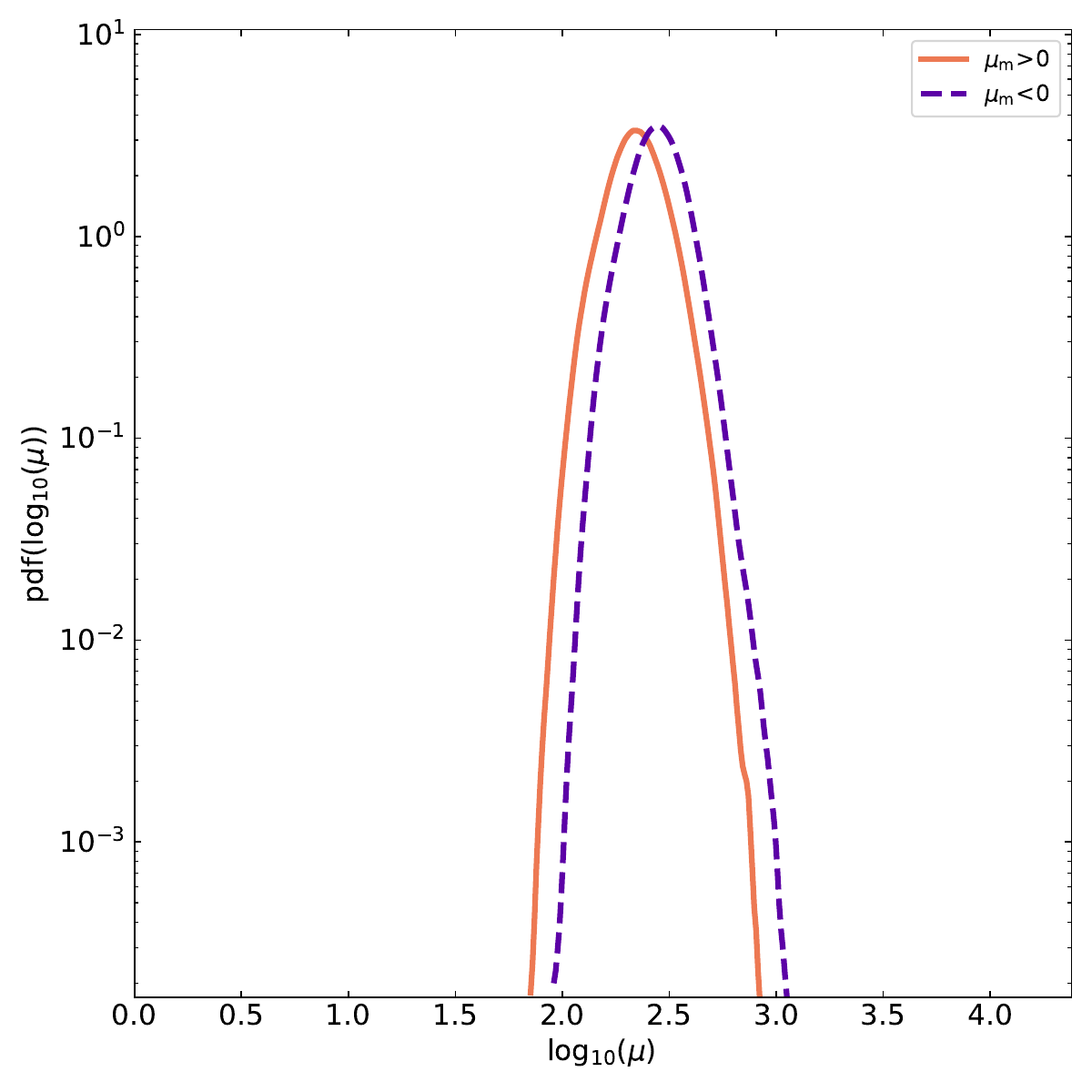}
  \caption[]{\label{fig:pos_neg_pdf_saturated} %
  Probability of magnification for two simulations with the same $|\mu_{\rm t}|$, $\mu_{\rm r}$, $\Sigma_{\ast}$ and different parities in the saturation regime $\Sigma_{\rm eff}/\Sigma_{\rm crit}\!=\!66.5$.
  }
\end{figure}

Fig. (\ref{fig:pos_neg_pdf_saturated}) depicts the presence of a saturation regime for $\Sigma_{\rm eff}/\Sigma_{\rm crit}\!\gtrsim\!30$ for the positive parity and $\Sigma_{\rm eff}/\Sigma_{\rm crit}\!\gtrsim\!25$ for the negative parity.
Within this regime, the magnification probability follows a log-normal distribution, with a shift in its central value as $\Sigma_{\rm eff}$ increases (the changes in width and amplitude are almost negligible).
It should be noted that the centre of the log-normal distribution requires subsequent correction, taking into account the value of $\mu_{\rm m}$.
At the highest simulated effective surface mass densities, both parities exhibit the same log-normal distribution shifted towards higher magnifications for the negative parity. This can be observed in Fig. (\ref{fig:pdfs_huge_sigma}).

\subsection{Constraints on PBHs}
In Section (\ref{sec:constrains}), we considered a hypothetical scenario where the microlenses, consisting of 25\% stars and 75\% primordial black holes (PBHs), constituted 1\% of the system's critical density.
The macro-model predicted a magnification of $\mu_{\rm m}\!=\!200$, resulting in an effective surface mass density twice the critical density value.
In this setup, background sources would undergo microlensing events following the probability distributions illustrated in Fig. (\ref{fig:analytic_pdf_2sigma_crit}).
We would observe these events when the product of the intrinsic luminosity of the source and the magnification ($L\mu$) exceeds the detection threshold of our telescopes.

The HST and the JWST have already observed numerous microlensing events, and the number of such observations will continue to increase over time.
Continuous monitoring of arcs located in the outskirts of large galaxy clusters, will eventually accumulate sufficient data to perform statistical inference on the surface mass density of microlenses in these regions.
Additionally, photometric measurements of the ICL in the same area will allow us to estimate the contribution of stars to the population of microlenses.
Our current methodology provides a rapid estimation of the magnification probability, which can be compared against observational data.
This allows us to determine the most likely value of $\Sigma_{\ast}$ based on the experimental results.
Moreover, since $\Sigma_{\rm b}$ can be measured independently, the dark matter surface mass density ($\Sigma_{\rm DM}$) can be obtained simply by taking the difference between the inferred total density (from microlensing flux fluctuations) and the contribution from baryonic matter (from SED fitting of the observed photometry).

The data release by the LIGO-Virgo-KAGRA collaboration in 2016 \citep{Abbott2016} has brought renewed interest in the in primordial black holes as a potential candidate for dark matter.
While observational searches for the extra-galactic $\gamma$-ray background \citep{Page1976}, have ruled out the mass range of PBHs with initial masses $M_{\rm PBH}\!\lesssim\!10^{-17}M_\sun$, CMB and dynamical constraints have ruled out massive PBHs ($M_{\rm PBH}\!\gtrsim\!100~M_\sun$) \citep{Serpico2020}\citep{Lu2021}.
The remaining window of PBH masses, where they could potentially constitute a significant fraction of the DM content in the Universe, is an elusive range where only gravitational interactions play a significant role in possible observations.

Several methods have been suggested to constrain the abundance of PBHs in this open window (see \cite{Carr2021} and \cite{Green2021} for an extended review), in particular gravitational microlensing.
The EROS-2 and MACHO surveys \citep{Tisserand2007}\citep{Alcock2000}, which focused on microlensing observations of stars in the Magellanic Clouds, and the OGLE survey \citep{Niikura2019b}, which conducted microlensing studies of the galactic bulge, have provided valuable constraints on the mass range of primordial black holes (PBHs) within the Milky Way. 
These studies have placed constraints on the presence of PBHs in the mass range of $10^{-6}\!\lesssim\!M_{\rm PBH}\!\lesssim\!10~M_\sun$ at the (1\,-\,0.1)\% level around 1$M_\sun$.
The Subaru Hyper Suprime-Cam has played a crucial role in further constraining the mass range of primordial black holes within the Milky Way and the Andromeda galaxy (M31) \citep{Niikura2019a}.
In particular, the Subaru Hyper Suprime-Cam has extended the constraints to even lower PBH masses, specifically for $M_{\rm PBH}\!\gtrsim\!10^{-11}~M_\sun$, through the microlensing observations of stars from M31.

Quasar microlensing involves a similar mechanism to extreme magnification events, where compact objects within a galaxy act as lenses that magnify the flux of background quasars images, resulting in flux changes in these images.
\cite{Mediavilla2017} used optical and X-ray data from 24 lensed quasars.
The results indicated that $\sim20\%$ of the mass of the lens galaxies consisted of compact objects within the mass range of $0.05~M_\sun\!\gtrsim\!M\!\gtrsim\!0.45~M_\sun$.
Similarly to quasar microlensing, extremely lensed events offer a novel approach to constraining the amount of compact dark matter present in galaxy clusters.
In 2018 \cite{Oguri2018} utilised the Icarus data \citep{Kelly2018} to constrain the abundance of compact DM in the presence of the intracluster light.

By combining the results obtained from the aforementioned studies and incorporating future data collected by space telescopes such as the JWST, it will be possible to establish new constraints on the fraction of dark matter composed of PBHs, potentially even setting lower limits on their abundance.
These ongoing investigations hold promise for advancing our understanding of the nature and contribution of PBHs to the overall dark matter content of the Universe.

\subsection{Early stars, GWs, and high redshift supernovas}
The analytic approximation of the magnification probability presented in this paper not only provides a valuable method for constraining compact dark matter, but it also opens up possibilities for constraining the abundances of lensed sources at higher redshifts.
In a recent study \cite{Diego2023} quantified the probability of observing a supergiant star due to the crossing of a micro-caustic, which amplifies the star's intrinsic luminosity above the detection threshold of the HST.
To achieve this, the authors employed both a macro-model and a micro-model of the lensing system, along with the characterisation of the background galaxy, the Spock arc.
This, and previous investigations relied on computationally expensive numerical simulations to account for the variability of the macro-model magnification and the contribution of microlenses within the arc.
However, with the development of our new tool, these types of analyses can be significantly accelerated, providing faster and more efficient estimations.
In their study, \cite{Diego2023} also presented forecasts for the constraints on these stars using the JWST.
By incorporating the improved magnification probability obtained from our analytic approach, future observations with the JWST will enable more precise constraints on the presence and characteristics of lensed supergiant stars.

The rate of extreme magnification events depends on both the number of background sources and the probability of magnification that produces detectable events.
Since we can now infer the later, by observing the number of events over a certain period of time, we can extract information about the abundance of the background sources.

Among the sources whose combined intrinsic luminosity and maximum magnifications achievable can lead to extreme magnification events, we discuss the following: high redshift stars, high redshift supernovae (SNe), gravitational waves (GWs) from binary black hole systems, and population III stars.

The discussion of microlensing effects on quasars (QSOs) will not be addressed here, as they have already been extensively investigated, and their typical sizes surpass those of the previously described sources, thereby limiting the maximum magnification to the order of $\mathcal{O}(10)$.
However, it is worth noting that certain regions within QSOs exhibit smaller sizes, which can enhance the maximum magnification.
The detection of such variations may be possible through multi-wavelength studies of these objects.

\subsubsection{Luminous stars at cosmological distances}
Although the probability of extreme magnification events for individual stars may be low, the vast number of stars in the Universe makes them ideal candidates to undergo such events.

The probability of a source intersecting a micro-caustic and experiencing extreme magnification increases with the number of background sources available.
When considering cosmological distances ($z\!>\!1$), it becomes challenging to detect stars unless they are particularly luminous.
Telescopes such as the HST and JWST are capable of observing the most massive and bright stars at these distances.

In general, the mass-luminosity relation for main sequence stars follows a relationship, where the luminosity $L$ is correlated with the stellar mass $M$ as follows:
\begin{equation}
    \frac{L}{L_\sun} = A\left(\frac{M}{M_\sun}\right)^\alpha,
\end{equation}
where the slope $\alpha$ typically ranges from 3 to 4, with slightly lower values observed for the most massive stars.
This mass-luminosity relation indicates that the most massive stars are capable of attaining the highest levels of luminosity within the main sequence.
Stars outside the main sequence exhibit different mass-luminosity relations.
However, the population of stars outside the main sequence is generally smaller in number compared to main sequence stars.
Therefore, they are expected to contribute to a lower rate of extremely lensed events.

Smaller stars have a higher maximum magnification due to their smaller radius, but their intrinsic luminosity is also lower compared to larger stars.
The luminosity of a star is generally correlated with its radius, and the exponent $a$ in the scaling relation $L\!\propto\!R^a$ is typically greater than 1.
As a result, the maximum luminosity $L_{\rm max}$ scales with the radius as $L_{\rm max}\!\propto\!R^{a-0.5}$.
This implies that larger stars are more likely to be observed in extreme magnification events because their intrinsic luminosity exceeds the decrease in maximum magnification with respect to smaller stars.
Therefore, as a general rule, larger and more massive stars are more commonly observed in caustic crossing events due to their higher intrinsic luminosity.

\subsubsection{Population III stars}
The existence and properties of Pop. III stars \citep{Bromm2001} are of great interest in astrophysics and cosmology.
These hypothetical stars, which are believed to have formed in the early universe and have very low or negligible metallicity, could provide crucial insights into the early stages of cosmic evolution.
In their paper \cite{Windhorst2018} discuss the luminosity-mass relation for Pop III stars as well as a similar population called "Pop II.5" stars, which have a modest metallicity.
They found that this relation is close to the Eddington limit, where the luminosity is proportional to the mass ($L\!\propto\!M$).

Due to their short lifespan, Pop. III stars are primarily expected to exist at high redshifts ($z\!\gtrsim\!6-7$).
The study of caustic crossing events with the JWST may lead to the first direct observations of Pop. III stars and significantly advance our understanding of the early Universe, including the re-ionization process and the formation of the first galaxies.

\subsubsection{GWs from binary black holes}
Gravitational waves can also experience deflections in their paths as they propagate through regions with large distributions of matter.
This gravitational lensing effect on GWs can lead to magnification or demagnification of the waves, similar to the lensing effect on light.

\cite{Broadhurst2018} proposes that a subset of the low-frequency GWs detected by the LIGO-Virgo collaboration could be attributed to GWs originating from sources at redshifts greater than one ($z\!>\!1$) and magnified by a factor greater $\mu\!>\!100$.
These magnified GWs are hypothesised to arise from binary black hole systems with chirp masses around $30~M_\sun$.
At these magnifications microlensing events will only be important for small sources, but the large wavelengths of the GWs suggest that wave optics might need to be considered.

\subsubsection{High redshift SNe}
SNe are extremely compact and luminous objects prone to microlensing effects.
If the progenitor is located within a few astronomical units (AU) from a micro-caustic, the expansion of its photosphere can lead to extreme magnification increasing over time (Typical rate of expansion $\sim$6 AU/day).
As the photosphere expands, it can intersect the micro-caustic, resulting in a maximum increase in the observed brightness of the SNe. 
However, as the photosphere continues to expand and move away from the micro-caustic, the magnification gradually decreases.
This effect will leave an in-print on the light curve of the SNe.

All SNe can be observed through this mechanism, but we will focus on type Ia SNe.
These SNe are extremely important in the field of cosmology, as they serve as standard candles for measuring astronomical distances.
The origin of the initial Type Ia SNe could potentially be traced back to the initial generation of white dwarfs.
Consequently, their abundance at high redshifts ($z\!\gtrsim\!6$) remains uncertain, given the time required for the formation of binary systems. 
The JWST has the capability to provide data of Type Ia SNe at redshifts ranging from $z\!=\!6$ to $12$, which would considerably augment our understanding of the abundance of these entities as well as their properties at high redshifts.

See \cite{Diego2019} for a more extensive analysis on the methodologies employed for measuring these phenomena and their corresponding likelihood of occurrence.

%% file: sections/conclusion.tex
Giant arcs intersecting a cluster's-CC provides an exceptional scenario for caustic crossing events to arise.
If small and bright sources happen to lay close to the cluster caustic, they will undergo extreme magnification boosting their fluxes by factors of hundreds or even thousands. This magnification can momentarily be even larger when the star crosses a microcaustic, providing a boost of $\mathcal{O}$(10) magnitudes and making a luminous star at redshift $z>1$ detectable with current telescopes. 
This new branch of gravitational lensing opens the door to answer some questions about the nature and abundances of the magnified sources, providing a natural telescope to detect them.
Moreover, monitoring these arcs will allow to inferred the characteristics of the underlying population of microlenses, constraining they surface mass density and possibly their nature. This is particularly interesting in the context of dark matter, where certain DM candidates, such as primordial black holes, can contribute to the microlensing signal.

This type of analyses require to know the probability distribution function of the magnification.
Previous works have relied on expensive numerical simulations, with one simulation per set of parameters to be tested.
Furthermore, accurate approximations of the magnification probability require large areas $A$ in the source plane (increased by a factor $\mu_{\rm m}$ in the lens plane) and small pixel scales.
These three effects combined result in thousands of CPU-hours per simulation, making it extremely difficult to carry out such analysis.
We have developed an approximation tool that enables a straightforward and efficient estimation of magnification probabilities by considering the scaling properties associated with the lens parameters.
To develop our magnification probability approximation tool, we conducted an extensive series of simulations on microlensing near macro-CCs, involving approximately 0.5 million CPU-hours.
These simulations explored the parameter space for both, macro- and micro-models.
We obtain the magnification probability distributions and subsequently derived analytical models to represent these distributions.
We derive the scaling relations between the model parameters and physical parameters such as the effective surface mass density of microlenses. These scaling relations allow us to quickly compute the probability of magnification in a wide range of scenarios.
As a result, we can now determine the magnification probability directly from the system's physical parameters, eliminating the need for extensive simulations, offering a viable alternative to time-consuming numerical simulations.
This analytical approximation can be obtained in less than one second, thereby enabling the aforementioned studies. 

During our modelling process, we observe that in extreme scenarios where $\Sigma_{\rm eff}\gg 1$ the magnification probability deviates from the well-known power-law distribution with an index of -3 (-2 when considering probabilities in logarithmic bins).
Instead, it exhibits a log-normal behaviour when the micro-caustics overlap in the source plane.
This effect becomes prominent when either the surface mass density of the microlenses increases or when approaching the cluster-CC as the microlens magnification, $\mu_{\rm m}$, grows.
Interestingly, we discover a degeneracy in the magnification probability between the surface mass density of microlenses and the tangential magnification.
Many parameters scale in terms of the effective surface mass density, $\Sigma_{\rm eff}=\Sigma_{\ast} \times \mu_{\rm t}$.
The aforementioned degeneracy can be broken by the radial magnification, denoted as $\mu_{\rm r}$.
%As a result, our calculations need to be corrected by a factor of $\mu_{\rm m}/1000$.
In particular, we find that two simulations with the same $\Sigma_{\rm eff}$ but different $\Sigma_{\ast}$ and $\mu_{\rm t}$ will have the same shape but will be shifted relative to each other.

Our simulations confirm the more is less effect discussed by \cite{Welch2022a}, as well as the concept of the effective microlens mass, $\mu_{\rm t}M$, mentioned by \cite{Diego2018}.

By delving into the study of the negative parity case, which has often been overlooked in the literature, we have made intriguing observations.
We have found that this regime exhibits similarities to the positive parity regime, but with two distinct differences:
1) A demagnification effect emerges relative to the most likely magnification value.
This probability of demagnification follows a negative power-law scaling ($\mu^{-1/2}$), which can even be observed for isolated lenses.
Additionally, there is a probability excess at the lower end of the magnification distribution, and its position is solely determined by the radial magnification, $\mu_{\rm r}$;
2) Although the probability of magnification at the most likely factor is smaller compared to its positive parity counterpart, the probability at the largest magnification values is slightly higher in the negative parity regime.

We have identified two distinct regimes based on the effective surface mass density.
When $\Sigma_{\rm eff}$ is smaller than the critical value of the system, the magnification probability follows a similar model as that of isolated lenses near macro-CCs.
The only additional component is the power-law transition to a log-normal distribution at the high magnification end of the probability distribution.
When $\Sigma_{\rm eff}$ is comparable to or significantly greater than its critical value, the magnification probability functions are modelled as the sum of two log-normal distributions for the positive parity case.
In this regime, the negative parity exhibits a power-law demagnification but no longer follows a fixed slope as observed in the low-$\Sigma$ regime.
As the effective surface mass density increases, the contribution of the power-law demagnification becomes negligible.
In the negative parity regime, when $3<\Sigma_{\rm eff}/\Sigma_{\rm crit}<9$, the magnification probability is effectively modelled by three log-normal distributions.

When the effective surface mass density exceeds a certain threshold of approximately 30-40 times the critical surface mass density the magnification probability enters the saturation regime.
In this regime, the probability distribution is predominantly described by a single log-normal distribution.
Interestingly, for values of $\Sigma_{\rm eff}/\Sigma_{\rm crit}$ above approximately 55, the parameters of the log-normal distribution coincide for both the positive and negative parity regimes.
However, there is a slight shift towards higher magnifications in the negative parity side of the main-CC.
At this point the perturbation from the microlenses have completely erased the main-CC and both parities are virtually indistinguishable.

%% file: appendixes/resolution_effects.tex
The inverse ray shooting technique, despite its effectiveness, is not without limitations.
In this appendix, we address the impact of resolution effects on our estimation of the magnification probability.
We found three key factors that have an impact on the derived magnification probability from the simulations. 
These factors include the pixel size, the masses of the microlenses, and the limited statistics at the lowest magnification factors.

\subsection{Pixel size}
\begin{figure}[tpb]
  \centering
  \includegraphics[width=.8\columnwidth]{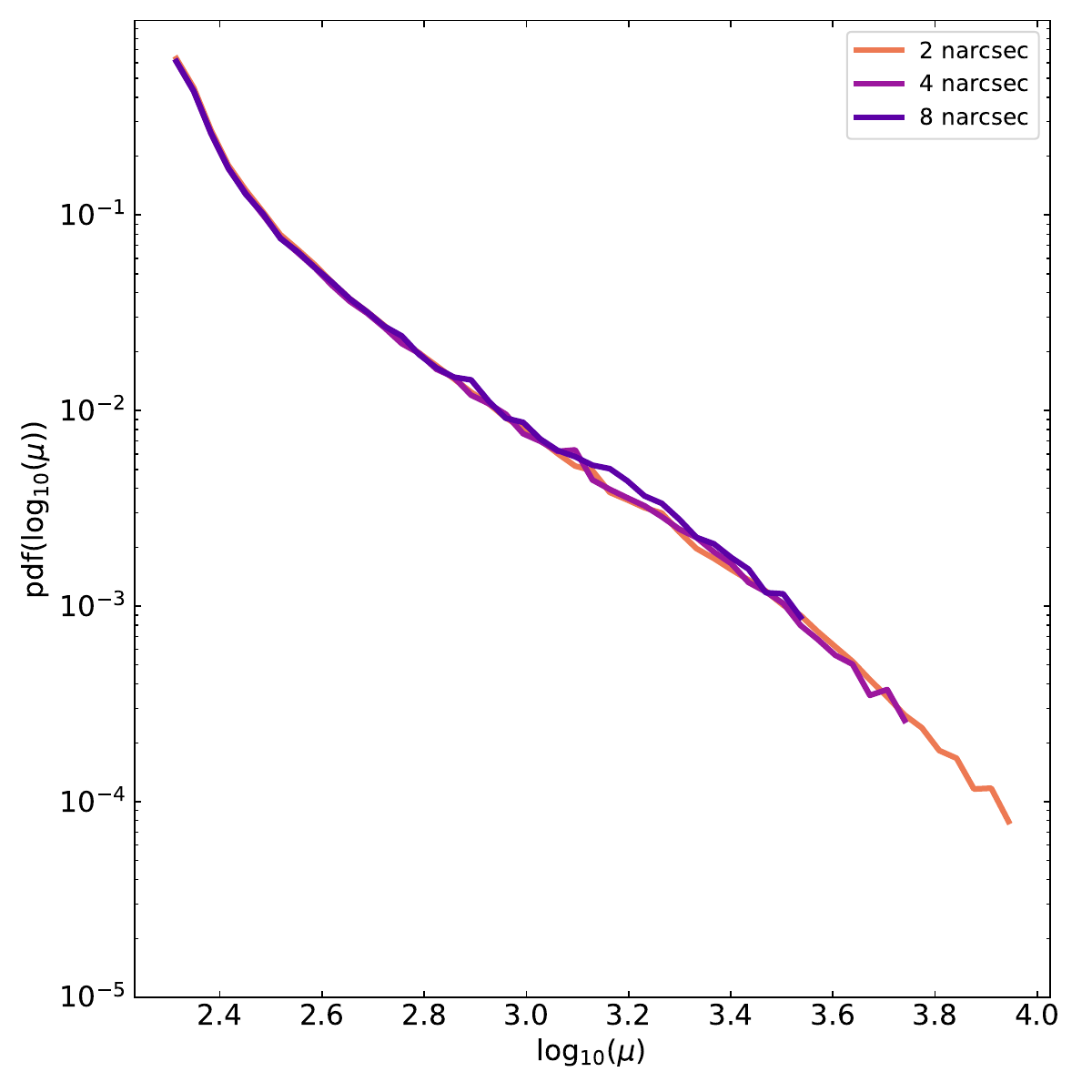}
  \caption[]{\label{fig:pixel_size_comp} %
  Magnification probability approximation for an inverse ray tracing simulation of an isolated microlens of 1$M_\sun$ near a macro CC.
  Each curve represents a different pixel size.
  Larger pixels reach smaller maximum magnifications.
  }
\end{figure}
When the pixel size in our simulations is increased, it imposes limitations on accurately representing the intricate structures within the micro-caustics web.
A larger pixel size mimics a larger source, which affects the maximum magnification experienced.
As discussed earlier, the maximum magnification is inversely proportional to the source size ($\mu_{\rm max}\!\propto\!1/\sqrt{R}$), meaning larger sources experience lower maximum magnifications.
Most of the peak magnification is mostly produced by the narrow tips of the micro-caustics.
With larger pixels, these specific regions cannot be properly sampled, smoothing out the extreme magnification values near the caustics. 
Consequently, the magnification values provided by larger pixels tend to reflect an average rather than capturing the enhanced magnification in the localised tips of the micro-caustics.
We expect to observe a decrease in the maximum magnification obtained in our simulations as the pixel size increases.

To quantitatively evaluate this effect, we conducted multiple simulations using identical parameters while varying only the pixel size.
By obtaining the approximated magnification probability for each simulation, we can compare the results at the high magnification end to analyse the impact of the pixel size.
In our study, we focused on simulating an isolated microlens that follows the well-established power-law behaviour, $A{>\mu} \propto \mu^{-2}$.
This choice allowed us to investigate the effect of pixel size and its influence on the resulting magnification probability.

Fig. (\ref{fig:pixel_size_comp}) illustrates the magnification probability obtained for three distinct pixel sizes.
It is evident that smaller pixel sizes yield better approximations of the magnification probability, particularly at higher magnifications. For our purposes, the most relevant finding is the fact that increasing the pixel size reduces the maximum magnification while maintaining the power-law behaviour up to the maximum magnification. This means that the power-law cutoff observed at extreme magnification values when $\Sigma_{\rm eff}$ approaches 1 is not a resolution effect, but a real property of the probability of magnification. \\

It is important to consider the computational time required for numerical simulations, which scales as $\mathcal{O}(N^2)$, where $N$ represents the number of pixels in one of the dimensions of the simulated region.
Reducing the pixel size by a factor of 1/2 increases the computational time by a factor of 4.
Therefore, a trade-off exists between the desired simulation time and the accuracy of the magnification probability.
After careful analysis, we have determined that a pixel size of 2~nas is well-suited for our specific task, of microlenses at cosmological distances and in realistic scenarios, striking a balance between computational efficiency and the required level of accuracy for the magnification estimates.

\subsection{Mass of the microlenses}
\begin{figure}[tpb]
  \centering
  \includegraphics[width=.8\columnwidth]{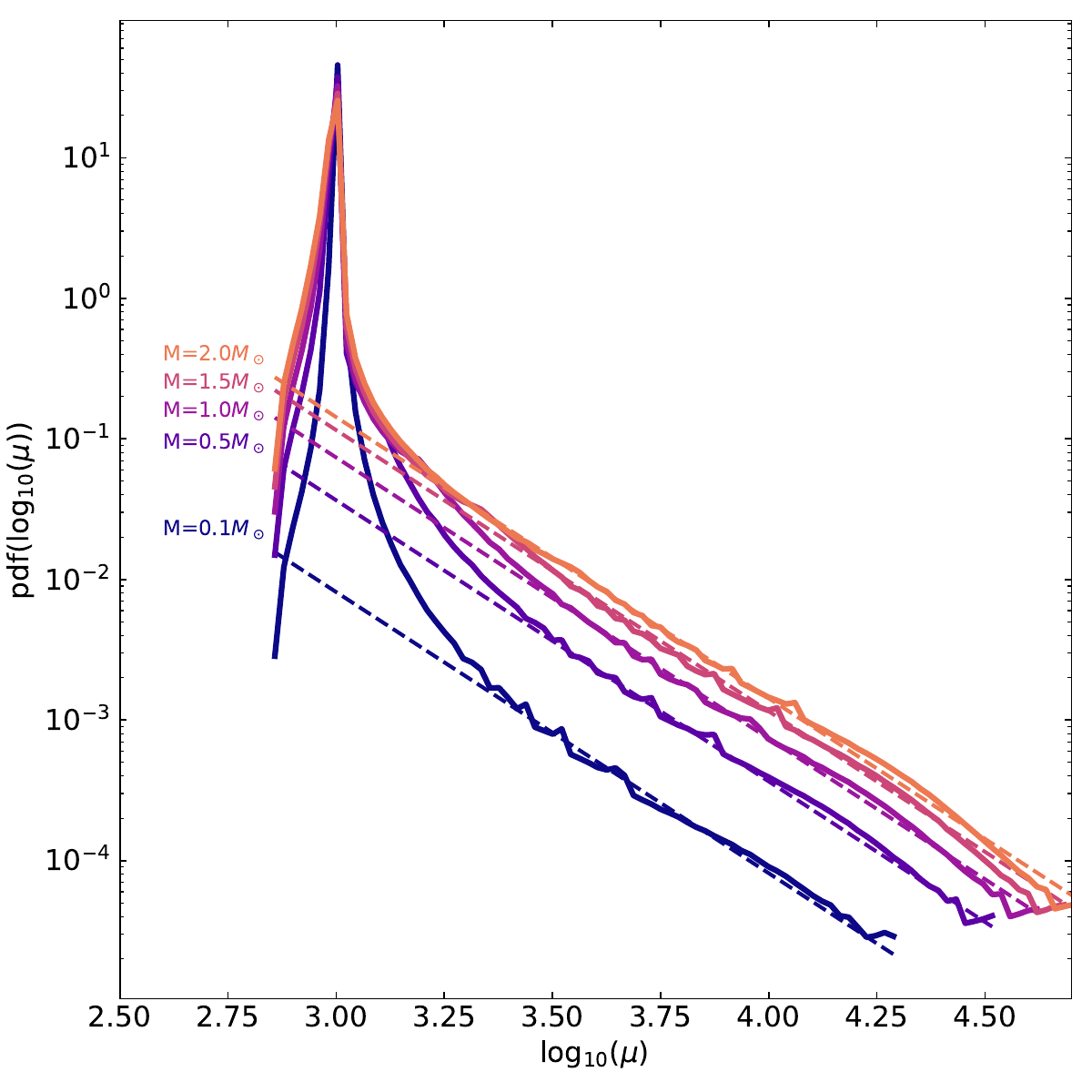}
  \caption[]{\label{fig:mass_reso} %
  Magnification probability approximation for an inverse ray shooting simulation of an isolated microlens near a macro-CC.
  Each curve represents a different microlens mass.
  For a fixed pixel size (2 nas), the maximum magnification we can measure increases with he mass of the microlens.
  The dashed lines represents the $\mu^{-2}$ power-law that govern the magnification probability at high magnification factor in this regime.
  The amplitude of this power-law increases cuadratically as seen in Sec. (\ref{sec:1lens}).
  }
\end{figure}
An additional resolution effect arises from the mass of the microlens itself.
Smaller microlenses exhibit smaller micro-caustics, as the size of the caustics scales linearly with the Einstein angle ($\theta_{\rm E}$), which is proportional to the mass ($M$) of the lens.
Consequently, lighter lenses have smaller caustics, and for a fixed pixel size, the maximum magnification achievable from these lenses is reduced.
In our simulations, we consider microlens masses obtained from a distribution derived by \cite{Spera2015}, which mimics an old stellar population with masses not exceeding 1.5 $M_\odot$, while remnants and ordinary stars are not lighter than 0.1 $M_\odot$.
Our primary focus is on the smallest lenses within this distribution.
Similar to our previous analysis, we simulate isolated masses ranging from 0.1 to 2 $M_\sun$ and analyse their respective magnification probabilities.
The results, presented in Fig. (\ref{fig:mass_reso}), confirm our expectations, with larger masses exhibiting higher maximum magnification values.
However, it is worth noting that even in the case of the smallest lenses from our simulations, this resolution effect does not significantly impact our findings.
The figure shows the scaling discussed in the main paper, where the amplitude of the power-law scales cuadratically with the mass of the microlens.

\subsection{Magnification approximation at the lowest values}
The inverse ray shooting technique employed in our study approximates the magnification in the source plane pixels based on the number of photons traced from the lens plane.
By utilising the deflection angle in the lens plane and solving the lens equation, we can determine the corresponding position in the source plane for each position in the lens plane.
This approach yields accurate results when the lens plane has a sufficiently large area and good resolution.
However, a limitation of this method arises from the fact that we can only obtain integer values for the magnification, while in reality, the magnification can be any real number.
Therefore, this technique effectively rounds the magnification value to the closest integer.
This limitation does not significantly impact our results when the magnification values differ by large factors (e.g., $10^{4}$ versus $10^{4}+1$).
However, it becomes more significant when the magnification values are of order 1, as the rounding leads to a noisier estimate of the probability of magnification.

In our positive parity simulations, the magnifications fall within the range of $\gtrsim\!\mu_{\rm m}$.
When the effective surface density increases, there is a slight reduction in this range.
However, since $\mu_{\rm m}$ is typically large, the minimum magnifications in our observations are not affected by this effect.

In simulations with negative parity, a demagnification effect relative to $\mu_{\rm m}$ appears.
This demagnification exhibits also a power-law behaviour, and its parameters can be constrained by studying larger magnifications where this effect is negligible.
Additionally, a distinctive feature in the form of a skewed log-normal bump emerges in the magnification distribution at the lowest magnification values.
The position of this bump decreases as $\mu_{\rm r}$ (as shown in Fig. (\ref{fig:scale_low_neg_extra_1})), and it is influenced by this demagnification effect.
It is important to note that characterising these peaks in the magnification distribution is challenging, and our accuracy in this regard may not be as precise as in other parts of the probability density functions.
Nevertheless, our fitting procedure successfully captures the main elements of this portion of the model.
In the super-critical simulations with negative parity, the combination of the demagnification effect and reduced statistics due to a smaller area in the source plane (scaled by a factor of $\mu_{\rm m}$, which is larger for these simulations) has hindered our ability to accurately model the low magnification excess observed in these specific cases.
Despite this limitation, we have observed an interesting trend: the fixed position of these peaks begins to shift towards larger magnification values as $\Sigma_{\rm eff}$ increases.

%% file: appendixes/peaks.tex
\begin{figure*}[tpb]
\centering
   \includegraphics[width=17cm]{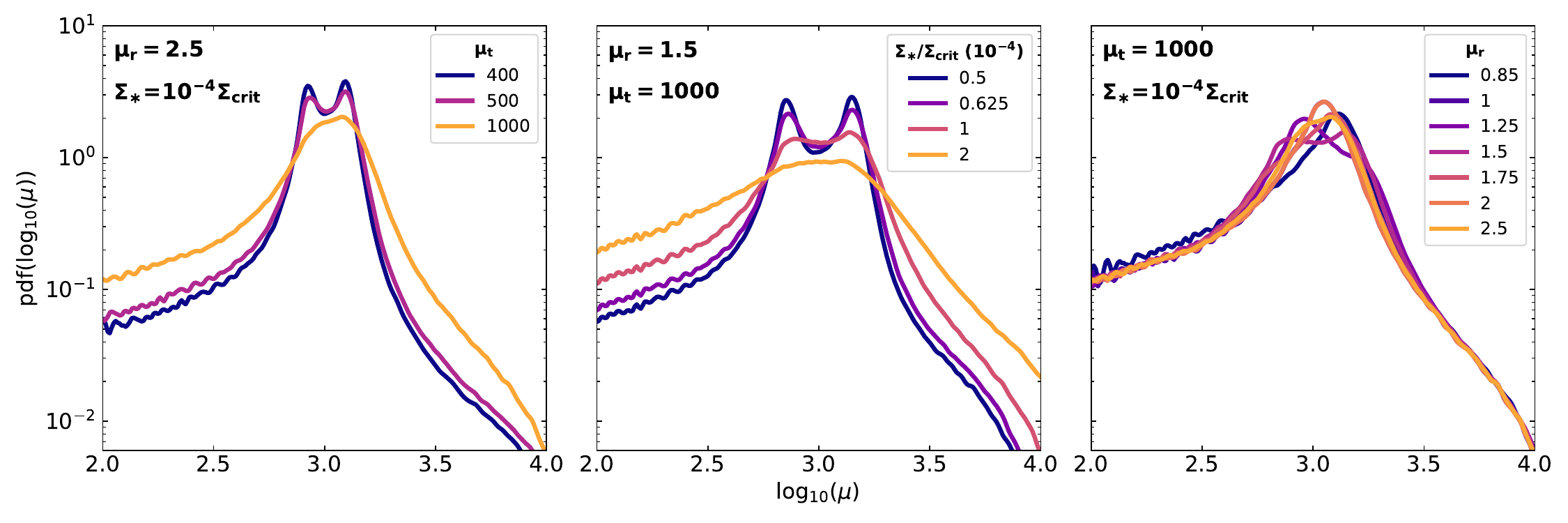}
     \caption{
     Zoom in on the magnification probability around the mode.
     We show the substructure of this peak and its changes as we fixed two parameters and allowed the remaining one to change.
     $\mu_{\rm t}$ ({\em left panel\/}) controls the position of each peak, while its width is only scaled with $\Sigma_{\ast}$ ({\em middle panel\/}).
     The ratio of peak amplitudes depends on $\mu_{\rm r}$ ({\em Right panel\/}).
     }
     \label{fig:peaks_substructure} %
\end{figure*}
In our simulations, we have discovered an intriguing feature characterised by the emergence of substructure around the mode of the magnification probability.
This effect is observed primarily at low values of $\Sigma_{\rm eff}$, and does not significantly impact the probability distribution at the largest magnifications, which is the driving force our current study.
However, it is important to note that this substructure may have implications in other cases or scenarios.
While we do not provide a detailed modelling of this substructure in the present paper, we offer a brief qualitative description.

In contrast to the previous description of a single peak characterised by a log-normal distribution, our recent observations reveal the presence of two distinct peaks around the expected value $\mu_{\rm m}$.
Moreover, the width, amplitude, and centre of these newly discovered peaks do not exhibit a clear scaling relationship with $\Sigma_{\rm eff}$ as observed in previous analyses, although we observe that this effect manifests itself only when $\Sigma_{\rm eff} < 0.3$,

To gain a better understanding of the observed substructure in the magnification probability, we conducted a series of simulations where two out of the three main model parameters, namely $\mu_{\rm t}$, $\mu_{\rm r}$, and $\Sigma_{\ast}$, were fixed while varying the third parameter.
The results of these simulations are presented in Fig. (\ref{fig:peaks_substructure}).

Our main findings can be summarised as follows:
\begin{itemize}
    \item The parameter $\mu_{\rm t}$ plays a crucial role in determining the positions of the peaks in the magnification probability.
    Increasing $\mu_{\rm t}$ leads to a shift of both peaks towards each other.
    However, the lower magnification peak experiences a larger shift compared to the other peak.
    The ratio between the heights or amplitudes of the peaks remains constant, although the amplitude of both peaks decreases as a result of the increased $\Sigma_{\rm eff}$.

    \item The parameter $\Sigma_{\ast}$ is correlated with the width of the peaks.
    Larger $\Sigma_{\rm eff}$ values result in broader peaks.
    The centres of the peaks remain constant, and the ratio of amplitudes also remains constant.
    Similar to the case of varying $\mu_{\rm t}$, the increase in $\Sigma_{\rm eff}$ leads to a reduction in the amplitude of both peaks.

    \item The ratio of amplitudes is controlled by the parameter $\mu_{\rm r}$.
    For lower radial magnifications, the peak with higher magnification dominates.
    The ratio between the amplitude of this peak and its lower magnification counterpart decreases as $\mu_{\rm r}$ increases.
    At $\mu_{\rm r}\approx1.5$, the ratio is approximately 1.
    For radial magnifications larger than 2.5, only one peak is observed.
    The position of the peaks may also be slightly influenced by $\mu_{\rm r}$.
\end{itemize}

The observed substructure in the magnification probability manifests itself primarily for low radial magnifications and low $\Sigma_{\rm eff}$ values, typically on the order of $\lesssim\!10\%$ of the critical surface density.
In our simulations, where the critical surface density is approximately $\Sigma_{\rm crit}\!\approx\!3100~M_\sun/\rm{pc}^2$, we found that this effect is at it highest for $\mu_{\rm t}\!=\!1000$, $\Sigma_{\ast}\!=\!0.31~M_\sun/\rm{pc}^2$, and $\mu_{\rm r}\!=\!1.5$.
For the sake of simplicity, we have presented the observed substructure effect only in the context of negative parity simulations.
However, it is important to note that we have also observed this effect in simulations with both positive and negative parity.

To the best of our knowledge, the substructure effect that we have observed in our simulations represents a newly discovered phenomenon whose modelling is left for future work.